\documentclass[11pt]{article}
\usepackage{epsfig}
\newcommand{\sect}[1]{ \section{#1} \setcounter{equation}{0} }

\newcommand{\pslash}{p \! \! \! /} 
\newcommand{\qslash}{q \! \! \! /}

\newcommand{\MSbar}{\overline{\mbox{MS}}}

\newcommand{\Nf}{N_{\!f}}

\begin{document}
\title{Off-shell two loop QCD vertices}
\author{J.A. Gracey, \\ Theoretical Physics Division, \\ 
Department of Mathematical Sciences, \\ University of Liverpool, \\ P.O. Box 
147, \\ Liverpool, \\ L69 3BX, \\ United Kingdom.} 
\date{}
\maketitle 

\vspace{5cm} 
\noindent 
{\bf Abstract.} We calculate the triple gluon, ghost-gluon and quark-gluon
vertex functions at two loops in the $\MSbar$ scheme in the chiral limit for an
arbitrary linear covariant gauge when the external legs are all off-shell.

\vspace{-14.5cm}
\hspace{13cm}
{\bf LTH 1011}

\newpage 

\sect{Introduction.}

In recent years there has been renewed interest in understanding the vertex
functions of Quantum Chromodynamics (QCD) which is the non-abelian gauge theory
describing the strong nuclear force. The main techniques used to study the 
vertex functions are lattice gauge theory and the Dyson-Schwinger method. In 
general they both aim to probe the non-perturbative structure of the $3$-point 
functions of QCD in order to gain insight into the properties of colour and 
quark confinement, dynamical symmetry breaking and the formation of bound 
states. For instance, see \cite{1,2,3} for reviews of Dyson-Schwinger 
techniques in this area. Such knowledge of the vertex functions can also inform
the construction of models which explore the hadronization of quarks into the 
bound states seen in nature, \cite{1}. In this respect providing analytic
perturbative information on the triple gluon, quark-gluon and ghost-gluon 
vertices is important. There have been several one loop perturbative studies of 
the $3$-point functions over several decades in a selection of momentum 
configurations of the external fields. Several of the original key articles in 
this instance are \cite{4,5}. In the early part of the vertex function 
programme Ball and Chiu, \cite{5}, examined the one loop off-shell functions 
for the trivalent vertices in the $\MSbar$ scheme with the aim of accessing the
singularity structure. Indeed \cite{5} provided an insight into properties of 
the vertices when the gauge fields were transverse. Moreover, the basis of 
tensors presented in \cite{5} with respect to which the vertices were 
decomposed has allowed Schwinger-Dyson studies to access certain properties of 
the vertices. For instance, these have produced certain ans\"{a}tze for the all
orders vertex functions which are used as input structures for the solution of 
the tower of the Schwinger-Dyson $n$-point functions. In certain cases results 
have emerged which match very closely with results from lattice gauge theory. 
One such example is the deep infrared behaviour of the gluon and Faddeev-Popov 
ghost propagators in the Landau gauge. For example, in recent years it has been
shown that the gluon propagator freezes to a non-zero finite value at zero 
momentum. Both lattice and Schwinger-Dyson analyses are in accord, 
\cite{6,7,8,9,10,11,12,13,14,15}. 

A parallel early work was the development of renormalization schemes as
alternatives to the $\MSbar$ scheme of \cite{16,17}. In \cite{4} the momentum 
subtraction (MOM) schemes were constructed. These were based on the three 
$3$-point vertices of QCD. Briefly the MOM scheme is such that the finite part 
of the $2$- and $3$-point functions are absorbed into the appropriate
renormalization constant at a particular subtraction point in addition to the
terms singular in the regulator. For the coupling constant renormalization this
is at the point where the square of the momenta of the three external fields 
are all equal which is known as the symmetric point. This produces three
separate MOM schemes each based on one of the $3$-point vertices. More recently
that one loop MOM renormalization of QCD was extended to two loops in 
\cite{18}. In respect of analytic work there have been various combinations of
computations subsequent to \cite{4} in a variety of gauges, with different 
external momenta configurations including some external legs on-shell and at 
one or two loops. A representative selection of such combinations are 
\cite{19,20,21,22,23,24,25,26,27}. More recently, a string inspired method has 
allowed all types of one loop off-shell vertex functions to be studied
simultaneously, \cite{28}. Such an approach has the potential to be extended 
to two loops. Another aspect that knowledge of the explicit structure of the 
vertex functions is required for and that is in relation to lattice matching. 
If one computes vertex structures on the lattice in say a certain momentum
regime, one always has to ensure that in the limit back to the high energy 
region the numerical results match smoothly onto the known explicit 
perturbative results. Such matching can aid estimation of errors, for instance.
Given these considerations the aim of this article is to provide the full two 
loop $\MSbar$ results for the three $3$-point vertices of QCD in an arbitrary
linear covariant gauge in the chiral limit where the three external fields are 
all off-shell. This extends the symmetric point analysis of \cite{18} at two
loops and like \cite{4,18} will be for a non-exceptional momentum 
configuration. It will be useful, for instance, in being a potential guide 
along the route envisaged in \cite{28}. 

That such a computation can be achieved is due to recent developments in 
multiloop perturbation theory. We mention the key ones briefly. The first
is the provision of the Laporta algorithm, \cite{29}. This is a mathematical
construction which allows one to determine algebraic relations between Feynman 
integrals within a graph of a Green's function. While this is not a major 
observation, what has been crucial is that such relations can be systematically
solved in terms of a small set of basis integrals known as masters, \cite{29}. 
The second aspect is that for $3$-point functions all such one and two loop
masters are known for off-shell legs. The main early work in this respect was 
given in \cite{30,31}, prior to the development of the Laporta algorithm. 
However, to obtain the complete result, several master integrals are required 
to an order in $\epsilon$ beyond that given in \cite{30,31}. This is because of
the presence of spurious poles in $\epsilon$ which appear in the solution of 
the algebraic relations. Here $\epsilon$ is the regularizing parameter of 
dimensional regularization which we use throughout where 
$d$~$=$~$4$~$-$~$2\epsilon$ is the spacetime dimension. More recently, 
techniques have been developed such as those given in \cite{32,33,34} founded 
on polylogarithms and their mathematical generalizations. However, we have used
a parallel set of masters constructed in a specific problem, \cite{35}, but for 
a less general external momentum configuration. Based on \cite{35} we have 
extended the basic master integrals required to the necessary order in 
$\epsilon$ for the {\em general} off-shell situation. Equipped with these two 
main ingredients the construction of the general off-shell two loop vertex 
functions is viable. In reporting our full vertex function results it is worth 
noting that we have endeavoured to be as general as possible. So, for instance,
one can extract the behaviour of the vertex functions in, say, the Feynman 
gauge or for an arbitrary gauge but at a specific external momentum 
configuration. One benefit of such a general approach is that aside from 
assisting lattice matching, it allows for studies of the vertices in relation 
to the various ans\"{at}ze Schwinger-Dyson methods used. For instance, one 
approximation which is sometimes used is to neglect graphs with quartic gluon 
vertices. As the full two loop perturbative vertex $3$-point functions cannot 
omit such vertices due, for instance, to requiring renormalizability, the 
behaviour of say the triple gluon vertex function for a range of momenta can be
compared with a Schwinger-Dyson analysis of the same vertex but with a 
particular ansatz. For example this could determine over which range of momenta
such ans\"{a}tze or truncation of the tower of equations the approximation is 
valid for. Indeed there have been recent studies in this direction in 
\cite{36,37}. For instance, in \cite{36} improvements have been made in a 
Dyson-Schwinger analysis of the triple gluon vertex, in the absence of quarks. 
One interesting aspect was to actually try and numerically quantify the effect 
of omitting those two loop diagrams with quartic gluon interactions, \cite{36}.
Our perturbative results should prove useful in future similar studies of other
vertices. 

The article is organized as follows. The computational setup together with a
description of the algorithm we use to complete the two loop analysis is given 
in section $2$. Explicit results for the ghost-gluon, quark-gluon and triple 
gluon vertices are given in sections $3$, $4$ and $5$. Concluding comments are 
given in section $6$. Two appendices provide the tensor basis for each vertex 
we consider and the projection matrices as well as a summary of the one and two
loop master integrals we needed.

\sect{Background.}

We begin be describing our computational setup. The overall procedure will be
similar to the symmetric point analysis of \cite{18} but the external momentum 
configuration for the $3$-point vertices will be different. Moreover, we will
use the same renormalization constant convention definitions as \cite{18}. If 
we denote the three external momenta by $p$, $q$ and $r$ then energy momentum 
conservation
\begin{equation}
p ~+~ q ~+~ r ~=~ 0
\end{equation}
simply implies that two momenta are independent. We take these to be $p$ and
$q$ and follow the notation of \cite{30,31} in defining new variables by
\begin{equation}
x ~=~ \frac{p^2}{r^2} ~~~,~~~
y ~=~ \frac{q^2}{r^2} ~~~,~~~
r^2 ~=~ -~ \mu^2 ~.
\label{conf1}
\end{equation}
Thus all three external legs on the vertex functions will be off-shell and at
distinct values unlike \cite{4,18}. Our final expressions will be functions of 
$x$ and $y$ and we carry out all our analysis in terms of the dimensionful 
scale $\mu$. Consequently, the scalar products are given by
\begin{equation}
pq ~=~ \frac{1}{2} \left[ x + y - 1 \right] \mu^2 ~~~,~~~
pr ~=~ \frac{1}{2} \left[ 1 + x - y \right] \mu^2 ~~~,~~~
qr ~=~ \frac{1}{2} \left[ 1 - x + y \right] \mu^2 
\label{conf2}
\end{equation}
and are set to these values at each stage of the calculations. To extract data
for the symmetric point configuration it is clear that this corresponds to
$x$~$=$~$y$~$=$~$1$. As we are considering vertex functions, to the two loop 
order we will be working to it transpires that each vertex function has a 
common colour group factor. At the outset it is best to factor this off and
focus on the colourless amplitudes by defining, \cite{18}, 
\begin{eqnarray}
\left\langle A^a_\mu(p) A^b_\nu(q) A^c_\sigma(-p-q) \right\rangle 
&=& f^{abc} \Sigma^{\mbox{\footnotesize{ggg}}}_{\mu \nu \sigma}(p,q)
\nonumber \\
\left\langle \psi^i(p) \bar{\psi}^j(q) A^c_\sigma(-p-q) \right\rangle 
&=& T^c_{ij} \Sigma^{\mbox{\footnotesize{qqg}}}_\sigma(p,q)
\nonumber \\
\left\langle c^a(p) \bar{c}^b(q) A^c_\sigma(-p-q) \right\rangle 
&=& f^{abc} \Sigma^{\mbox{\footnotesize{ccg}}}_\sigma(p,q) 
\label{vertdef}
\end{eqnarray}
where the labels $ggg$, $qqg$ and $ccg$ respectively denote the triple gluon,
quark-gluon and ghost-gluon vertices. In (\ref{vertdef}) and analogous
functions later they can be regarded as being evaluated for the momentum
configuration defined in (\ref{conf1}) and (\ref{conf2}). To proceed we have to
write the colourless amplitudes in terms of Lorentz scalars. This requires 
choosing a set of basis tensors for each vertex. While the Ball-Chiu analysis 
\cite{5} was one of the first decompositions considered, we have chosen to 
follow those given for the symmetric point in \cite{18}. One reason for this is
that we retain one basic tensor structure per basis element, rather than 
combine them at the outset. In \cite{5} the basis choice was motivated in part 
by particular structures the overall vertex function should satisfy. This could 
slow down the computer algebraic computations. A second reason is to have a 
smooth way of checking that the results of \cite{18} emerge in the symmetric 
point limit of the full expression. Moreover, in \cite{18} no assumption was 
made to simplify the tensor basis due to the completely symmetric nature of the 
analysis. For instance, it was noted a postiori that the triple gluon vertex 
function could be written in terms of {\em three} independent tensor 
structures. However, for the general case one cannot make any such assumptions 
about the final structure of this or either of the other two vertex functions. 
Therefore, we will consider the most general Lorentz tensor decomposition. In 
other words we will use a basis built from the two independent external 
momenta, $p$ and $q$, the metric tensor, $\eta_{\mu\nu}$, and in the case of 
the quark-gluon vertex we will append tensors deriving from the 
$\gamma$-algebra. 

First, we define in general terms the decomposition into Lorentz scalar
amplitudes. For each vertex function we set
\begin{eqnarray}
\Sigma^{\mbox{\footnotesize{ccg}}}_\sigma(p,q)
&=& \sum_{k=1}^{2}
{\cal P}^{\mbox{\footnotesize{ccg}}}_{(k) \, \sigma }(p,q) \,
\Sigma^{\mbox{\footnotesize{ccg}}}_{(k)}(p,q) \nonumber \\
\Sigma^{\mbox{\footnotesize{qqg}}}_\sigma(p,q)
&=& \sum_{k=1}^{6}
{\cal P}^{\mbox{\footnotesize{qqg}}}_{(k) \, \sigma }(p,q) \,
\Sigma^{\mbox{\footnotesize{qqg}}}_{(k)}(p,q) \nonumber \\
\Sigma^{\mbox{\footnotesize{ggg}}}_{\mu \nu \sigma}(p,q)
&=& \sum_{k=1}^{14}
{\cal P}^{\mbox{\footnotesize{ggg}}}_{(k) \, \mu \nu \sigma }(p,q) \,
\Sigma^{\mbox{\footnotesize{ggg}}}_{(k)}(p,q) 
\end{eqnarray}
where $\Sigma^i_{(k)}(p,q)$ are the amplitudes and 
${\cal P}^i_{(k) \, \mu_1 \ldots \mu_n }(p,q)$ is the associated Lorentz
tensor. We use the same labelling convention as \cite{18} and $n$~$=$~$1$ for 
the ghost-gluon and quark-gluon vertices and $3$ for the triple gluon case. The 
explicit expressions for each situation is given in Appendix A. To disentangle
the amplitudes in order to evaluate each one individually we extend the method
of \cite{18} to the general momentum point. This involves projecting the 
$3$-point function with a specific Lorentz tensor for each label $k$. This 
projection matrix ${\cal M}^i_{kl}$ where $i$ labels the vertex, is defined as 
the inverse of the symmetric matrix ${\cal N}^i_{kl}$ whose elements are 
defined by the product of the $k$th and $l$th basis tensor,
\begin{equation}
{\cal N}^i_{kl} ~=~ {\cal P}^i_{(k) \, \mu_1 \ldots \mu_{n_i}}(p,q) 
{\cal P}^{i ~\, \mu_1 \ldots \mu_{n_i}}_{(l)}(p,q) 
\end{equation}
where there is no sum over $i$. We have provided the projection matrix for each
of the three vertices in Appendix A as their explicit forms are too involved 
for the main discussion due to the fact that we are in a general momentum 
configuration. Hence the $k$th amplitude of each of the three vertices is given
by 
\begin{eqnarray}
f^{abc} \Sigma^{\mbox{\footnotesize{ccg}}}_{(k)}(p,q) &=&
{\cal M}^{\mbox{\footnotesize{ccg}}}_{kl} \left(
{\cal P}^{\mbox{\footnotesize{ccg}} \, \sigma}_{(l)}(p,q)
\left\langle c^a(p) \bar{c}^b(q) A^c_\sigma(-p-q)
\right\rangle \right) 
\nonumber \\
T^c_{ij} \Sigma^{\mbox{\footnotesize{qqg}}}_{(k)}(p,q) &=&
{\cal M}^{\mbox{\footnotesize{qqg}}}_{kl} \left(
{\cal P}^{\mbox{\footnotesize{qqg}} \, \sigma}_{(l)}(p,q)
\left\langle \psi^i(p) \bar{\psi}^j(q) A^c_\sigma(-p-q)
\right\rangle \right) 
\nonumber \\
f^{abc} \Sigma^{\mbox{\footnotesize{ggg}}}_{(k)}(p,q) &=&
{\cal M}^{\mbox{\footnotesize{ggg}}}_{kl} \left(
{\cal P}^{\mbox{\footnotesize{ggg}} \, \mu \nu \sigma}_{(l)}(p,q)
\left\langle A^a_\mu(p) A^b_\nu(q) A^c_\sigma(-p-q)
\right\rangle \right ) ~.
\end{eqnarray}
For this decomposition there is an additional aspect for the quark-gluon 
vertex. As the external quarks carry spinor indices one has to allow for this
in the projection. So a trace over these indices is also included implicitly
within our construction. In addition we use the generalized $\gamma$-matrices
of \cite{38,39,40,41,42} denoted by $\Gamma_{(n)}^{\mu_1\ldots\mu_n}$ which are
defined by
\begin{equation}
\Gamma_{(n)}^{\mu_1 \ldots \mu_n} ~=~ \gamma^{[\mu_1} \ldots \gamma^{\mu_n]} 
\end{equation}
which is totally antisymmetric in its Lorentz indices and a factor of $1/n!$ is
understood in the definition. Although one could use a different construction
to accommodate the potential tensor basis element $\gamma^\mu \pslash \qslash$ 
the benefit of using $\Gamma_{(3)}^{\mu\nu\sigma}p_\nu q_\sigma$ is that there
is a partitioning of the projection matrix. 

One of the reasons for discussing the projection into Lorentz scalar amplitudes 
is that this is necessary for the way we perform the explicit loop
computations. This is based on the Laporta algorithm, \cite{29}, which is a 
method to systematically construct all the relevant integration by parts 
relations between Feynman integrals within a specific defining topology. Such 
integrals are necessarily scalar for the application of the algorithm. The 
outcome of the procedure is to solve the set of of relations for all the 
integrals in the problem at hand in terms of a basis set of what is termed 
master integrals, \cite{29}. Their value is determined by direct computation by 
non-integration by parts methods. For $3$-point functions in the general 
momentum configuration to two loops there are $1$ one loop and $4$ two loop 
basic topologies. For the latter there is one non-planar topology and three 
ladder integrals which are each formal rotations of each other but different 
external momenta. Once one projects the vertex function to scalar integrals 
then one writes the numerator momentum scalar products in terms of the topology
propagators. For the one loop topology this is always possible but for the $4$ 
two loop cases one has to append an additional propagator within the Laporta 
context to handle an irreducible numerator scalar product. Once such 
rearrangements have been carried out one constructs the integration by parts 
relations and solves the system in order to write each vertex amplitude in 
terms of the master integral basis. For our problem the application of the 
Laporta algorithm can only be performed with a computer package and we use the 
{\sc Reduze} encoding, \cite{43}, which is written in C++ and uses {\sc GiNaC},
\cite{44}. In this version a database of relations is constructed and the 
required integrals can be readily extracted. Such expressions are converted 
into the symbolic manipulation language {\sc Form}, \cite{45}. We use this and 
its multi-threaded version {\sc Tform}, \cite{46}, to handle the large amounts 
of intermediate algebra. Indeed the whole computation for each of the three 
vertices was performed automatically in {\sc Form}. The Feynman graphs were 
generated in {\sc Qgraf}, \cite{47}, and converted into {\sc Form} notation. 
Overall there are $2$ one loop and $33$ two loop graphs for each of the 
ghost-gluon and quark-gluon vertices. In the case of the triple gluon vertex 
there were $8$ one loop and $106$ two loop graphs to compute. With its $14$ 
amplitudes to determine this was by far the largest of our three vertex 
functions.

A final part of the computation is the substitution of the master integrals. 
For these we have used the values as they are already recorded in various 
articles. For the most part they were provided in \cite{30,31}. However, when 
one uses integration by parts, such as in the Laporta approach, there is a
potential that spurious poles in $\epsilon$ can appear. Hence one may require 
the master integrals to various finite powers in $\epsilon$ in order to extract
the finite parts for the vertex functions. That is the case here and we have 
had to extend several of the master integrals of \cite{30,31} to 
$O(\epsilon^2)$. This extends the same observation of \cite{35} for the 
interpolating configuration. It is so called since it depends on a parameter 
which links the fully symmetric point with the asymmetric point where one of 
the external momenta is zero. Though this latter point is an exceptional 
momentum configuration we note that (\ref{conf1}) is non-exceptional and hence 
not subject to any infrared problems. We have devolved the technical discussion
of extracting the necessary $O(\epsilon^2)$ terms of the various master 
integrals to Appendix B. Moreover, for completeness we also give there the 
explicit expressions for the functions which will appear in our final vertex 
amplitudes which were derived in \cite{30,31} in terms of the polylogarithm 
function. Though we note that as a check on our computational setup we have 
reproduced the full calculation of \cite{35} for that interpolating momentum 
configuration. In \cite{35} the Green's function under consideration was a 
quark $2$-point function with the quark mass operator inserted with a non-zero 
external momentum flowing through it.

Finally, we note that as we are performing an automatic computation we use the
procedure of \cite{48} to implement the renormalization procedure. We calculate 
in terms of bare parameters such as the coupling constant and the gauge 
parameter. Once all the diagrams have been compiled and added together the bare
parameters are replaced by their renormalized counterparts with the 
corresponding renormalization constant as the constant of proportionality. This
automatically introduces the requisite counterterms. To handle the divergences 
we use the $\MSbar$ scheme. This provides a minor check on our construction. As
these renormalization constants have been established for a long time, 
\cite{49,50,51,52,53,54}, we must obtain the same values for the coupling 
constant renormalization to two loops for example. Thus we should not have a 
pole in $\epsilon$ whose residue is a function of $x$ or $y$ or both momentum 
ratio parameters. Such poles would contradict the renormalizability of QCD. 
This is a non-trivial observation since one has spurious poles within the main 
part of the integration. So equally one should not find quartic and triple 
poles in $\epsilon$ with $x$ or $y$ dependent residues either. In this respect 
we note that this is partially as a consequence of functional relations between
some of the basic functions which appear in the master integrals such as those 
given in (\ref{funrel}). In carrying out an $\MSbar$ renormalization we note 
that we subtract at the point given by (\ref{conf1}). Moreover, all the 
explicit finite expressions for the vertex function presented later in the 
article are at that point after renormalization. Therefore, the scale which 
appears in the running coupling constant, for instance, is the parameter $\mu$ 
of (\ref{conf1}). In other words we use $\mu$ as the mass scale which is 
introduced in to ensure that the coupling constant is dimensionless in 
$d$-dimensions. Choosing an independent scale, say $\tilde{\mu}$, is 
straightforward to incorporate. In the context of the article this is possible
to achieve as electronic versions of all these final expressions are provided 
in an attached data file.  In addition we have also included the expressions 
for each of the three $3$-point functions {\em prior} to any wave function,
coupling constant or gauge parameter renormalization. In other words these 
correspond to the bare vertex functions. This will allow, for example, an 
interested reader to extract results for other schemes, such as the momentum
subtraction (MOM) schemes given in \cite{4}. In order to achieve this the bare 
expressions are given to the requisite order in $\epsilon$ at one loop. These
terms are necessary when considering schemes which absorb finite parts into
some or all of the renormalization constants. The counterterms affecting the 
one loop part will contribute to the finite part of the vertex function at two 
loops. The benefit of providing this bare data is that it also allows one to 
perform the automatic renormalization using the method of \cite{48}.

\sect{Ghost-gluon vertex.}

In this section we discuss the amplitudes for the ghost-gluon vertex. As the
full gauge dependent result is large we present the Landau gauge expression
for one amplitude after renormalizing in the $\MSbar$ scheme. This is to
illustrate the complexity of the final expressions and to discuss several
points concerning their derivation. However, to appreciate the effect of the 
two loop corrections in relation to the one loop part, we will compare the 
amplitudes graphically for various cross-sections of the function domain. The 
reason for concentrating on the Landau gauge in presenting results is that it 
is the one of intense lattice and Schwinger-Dyson interest in recent years. 
Thus the channel one amplitude for the ghost-gluon vertex in the $\MSbar$ 
scheme is
\begin{eqnarray}
\Sigma^{\mbox{\footnotesize{ccg}},\alpha=0}_{(1)}(p,q) &=&
-~ 1 \nonumber \\
&& +~ \left[
- \frac{9}{4} \Phi_1(x,y) y^2 \Delta_G^{-1}
- \frac{15}{16} \Phi_1(x,y) y
- \frac{3}{4} \Phi_1(x,y) x y \Delta_G^{-1}
- \frac{1}{2}
- \frac{1}{4} y
\right. \nonumber \\
&& \left. ~~~~
- \frac{1}{4} \ln(x) y \Delta_G^{-1}
- \frac{1}{4} \ln(x) x \Delta_G^{-1}
- \frac{1}{8} \ln(x) y
- \frac{1}{8} \ln(x) x
+ \frac{1}{16} \Phi_1(x,y) x
\right. \nonumber \\
&& \left. ~~~~
+ \frac{1}{8} \ln(y)
+ \frac{1}{8} \ln(y) y
+ \frac{1}{8} \ln(y) x
+ \frac{3}{16} \Phi_1(x,y)
+ \frac{1}{4} x
+ \frac{1}{4} \ln(x)
\right. \nonumber \\
&& \left. ~~~~
+ \frac{1}{4} \ln(x) \Delta_G^{-1}
+ \frac{5}{4} \Phi_1(x,y) y \Delta_G^{-1}
+ \frac{3}{2} \ln(y) y \Delta_G^{-1}
- \ln(y) y^2 \Delta_G^{-1}
\right. \nonumber \\
&& \left. ~~~~
- \ln(y) x y \Delta_G^{-1}
- \Phi_1(x,y) x y^2 \Delta_G^{-1}
+ \Phi_1(x,y) y^3 \Delta_G^{-1}
\right. \nonumber \\
&& \left. ~~~~
+ 2 \ln(x) x y \Delta_G^{-1}
\right] C_A a
\nonumber \\
&&
+~ \left[
- \frac{149}{18} \ln(x) x y \Delta_G^{-1}
- \frac{59}{12} \ln(y) y \Delta_G^{-1}
- \frac{163}{36} \Phi_1(x,y) y \Delta_G^{-1}
\right. \nonumber \\
&& \left. ~~~~
- \frac{149}{36} \Phi_1(x,y) y^3 \Delta_G^{-1}
- \frac{19}{12} \ln(y) \Phi_1(x,y) y^2 \Delta_G^{-1}
- \frac{3}{2} \ln(x) \Phi_1(x,y) y^2 \Delta_G^{-1}
\right. \nonumber \\
&& \left. ~~~~
- \frac{47}{36} \ln(x)
- \frac{25}{24} \Phi_1(x,y)
- \frac{11}{12} \ln^2(y) y^2 \Delta_G^{-1}
- \frac{11}{12} \ln(y) \Phi_1(x,y) x y^2 \Delta_G^{-1}
\right. \nonumber \\
&& \left. ~~~~
- \frac{19}{24} \ln(x) \Phi_1(x,y) y
- \frac{2}{3} \ln(x) \Phi_1(x,y) x y \Delta_G^{-1}
- \frac{2}{3} \ln^2(y) x y \Delta_G^{-1}
\right. \nonumber \\
&& \left. ~~~~
- \frac{11}{18} \ln(y)
- \frac{1}{2} \ln(x) \Phi_1(x,y) x y^2 \Delta_G^{-1}
- \frac{1}{2} \ln(y) \Phi_1(x,y) x y \Delta_G^{-1}
\right. \nonumber \\
&& \left. ~~~~
- \frac{11}{24} \ln(y) \Phi_1(x,y) y
- \frac{7}{18} \ln(x) \Delta_G^{-1}
- \frac{11}{36} x
- \frac{1}{4} \ln^2(x) y^2 \Delta_G^{-1}
\right. \nonumber \\
&& \left. ~~~~
- \frac{1}{4} \ln(x) \ln(y) y^2 \Delta_G^{-1}
- \frac{5}{24} \ln(x) \Phi_1(x,y) x
- \frac{1}{6} \ln^2(x) y
\right. \nonumber \\
&& \left. ~~~~
- \frac{1}{6} \ln(x) \ln(y) y^3 \Delta_G^{-1}
- \frac{1}{6} \ln(x) \ln(y) x \Delta_G^{-1}
- \frac{1}{6} \Omega_2\left(\frac{y}{x},\frac{1}{x}\right) x
\right. \nonumber \\
&& \left. ~~~~
- \frac{1}{6} \Omega_2\left(\frac{x}{y},\frac{1}{y}\right) x
- \frac{1}{6} \Phi_1(x,y) y^2
- \frac{1}{12} \ln^2(x) x \Delta_G^{-1}
\right. \nonumber \\
&& \left. ~~~~
- \frac{1}{12} \ln^2(x) x y^2 \Delta_G^{-1}
- \frac{1}{12} \ln(x) \ln(y) x
- \frac{1}{12} \ln(x) \Phi_1(x,y) \Delta_G
\right. \nonumber \\
&& \left. ~~~~
- \frac{1}{12} \ln^2(y) y
- \frac{1}{12} \ln^2(y) x y^2 \Delta_G^{-1}
- \frac{1}{36} \ln(y) y
- \frac{1}{36} \ln(y) x
+ \frac{1}{36} \ln(x) y
\right. \nonumber \\
&& \left. ~~~~
+ \frac{1}{36} \ln(x) x
+ \frac{1}{24} \ln(y) \Phi_1(x,y) x
+ \frac{1}{12} \ln^2(x) \Delta_G^{-1}
+ \frac{1}{12} \ln^2(x) y \Delta_G^{-1}
\right. \nonumber \\
&& \left. ~~~~
+ \frac{1}{12} \ln^2(x) y^3 \Delta_G^{-1}
+ \frac{1}{12} \ln^2(y)
+ \frac{1}{12} \ln^2(y) y^3 \Delta_G^{-1}
+ \frac{1}{12} \ln^2(y) x
\right. \nonumber \\
&& \left. ~~~~
+ \frac{1}{12} \ln(y) \Phi_1(x,y) \Delta_G
+ \frac{1}{12} \Phi_1(x,y) \Delta_G
+ \frac{1}{8} \ln(y) \Phi_1(x,y)
\right. \nonumber \\
&& \left. ~~~~
+ \frac{1}{6} \ln(x) \ln(y) \Delta_G^{-1}
+ \frac{1}{6} \ln(x) \ln(y) x y^2 \Delta_G^{-1}
+ \frac{1}{6} \Omega_2\left(\frac{y}{x},\frac{1}{x}\right)
\right. \nonumber \\
&& \left. ~~~~
+ \frac{1}{6} \Omega_2\left(\frac{y}{x},\frac{1}{x}\right) y
+ \frac{1}{6} \Omega_2\left(\frac{x}{y},\frac{1}{y}\right)
+ \frac{1}{6} \Omega_2\left(\frac{x}{y},\frac{1}{y}\right) y
+ \frac{1}{6} \Phi_1(x,y) x y
\right. \nonumber \\
&& \left. ~~~~
+ \frac{1}{4} \ln^2(x)
+ \frac{1}{4} \ln(x) \ln(y) y
+ \frac{7}{24} \ln(x) \Phi_1(x,y)
+ \frac{11}{36} y
+ \frac{3}{8} \ln(x) \ln(y)
\right. \nonumber \\
&& \left. ~~~~
+ \frac{7}{18} \ln(x) y \Delta_G^{-1}
+ \frac{7}{18} \ln(x) x \Delta_G^{-1}
+ \frac{41}{72} \Phi_1(x,y) x
+ \frac{7}{12} \ln(x) \ln(y) y \Delta_G^{-1}
\right. \nonumber \\
&& \left. ~~~~
+ \frac{2}{3} \ln(x) \Phi_1(x,y) y^3 \Delta_G^{-1}
+ \frac{3}{4} \ln(x) \ln(y) x y \Delta_G^{-1}
+ \frac{3}{4} \ln(y) \Phi_1(x,y) y^3 \Delta_G^{-1}
\right. \nonumber \\
&& \left. ~~~~
+ \frac{5}{6} \ln(x) \Phi_1(x,y) y \Delta_G^{-1}
+ \frac{5}{6} \ln(y) \Phi_1(x,y) y \Delta_G^{-1}
+ \frac{31}{36}
+ \frac{4}{3} \ln^2(x) x y \Delta_G^{-1}
\right. \nonumber \\
&& \left. ~~~~
+ \frac{83}{24} \Phi_1(x,y) y
+ \frac{15}{4} \Phi_1(x,y) x y \Delta_G^{-1}
+ \frac{149}{36} \ln(y) y^2 \Delta_G^{-1}
\right. \nonumber \\
&& \left. ~~~~
+ \frac{149}{36} \ln(y) x y \Delta_G^{-1}
+ \frac{149}{36} \Phi_1(x,y) x y^2 \Delta_G^{-1}
+ \frac{26}{3} \Phi_1(x,y) y^2 \Delta_G^{-1}
\right. \nonumber \\
&& \left. ~~~~
+ \ln^2(y) y \Delta_G^{-1}
\right] C_A T_F \Nf a^2 
\nonumber \\
&&
+~ \left[
-~ \frac{1159}{48} \Phi_1(x,y) y^2 \Delta_G^{-1}
- \frac{4315}{384} \Phi_1(x,y) y
- \frac{3203}{288} \ln(y) y^2 \Delta_G^{-1}
\right. \nonumber \\
&& \left. ~~~~
- \frac{3203}{288} \ln(y) x y \Delta_G^{-1}
- \frac{3203}{288} \Phi_1(x,y) x y^2 \Delta_G^{-1}
- \frac{295}{32} \Phi_1(x,y) x y \Delta_G^{-1}
\right. \nonumber \\
&& \left. ~~~~
- \frac{1351}{384} \ln^2(x) x y \Delta_G^{-1}
- \frac{901}{288}
- \frac{377}{192} \ln(x) \Phi_1(x,y) y \Delta_G^{-1}
- \frac{137}{72} \ln(x) y \Delta_G^{-1}
\right. \nonumber \\
&& \left. ~~~~
- \frac{137}{72} \ln(x) x \Delta_G^{-1}
- \frac{237}{128} \ln^2(y) y \Delta_G^{-1}
- \frac{57}{32} \zeta(3) y^2 \Delta_G^{-1}
- \frac{455}{288} y
\right. \nonumber \\
&& \left. ~~~~
- \frac{277}{192} \ln(x) \Phi_1(x,y) y^3 \Delta_G^{-1}
- \frac{45}{32} \ln(y) \Phi_1(x,y) y^3 \Delta_G^{-1}
\right. \nonumber \\
&& \left. ~~~~
- \frac{511}{384} \ln(y) \Phi_1(x,y) y \Delta_G^{-1}
- \frac{223}{192} \Phi_1(x,y) x y
- \frac{73}{64} \ln^2(x)
\right. \nonumber \\
&& \left. ~~~~
- \frac{55}{64} \Phi_2\left(\frac{1}{y},\frac{x}{y}\right) \frac{1}{y} \Delta_G^{-1}
- \frac{155}{192} \ln(x) \ln(y) y \Delta_G^{-1}
- \frac{51}{64} \Phi_2\left(\frac{y}{x},\frac{1}{x}\right) \frac{y^2}{x} \Delta_G^{-1}
\right. \nonumber \\
&& \left. ~~~~
- \frac{11}{16} \Phi_2\left(\frac{1}{y},\frac{x}{y}\right) y^2 \Delta_G^{-1}
- \frac{41}{64} \Omega_2\left(\frac{1}{y},\frac{x}{y}\right) x y \Delta_G^{-1}
- \frac{41}{64} \Omega_2(y,x) x y \Delta_G^{-1}
\right. \nonumber \\
&& \left. ~~~~
- \frac{61}{96} \ln^2(x) y \Delta_G^{-1}
- \frac{19}{32} \Phi_2\left(\frac{y}{x},\frac{1}{x}\right) \frac{y^3}{x} \Delta_G^{-1}
- \frac{149}{256} \Omega_2\left(\frac{1}{y},\frac{x}{y}\right) y \Delta_G^{-1}
\right. \nonumber \\
&& \left. ~~~~
- \frac{149}{256} \Omega_2(y,x) y \Delta_G^{-1}
- \frac{223}{384} \Phi_1(x,y) \Delta_G
- \frac{9}{16} \zeta(3) x y^2 \Delta_G^{-1}
\right. \nonumber \\
&& \left. ~~~~
- \frac{25}{48} \ln(x) \ln(y) \Delta_G^{-1}
- \frac{33}{64} \zeta(3) y
- \frac{117}{256} \Omega_2\left(\frac{1}{y},\frac{x}{y}\right)
- \frac{117}{256} \Omega_2(y,x)
\right. \nonumber \\
&& \left. ~~~~
- \frac{85}{192} \ln(x) \Phi_1(x,y)
- \frac{27}{64} \Phi_2(x,y) y \Delta_G^{-1}
- \frac{289}{768} \ln(y) \Phi_1(x,y) x
\right. \nonumber \\
&& \left. ~~~~
- \frac{115}{384} \ln^2(y) x
- \frac{17}{64} \Phi_2\left(\frac{y}{x},\frac{1}{x}\right) \frac{y}{x} \Delta_G^{-1}
- \frac{33}{128} \Omega_2\left(\frac{y}{x},\frac{1}{x}\right) x y \Delta_G^{-1}
\right. \nonumber \\
&& \left. ~~~~
- \frac{33}{128} \Omega_2\left(\frac{x}{y},\frac{1}{y}\right) x y \Delta_G^{-1}
- \frac{59}{256} \Omega_2\left(\frac{y}{x},\frac{1}{x}\right) x \Delta_G^{-1}
- \frac{59}{256} \Omega_2\left(\frac{x}{y},\frac{1}{y}\right) x \Delta_G^{-1}
\right. \nonumber \\
&& \left. ~~~~
- \frac{55}{256} \Omega_2\left(\frac{1}{x},\frac{y}{x}\right) \Delta_G^{-1}
- \frac{55}{256} \Omega_2(x,y) \Delta_G^{-1}
- \frac{61}{288} \ln(x) x
\right. \nonumber \\
&& \left. ~~~~
- \frac{157}{768} \Omega_2\left(\frac{y}{x},\frac{1}{x}\right)
- \frac{157}{768} \Omega_2\left(\frac{x}{y},\frac{1}{y}\right)
- \frac{13}{64} \Phi_2(x,y) y^3 \Delta_G^{-1}
\right. \nonumber \\
&& \left. ~~~~
- \frac{51}{256} \Omega_2\left(\frac{1}{y},\frac{x}{y}\right) x \Delta_G^{-1}
- \frac{51}{256} \Omega_2(y,x) x \Delta_G^{-1}
- \frac{73}{384} \ln^2(y)
\right. \nonumber \\
&& \left. ~~~~
- \frac{205}{1152} \Phi_1(x,y) x
- \frac{17}{96} \ln^2(y) y
- \frac{9}{64} \ln(x) \ln(y) x y \Delta_G^{-1}
- \frac{9}{64} \Phi_2(x,y)
\right. \nonumber \\
&& \left. ~~~~
- \frac{33}{256} \ln(y) \Phi_1(x,y)
- \frac{31}{256} \Omega_2\left(\frac{y}{x},\frac{1}{x}\right) y \Delta_G^{-1}
- \frac{31}{256} \Omega_2\left(\frac{x}{y},\frac{1}{y}\right) y \Delta_G^{-1}
\right. \nonumber \\
&& \left. ~~~~
- \frac{17}{144} \ln(x) y
- \frac{19}{192} \Omega_2\left(\frac{y}{x},\frac{1}{x}\right) y
- \frac{19}{192} \Omega_2\left(\frac{x}{y},\frac{1}{y}\right) y
\right. \nonumber \\
&& \left. ~~~~
- \frac{37}{384} \ln(y) \Phi_1(x,y) \Delta_G
- \frac{3}{32} \Phi_2(x,y) y^4 \Delta_G^{-1}
- \frac{11}{128} \ln(x) \ln(y) y
\right. \nonumber \\
&& \left. ~~~~
- \frac{29}{384} \ln^2(x) x \Delta_G^{-1}
- \frac{7}{96} \ln(x) \ln(y) y^3 \Delta_G^{-1}
- \frac{9}{128} \ln(y) \Phi_1(x,y) x y
\right. \nonumber \\
&& \left. ~~~~
- \frac{1}{16} \Phi_2\left(\frac{y}{x},\frac{1}{x}\right) \frac{1}{x} \Delta_G^{-1}
- \frac{1}{16} \Phi_2\left(\frac{y}{x},\frac{1}{x}\right) x \Delta_G^{-1}
- \frac{3}{64} \zeta(3) x
\right. \nonumber \\
&& \left. ~~~~
- \frac{3}{64} \Omega_2\left(\frac{1}{x},\frac{y}{x}\right) y
- \frac{3}{64} \Omega_2\left(\frac{1}{x},\frac{y}{x}\right) x
- \frac{3}{64} \Omega_2\left(\frac{y}{x},\frac{1}{x}\right) y^3 \Delta_G^{-1}
\right. \nonumber \\
&& \left. ~~~~
- \frac{3}{64} \Omega_2\left(\frac{1}{y},\frac{x}{y}\right) y^2 \Delta_G^{-1}
- \frac{3}{64} \Omega_2\left(\frac{x}{y},\frac{1}{y}\right) y^3 \Delta_G^{-1}
- \frac{3}{64} \Omega_2(x,y) y
\right. \nonumber \\
&& \left. ~~~~
- \frac{3}{64} \Omega_2(x,y) x
- \frac{3}{64} \Omega_2(y,x) y^2 \Delta_G^{-1}
- \frac{7}{192} \ln^2(x) x y^2 \Delta_G^{-1}
\right. \nonumber \\
&& \left. ~~~~
- \frac{7}{192} \ln^2(y) x y^2 \Delta_G^{-1}
- \frac{3}{128} \ln(x) \Phi_1(x,y) x y
+ \frac{1}{32} \Phi_2\left(\frac{y}{x},\frac{1}{x}\right)
\right. \nonumber \\
&& \left. ~~~~
+ \frac{7}{192} \ln^2(x) y^3 \Delta_G^{-1}
+ \frac{7}{192} \ln^2(y) y^3 \Delta_G^{-1}
+ \frac{3}{64} \ln(x) \Phi_1(x,y) y^2
\right. \nonumber \\
&& \left. ~~~~
+ \frac{3}{64} \ln(y) \Phi_1(x,y) y^2
+ \frac{3}{64} \Omega_2\left(\frac{y}{x},\frac{1}{x}\right) x y^2 \Delta_G^{-1}
+ \frac{3}{64} \Omega_2\left(\frac{1}{y},\frac{x}{y}\right) y
\right. \nonumber \\
&& \left. ~~~~
+ \frac{3}{64} \Omega_2\left(\frac{1}{y},\frac{x}{y}\right) x
+ \frac{3}{64} \Omega_2\left(\frac{x}{y},\frac{1}{y}\right) x y^2 \Delta_G^{-1}
+ \frac{3}{64} \Omega_2(y,x) y
\right. \nonumber \\
&& \left. ~~~~
+ \frac{3}{64} \Omega_2(y,x) x
+ \frac{3}{64} \Phi_2(x,y) x y
+ \frac{19}{384} \ln(x) \Phi_1(x,y) \Delta_G
\right. \nonumber \\
&& \left. ~~~~
+ \frac{7}{96} \ln(x) \ln(y) x y^2 \Delta_G^{-1}
+ \frac{29}{384} \ln^2(x) \Delta_G^{-1}
+ \frac{3}{32} \zeta(3) x y \Delta_G^{-1}
\right. \nonumber \\
&& \left. ~~~~
+ \frac{3}{32} \Phi_2\left(\frac{y}{x},\frac{1}{x}\right) x y \Delta_G^{-1}
+ \frac{3}{32} \Phi_2(x,y) y^2
+ \frac{3}{32} \Phi_2(x,y) x y^3 \Delta_G^{-1}
\right. \nonumber \\
&& \left. ~~~~
+ \frac{27}{256} \Omega_2\left(\frac{1}{x},\frac{y}{x}\right)
+ \frac{27}{256} \Omega_2(x,y)
+ \frac{7}{64} \ln(x) \ln(y)
+ \frac{7}{64} \Phi_2(x,y) x
\right. \nonumber \\
&& \left. ~~~~
+ \frac{17}{144} \ln(y) x
+ \frac{1}{8} \Phi_2\left(\frac{y}{x},\frac{1}{x}\right) \Delta_G^{-1}
+ \frac{9}{64} \ln^2(x) x
+ \frac{7}{48} \Omega_2\left(\frac{y}{x},\frac{1}{x}\right) x
\right. \nonumber \\
&& \left. ~~~~
+ \frac{7}{48} \Omega_2\left(\frac{x}{y},\frac{1}{y}\right) x
+ \frac{61}{384} \ln(x) \ln(y) x
+ \frac{51}{256} \Omega_2\left(\frac{1}{y},\frac{x}{y}\right) \Delta_G^{-1}
\right. \nonumber \\
&& \left. ~~~~
+ \frac{51}{256} \Omega_2(y,x) \Delta_G^{-1}
+ \frac{61}{288} \ln(y) y
+ \frac{55}{256} \Omega_2\left(\frac{1}{x},\frac{y}{x}\right) x \Delta_G^{-1}
\right. \nonumber \\
&& \left. ~~~~
+ \frac{55}{256} \Omega_2(x,y) x \Delta_G^{-1}
+ \frac{59}{256} \Omega_2\left(\frac{y}{x},\frac{1}{x}\right) \Delta_G^{-1}
+ \frac{59}{256} \Omega_2\left(\frac{x}{y},\frac{1}{y}\right) \Delta_G^{-1}
\right. \nonumber \\
&& \left. ~~~~
+ \frac{101}{384} \ln^2(x) y
+ \frac{37}{128} \ln(y) \Phi_1(x,y) x y \Delta_G^{-1}
+ \frac{19}{64} \Omega_2\left(\frac{1}{x},\frac{y}{x}\right) y^2 \Delta_G^{-1}
\right. \nonumber \\
&& \left. ~~~~
+ \frac{19}{64} \Omega_2(x,y) y^2 \Delta_G^{-1}
+ \frac{81}{256} \Omega_2\left(\frac{1}{x},\frac{y}{x}\right) y \Delta_G^{-1}
+ \frac{81}{256} \Omega_2(x,y) y \Delta_G^{-1}
\right. \nonumber \\
&& \left. ~~~~
+ \frac{11}{32} \Phi_2\left(\frac{1}{y},\frac{x}{y}\right) \frac{x}{y}
+ \frac{23}{64} \ln(x) \ln(y) y^2 \Delta_G^{-1}
+ \frac{23}{64} \Phi_2\left(\frac{y}{x},\frac{1}{x}\right) y \Delta_G^{-1}
\right. \nonumber \\
&& \left. ~~~~
+ \frac{23}{64} \Phi_2(x,y) x y \Delta_G^{-1}
+ \frac{25}{64} \Omega_2\left(\frac{1}{x},\frac{y}{x}\right) x y \Delta_G^{-1}
+ \frac{25}{64} \Omega_2(x,y) x y \Delta_G^{-1}
\right. \nonumber \\
&& \left. ~~~~
+ \frac{51}{128} \Omega_2\left(\frac{y}{x},\frac{1}{x}\right) y^2 \Delta_G^{-1}
+ \frac{51}{128} \Omega_2\left(\frac{x}{y},\frac{1}{y}\right) y^2 \Delta_G^{-1}
+ \frac{323}{768} \ln(x) \Phi_1(x,y) x
\right. \nonumber \\
&& \left. ~~~~
+ \frac{1}{2} \Phi_2\left(\frac{y}{x},\frac{1}{x}\right) y^2 \Delta_G^{-1}
+ \frac{1}{2} \Phi_2\left(\frac{1}{y},\frac{x}{y}\right)
+ \frac{25}{48} \ln(x) \ln(y) x \Delta_G^{-1}
\right. \nonumber \\
&& \left. ~~~~
+ \frac{67}{128} \ln^2(x) y^2 \Delta_G^{-1}
+ \frac{35}{64} \Phi_2\left(\frac{1}{y},\frac{x}{y}\right) y \Delta_G^{-1}
+ \frac{35}{64} \Phi_2(x,y) y
+ \frac{9}{16} \zeta(3)
\right. \nonumber \\
&& \left. ~~~~
+ \frac{9}{16} \zeta(3) y^3 \Delta_G^{-1}
+ \frac{43}{64} \Phi_2(x,y) x y^2 \Delta_G^{-1}
+ \frac{11}{16} \Phi_2\left(\frac{1}{y},\frac{x}{y}\right) x y \Delta_G^{-1}
\right. \nonumber \\
&& \left. ~~~~
+ \frac{23}{32} \Phi_2(x,y) y^2 \Delta_G^{-1}
+ \frac{311}{384} \ln^2(y) x y \Delta_G^{-1}
+ \frac{55}{64} \Phi_2\left(\frac{1}{y},\frac{x}{y}\right) \frac{x}{y} \Delta_G^{-1}
\right. \nonumber \\
&& \left. ~~~~
+ \frac{15}{16} \Phi_2\left(\frac{1}{y},\frac{x}{y}\right) \frac{1}{y}
+ \frac{361}{384} \ln(y) \Phi_1(x,y) y
+ \frac{223}{192} \Phi_1(x,y) y^2
\right. \nonumber \\
&& \left. ~~~~
+ \frac{39}{32} \zeta(3) y \Delta_G^{-1}
+ \frac{4}{3} \ln(y) \Phi_1(x,y) x y^2 \Delta_G^{-1}
+ \frac{289}{192} \ln(x) \Phi_1(x,y) y
\right. \nonumber \\
&& \left. ~~~~
+ \frac{97}{64} \ln(x) \Phi_1(x,y) x y^2 \Delta_G^{-1}
+ \frac{455}{288} x
+ \frac{137}{72} \ln(x) \Delta_G^{-1}
\right. \nonumber \\
&& \left. ~~~~
+ \frac{755}{384} \ln^2(y) y^2 \Delta_G^{-1}
+ \frac{767}{384} \Phi_1(x,y)
+ \frac{151}{72} \ln(y)
\right. \nonumber \\
&& \left. ~~~~
+ \frac{203}{96} \ln(x) \Phi_1(x,y) x y \Delta_G^{-1}
+ \frac{1051}{384} \ln(y) \Phi_1(x,y) y^2 \Delta_G^{-1}
+ \frac{941}{288} \ln(x)
\right. \nonumber \\
&& \left. ~~~~
+ \frac{109}{32} \ln(x) \Phi_1(x,y) y^2 \Delta_G^{-1}
+ \frac{237}{64} \Phi_2\left(\frac{1}{y},\frac{x}{y}\right) x \Delta_G^{-1}
\right. \nonumber \\
&& \left. ~~~~
+ \frac{3203}{288} \Phi_1(x,y) y^3 \Delta_G^{-1}
+ \frac{3751}{288} \Phi_1(x,y) y \Delta_G^{-1}
+ \frac{1433}{96} \ln(y) y \Delta_G^{-1}
\right. \nonumber \\
&& \left. ~~~~
+ \frac{3203}{144} \ln(x) x y \Delta_G^{-1}
+ \Phi_2\left(\frac{1}{y},\frac{x}{y}\right) \Delta_G^{-1}
\right] C_A^2 a^2 ~+~ O(a^3) ~.
\label{ghost1}
\end{eqnarray} 
This channel corresponds to the Feynman rule of the original vertex itself. 
Here $\zeta(z)$ is the Riemann zeta function, $C_A$, $C_F$ and $T_F$ are the 
usual colour group Casimirs, $\Nf$ is the number of massless flavours, 
$a$~$=$~$g^2/(16\pi^2)$ and $g$ is the coupling constant. We have introduced 
the Gram determinant $\Delta_G$ which depends on $x$ and $y$ and is given by 
\begin{equation}
\Delta_G(x,y) ~=~ x^2 ~-~ 2 x y ~+~ y^2 ~-~ 2 x ~-~ 2 y ~+~ 1 ~.
\end{equation}
This arises from the determinant of the matrix which produces the projection
matrix. In the above expression we have endeavoured to simplify the tedious 
algebra as far as possible using {\sc Form}. However, due to the presence of 
$\Delta_G^{-1}$ it was not always possible to do so in a symmetric way for 
cases where the expression has say an $x$ and $y$ interchange symmetry. This is
the reason for the absence of any obvious underlying symmetries. However, as a 
check on our all the amplitudes we have computed we have taken the limit to the
full symmetric point and compared with the results of \cite{18}. We found exact
agreement for all $\alpha$. In order to assist others with this check we note 
the limits of the various functions as $x$~$\rightarrow$~$1$ and 
$y$~$\rightarrow$~$1$ in the notation of \cite{55}. We have
\begin{eqnarray}
\Phi_1(1,1) &=& \frac{2}{3} \psi^\prime \left( \frac{1}{3} \right) ~-~ 
\frac{4\pi^2}{9} ~~~,~~~ 
\Psi_1(1,1) ~=~ \frac{\pi}{4\sqrt{3}} \ln^2(3) ~-~ 12 s_3 
\left( \frac{\pi}{6} \right) ~+~ \frac{35\pi^3}{108\sqrt{3}} \nonumber \\
\Phi_2(1,1) &=& \frac{1}{36} \psi^{\prime\prime\prime} 
\left( \frac{1}{3} \right) ~-~ \frac{2\pi^4}{27} \nonumber \\
\Omega_2(1,1) &=& \frac{8\pi^2}{3} ~+~ 4 \zeta(3) 
~-~ 24 s_2 \left( \frac{\pi}{6} \right) 
~+~ 48 s_2 \left( \frac{\pi}{2} \right)
~+~ 40 s_3 \left( \frac{\pi}{6} \right)
~-~ 32 s_3 \left( \frac{\pi}{2} \right) \nonumber \\
&& -~ 4 \psi^\prime \left( \frac{1}{3} \right)
~+~ \frac{29\pi^3}{162\sqrt{3}}
~+~ \frac{2\pi}{\sqrt{3}} \ln(3)
~-~ \frac{\pi}{6\sqrt{3}} \ln^2(3) \nonumber \\
\chi_1(1,1) &=& \frac{\pi^4}{27} 
~-~ \frac{\pi^2}{18} \psi^\prime \left( \frac{1}{3} \right)
~-~ {\cal H}^{(2)}_{31} ~~~,~~~ 
\chi_3(1,1) ~=~ \frac{\pi^4}{27} 
~-~ \frac{\pi^2}{18} \psi^\prime \left( \frac{1}{3} \right)
~+~ {\cal H}^{(2)}_{43} \nonumber \\
\chi_2(1,1) &=&
12 s_2 \left( \frac{\pi}{6} \right)
~-~ 24 s_2 \left( \frac{\pi}{2} \right)
~-~ 8 s_3 \left( \frac{\pi}{6} \right)
~+~ 16 s_3 \left( \frac{\pi}{2} \right) 
~-~ \frac{4}{3} \pi^2 
~-~ 4 \zeta(3) \nonumber \\
&& +~ 2 \psi^\prime \left( \frac{1}{3} \right)
~-~ \frac{67\pi^3}{162\sqrt{3}}
~-~ \frac{\pi}{\sqrt{3}} \ln(3)
~-~ \frac{\pi}{6\sqrt{3}} \ln^2(3) 
\end{eqnarray}
where $\psi(z)$ is the derivative of the logarithm of the Euler Gamma function,
\begin{equation}
s_n(z) ~=~ \frac{1}{\sqrt{3}} \Im \left[ \mbox{Li}_n \left(
\frac{e^{iz}}{\sqrt{3}} \right) \right]
\end{equation}
and $\mbox{Li}_n(z)$ is the polylogarithm function. Also ${\cal H}^{(2)}_{31}$ 
and ${\cal H}^{(2)}_{43}$ are two harmonic polylogarithms, \cite{55}, but in a 
symmetric point computation they always appear in the combination
\begin{equation}
\Sigma ~=~ {\cal H}^{(2)}_{31} ~+~ {\cal H}^{(2)}_{43} ~.
\end{equation}
Although $\Psi_1(x,y)$ does not appear in our final expressions we have
included its symmetric point value for any intermediate checks. In addition we
have provided the values for the intermediate functions, $\chi_i(x,y)$ for
$i$~$=$~$1$, $2$ and $3$, which appear in the derivation of the master 
integrals at $O(\epsilon^2)$ and discussed in Appendix B. In performing this 
symmetric limit check it should be noted that in \cite{18} the master integrals
used for that computation were based on those given in \cite{55}. There it was 
assumed that two expressions were distinct. However, it transpires that 
$\Sigma$ is not independent and is related to another combination of numbers in
a master integral\footnote{We are particularly indebted to Dr M. Gorbahn on
this point.}. Specifically,
\begin{equation}
\Sigma ~=~ \frac{1}{36} \psi^{\prime\prime\prime}\left( \frac{1}{3} \right)
- \frac{2\pi^4}{27} ~.
\label{sigrel}
\end{equation} 
Though the final expressions and numerical values for the results of \cite{18} 
are still valid since (\ref{sigrel}) only simplifies the results. We have also
checked all our general expressions against another computation, \cite{56}, of 
each of the three vertices. That independent calculation was for the momentum
configuration used in \cite{35} and in this instance is a non-trivial check. 
The reason for this is that to compare with \cite{35} the definition of the 
overall mass scale $\tilde{\mu}$ used there, \cite{56}, is different from the 
scale $\mu$ used here. Therefore, to effect the comparison we have to allow for
the effect of extra logarithms deriving from the ratio 
$(\tilde{\mu}^2/\mu^2)^\epsilon$ which multiplies each loop integral. Once this
is included we find {\em precise} agreement. Indeed as part of this process we 
used the bare expressions provided in the attached data file for the 
verification. Hence, that non-trivial check ensures that those expressions are
correct and hence can be used to deduce the general vertex functions when 
$r^2$ is not tied to $\mu^2$. 

{\begin{figure}[ht]
\includegraphics[width=7.6cm,height=6cm]{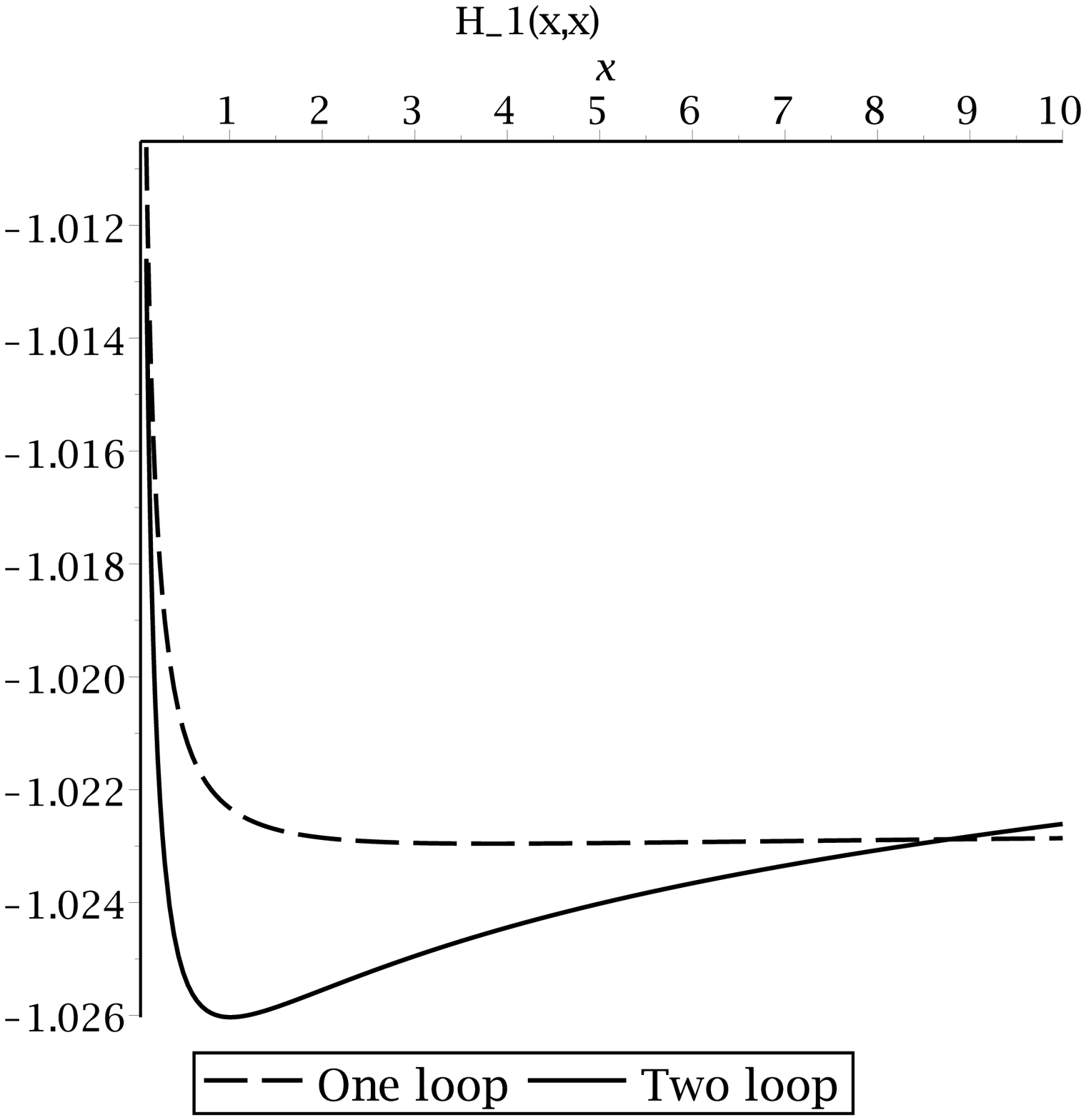}
\quad
\includegraphics[width=7.6cm,height=6cm]{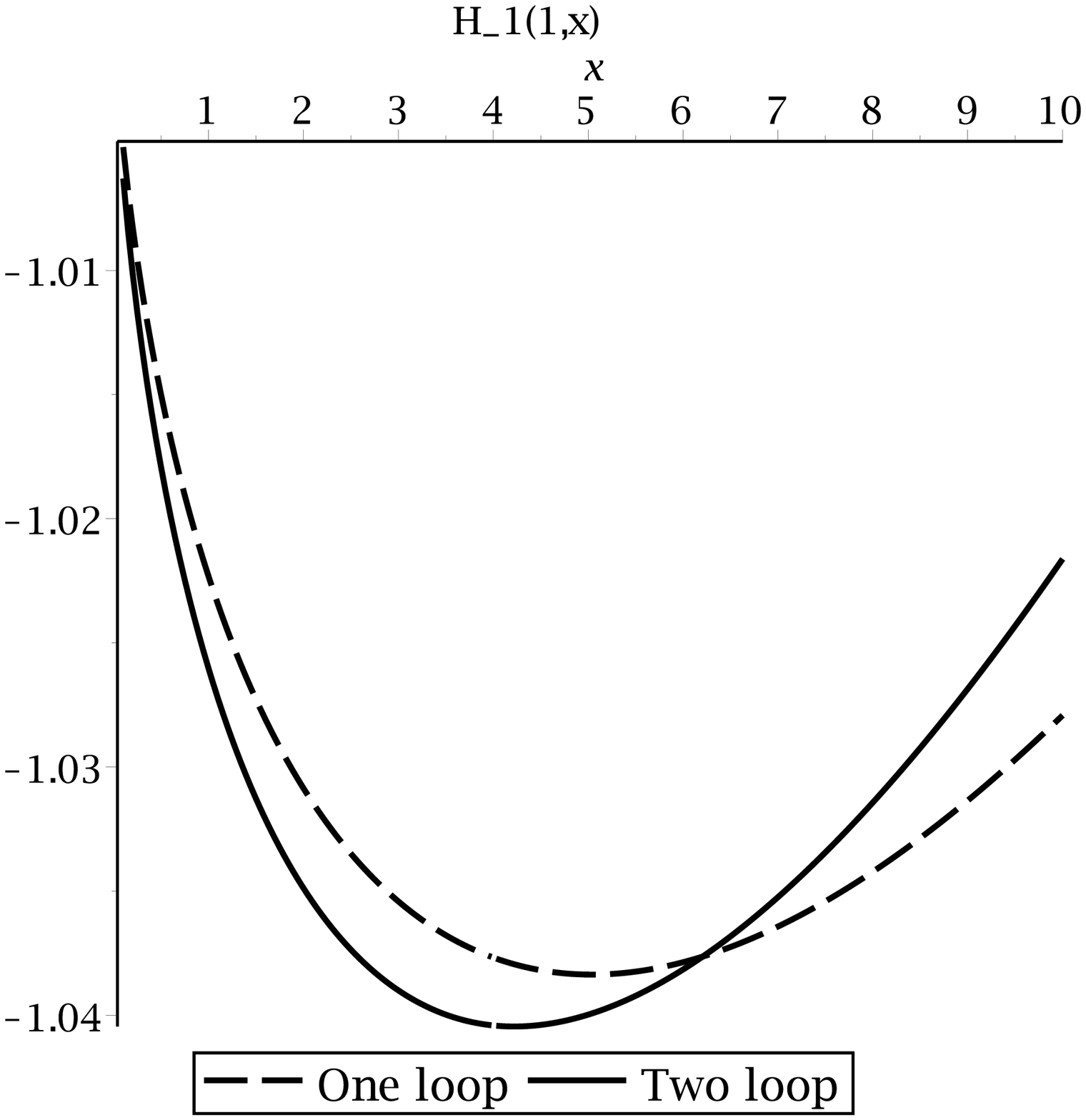}

\vspace{0.8cm}
\includegraphics[width=7.6cm,height=6cm]{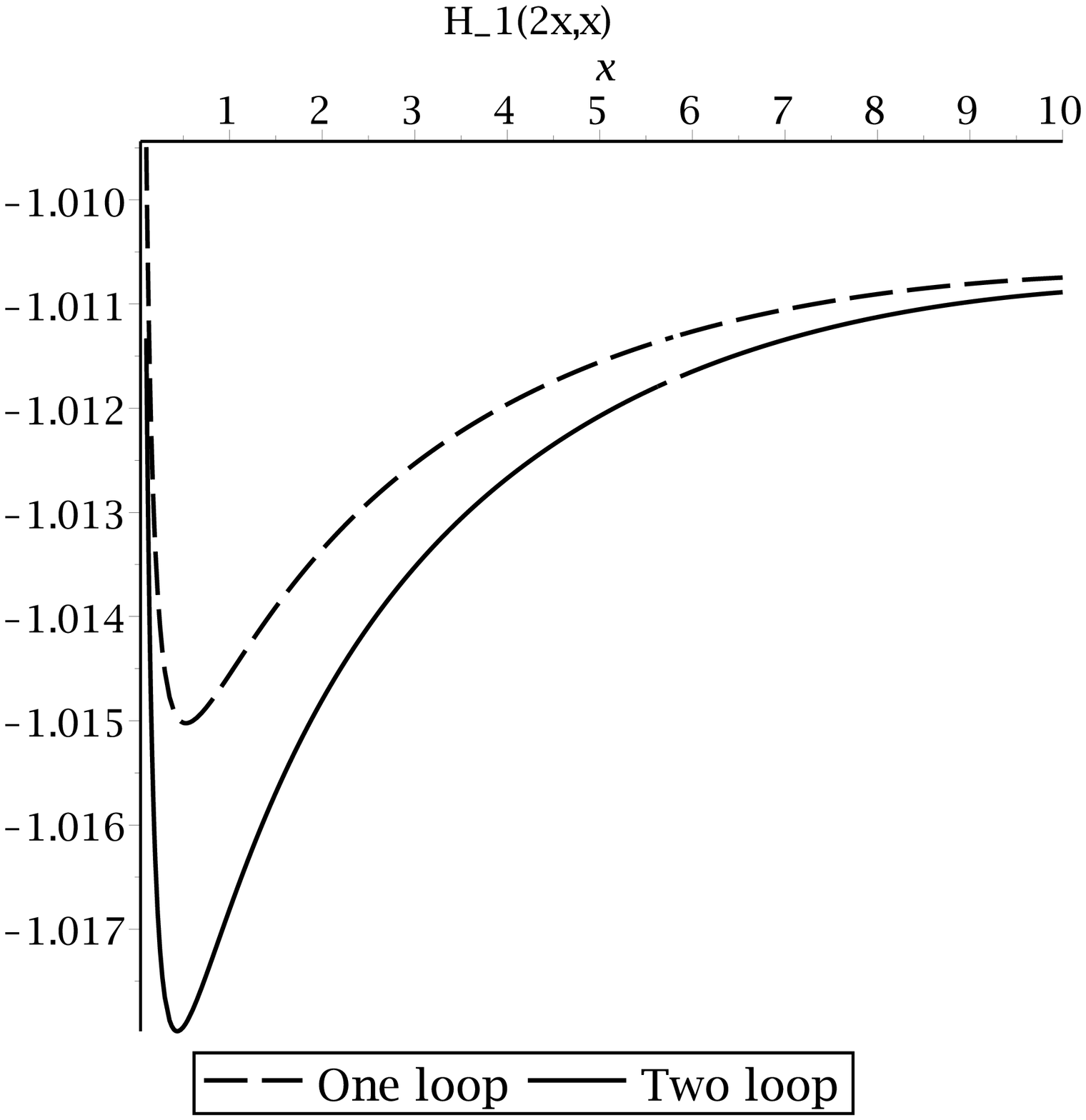}
\quad
\includegraphics[width=7.6cm,height=6cm]{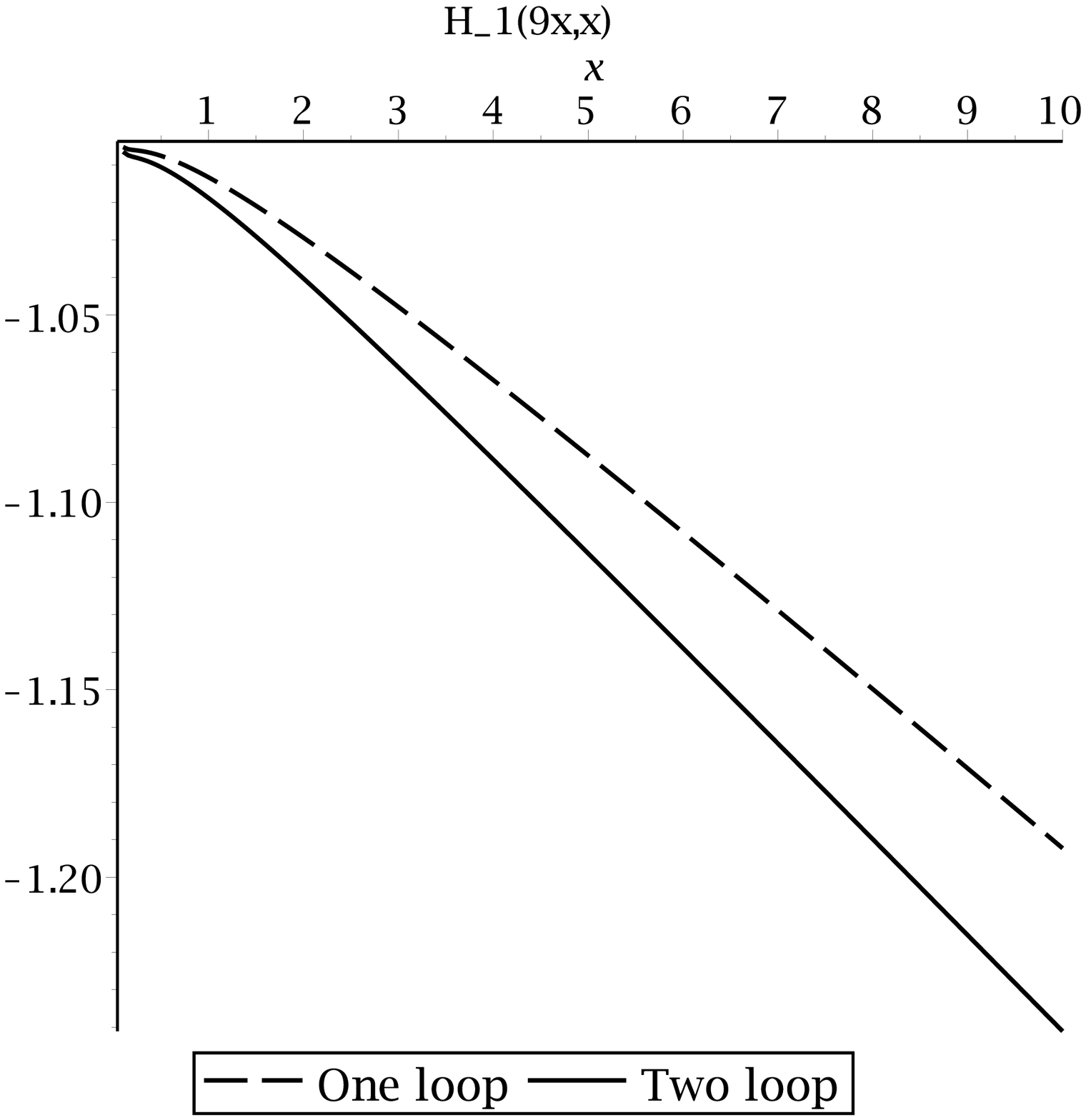}
\caption{Various sections of the Landau gauge channel $1$ ghost-gluon vertex
amplitude for $\alpha_s$~$=$~$0.125$.}
\end{figure}}

To appreciate the effect the two loop corrections have we have provided graphs
of the amplitudes for both channels in Figures $1$ and $2$. In these we use the
reduced notation
\begin{equation}
H_k(x,y) ~=~ \Sigma^{\mbox{\footnotesize{ccg}}}_{(k)}(p,q) 
\end{equation}
where $x$ and $y$ are the variables defined in (\ref{conf1}), and plot the 
shape of the functions along various lines on the $(x,y)$-plane. Though we will
use $x$ as the plotting variable in the argument of sections. Also our graphs
will all be in the Landau gauge and for a particular value of the coupling 
constant which is $\alpha_s$~$=$~$0.125$ where $\alpha_s$~$=$~$g^2/(4\pi)$. For
channel $1$ it is clear that the two loop corrections are not significantly 
different for a reasonably large range of the variables. Though, of course, 
this is for a specific choice of coupling constant and a much larger value 
would lead to the two loop contributions dominating. One interesting aspect of 
the two loop correction is that for such a large expression, (\ref{ghost1}), 
there is not a significant deviation from the one loop result for this channel.
A similar feature emerges for other channels and vertices. For instance, for 
channel $2$ the two loop correction appears to have a smaller deviation from 
the one loop part in comparison. For two of the graphs in Figure $2$ there 
appear to be gaps around $4$ and $6$. These arise from singularities due to the
Gram determinant as is apparent in (\ref{ghost1}). For the ghost-gluon vertex 
this is a simple pole. However, as is evident in the full expressions given in 
the data file, for the other vertices the singularity is more severe being a 
double pole for the quark-gluon vertex and a triple pole for the triple gluon 
vertex.

{\begin{figure}[hb]
\includegraphics[width=7.6cm,height=6cm]{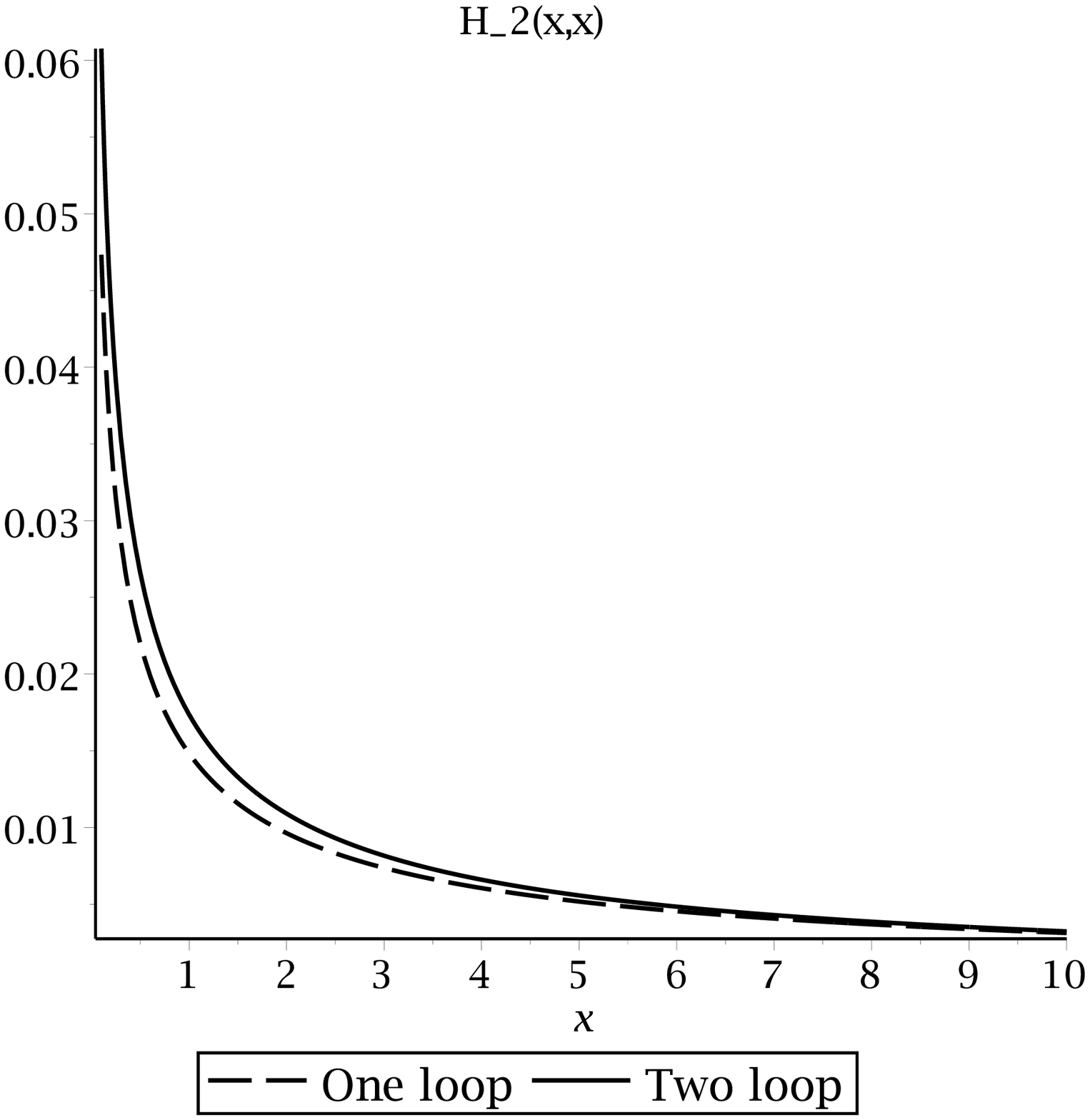}
\quad
\includegraphics[width=7.6cm,height=6cm]{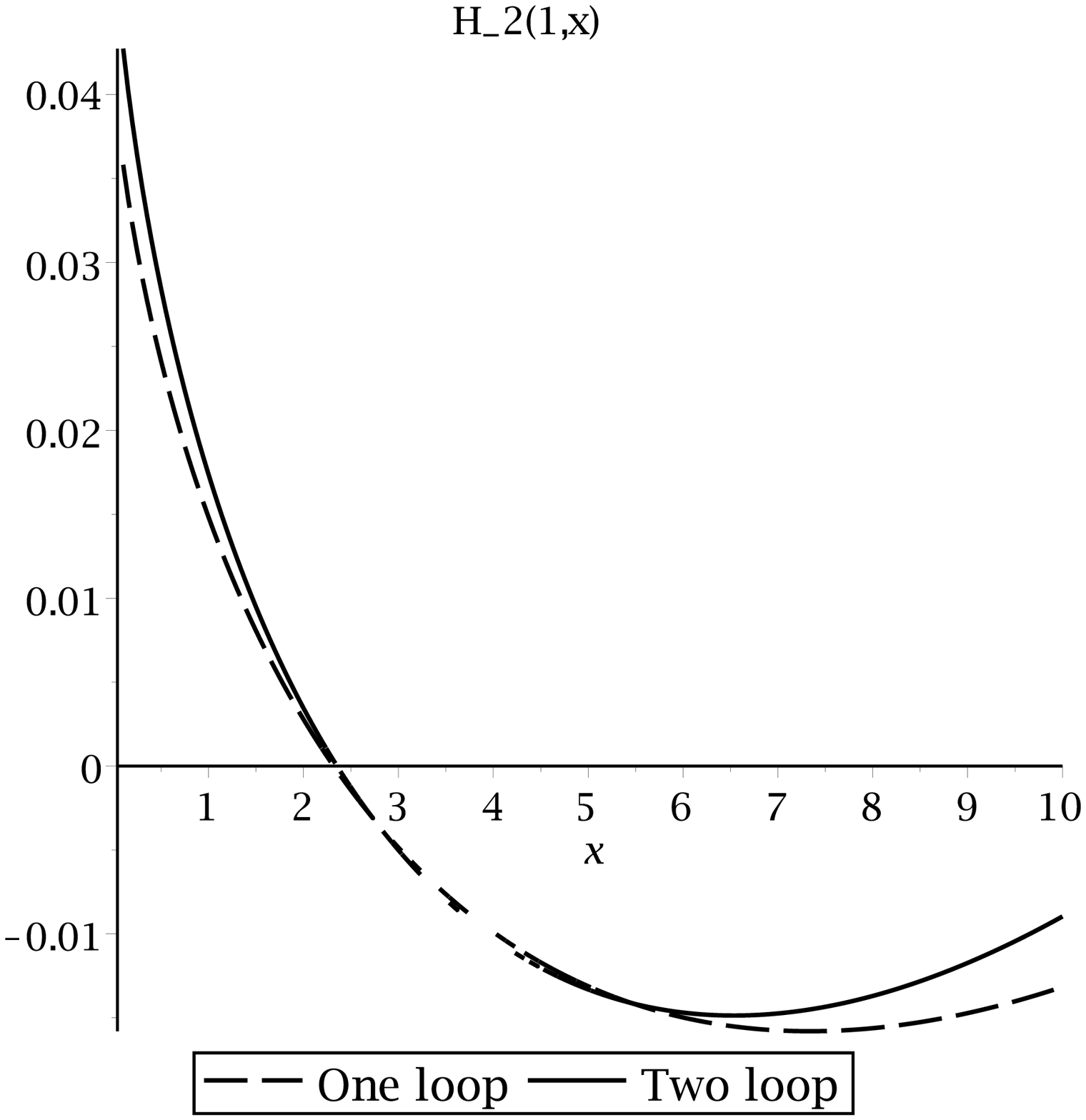}

\vspace{0.8cm}
\includegraphics[width=7.6cm,height=6cm]{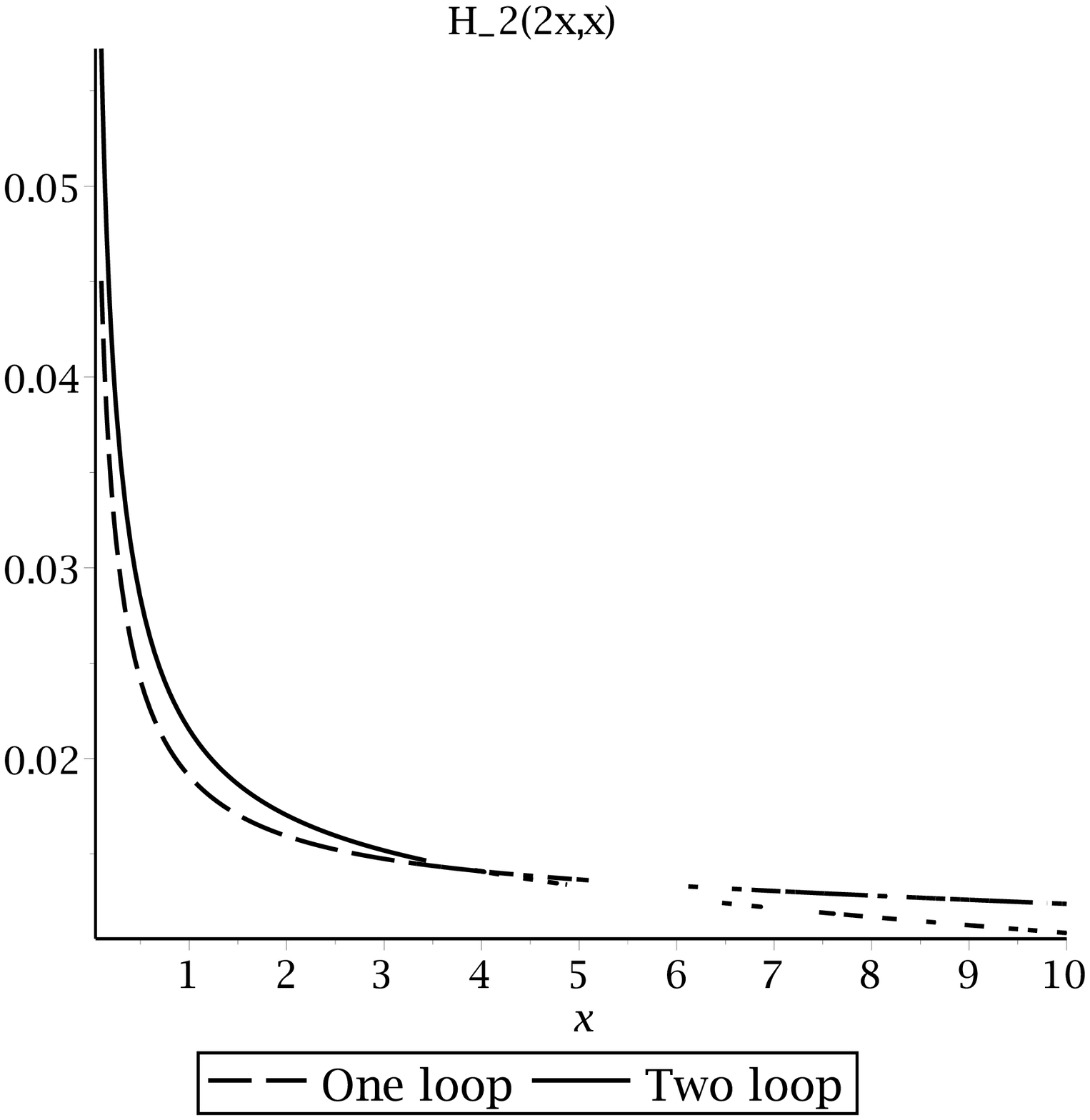}
\quad
\includegraphics[width=7.6cm,height=6cm]{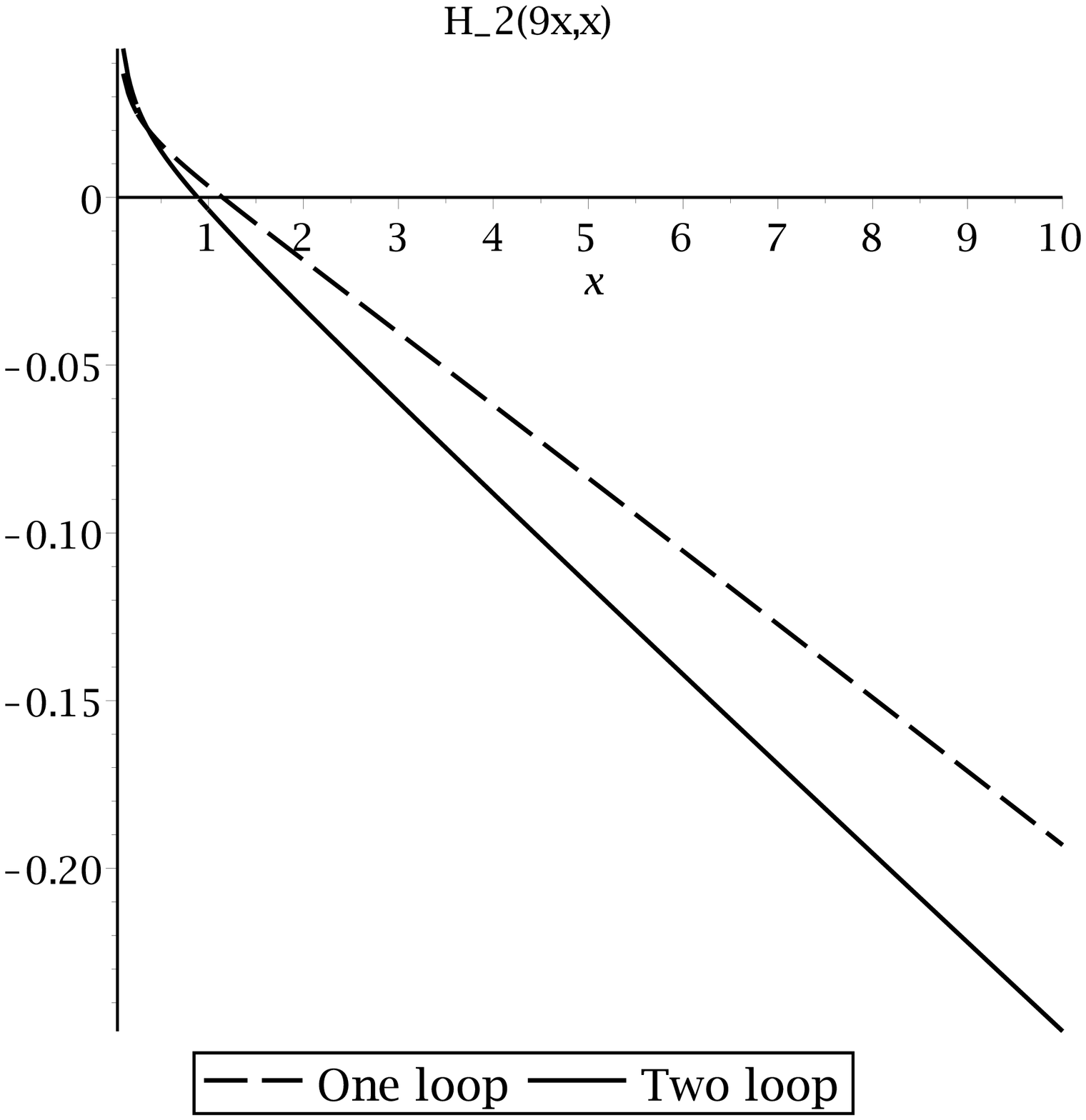}
\caption{Various sections of the Landau gauge channel $2$ ghost-gluon vertex
for $\alpha_s$~$=$~$0.125$.}
\end{figure}}

\sect{Quark-gluon vertex.}

We continue the presentation of the results by focusing on the quark-gluon
vertex in this section. Again we have checked that the correct symmetric point 
$\MSbar$ expressions emerge in comparison with \cite{18} for arbitrary $\alpha$
as well as performing the other checks discussed for the ghost-gluon vertex. 
Similarly to the previous case we focus on graphical illustrations. For the 
figures relating to the quark-gluon vertex amplitudes our syntax is 
\begin{equation}
Q_k(x,y) ~=~ \Sigma^{\mbox{\footnotesize{qqg}}}_{(k)}(p,q) ~.
\end{equation}
In Figure $3$ we show the same section of all six amplitudes for this vertex in
order to compare. Clearly, the two loop corrections for this specific value of
the coupling constant are effectively negligible. The same is true for other
sections as is clear in Figure $4$ and this indicates the more severe nature
of the singularity.

{\begin{figure}
\includegraphics[width=7.6cm,height=6cm]{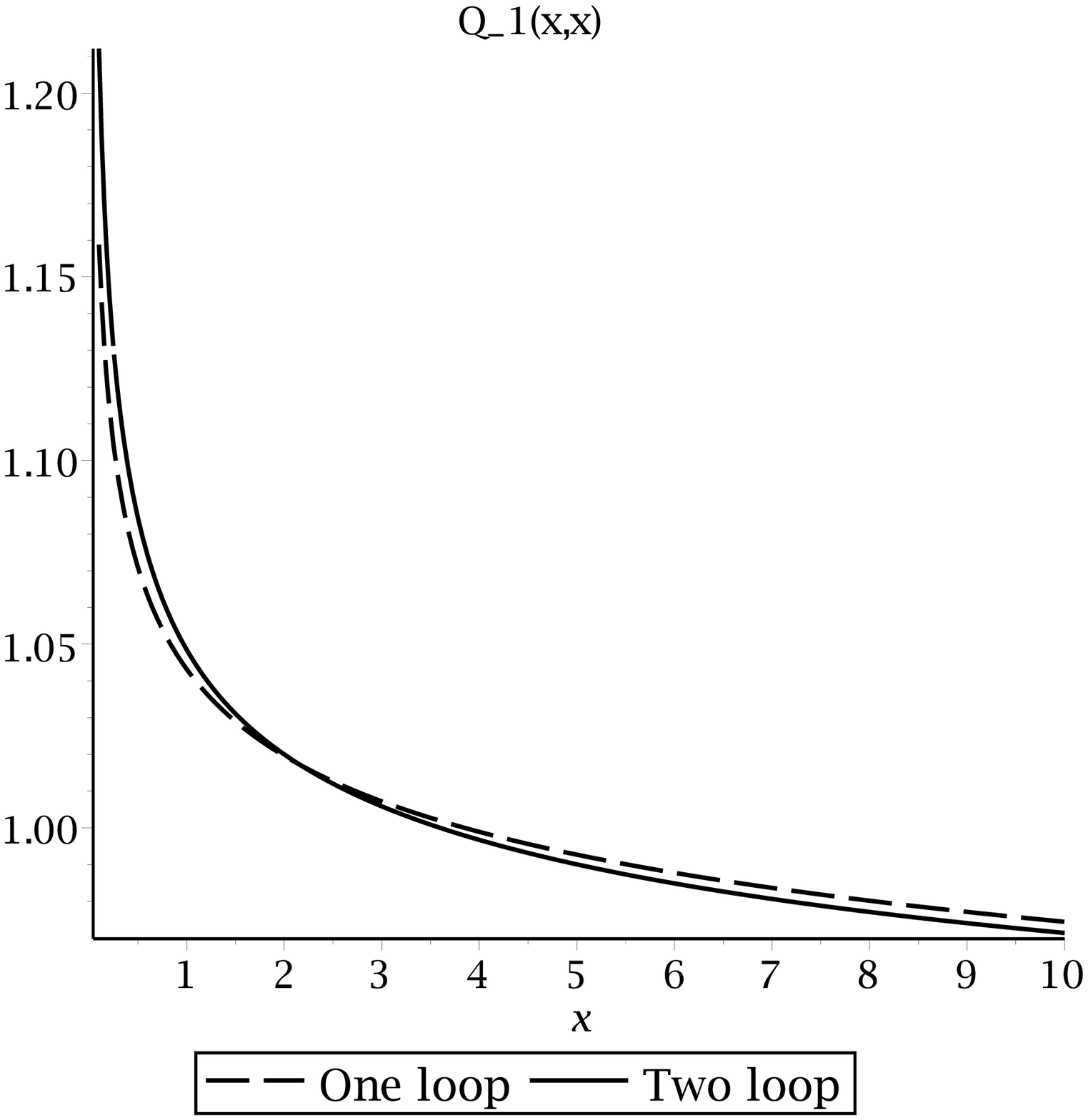}
\quad
\includegraphics[width=7.6cm,height=6cm]{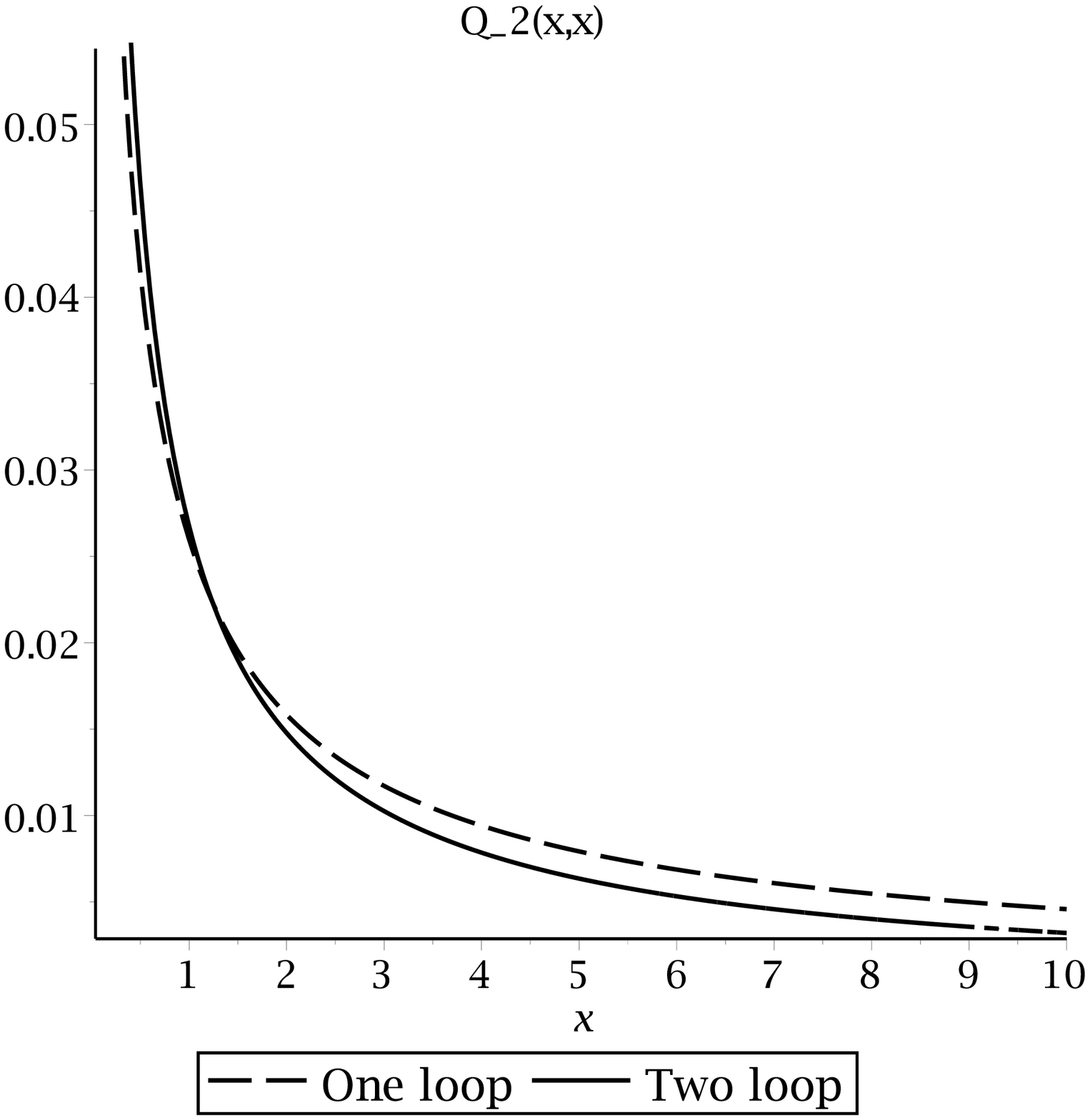}

\vspace{0.8cm}
\includegraphics[width=7.6cm,height=6cm]{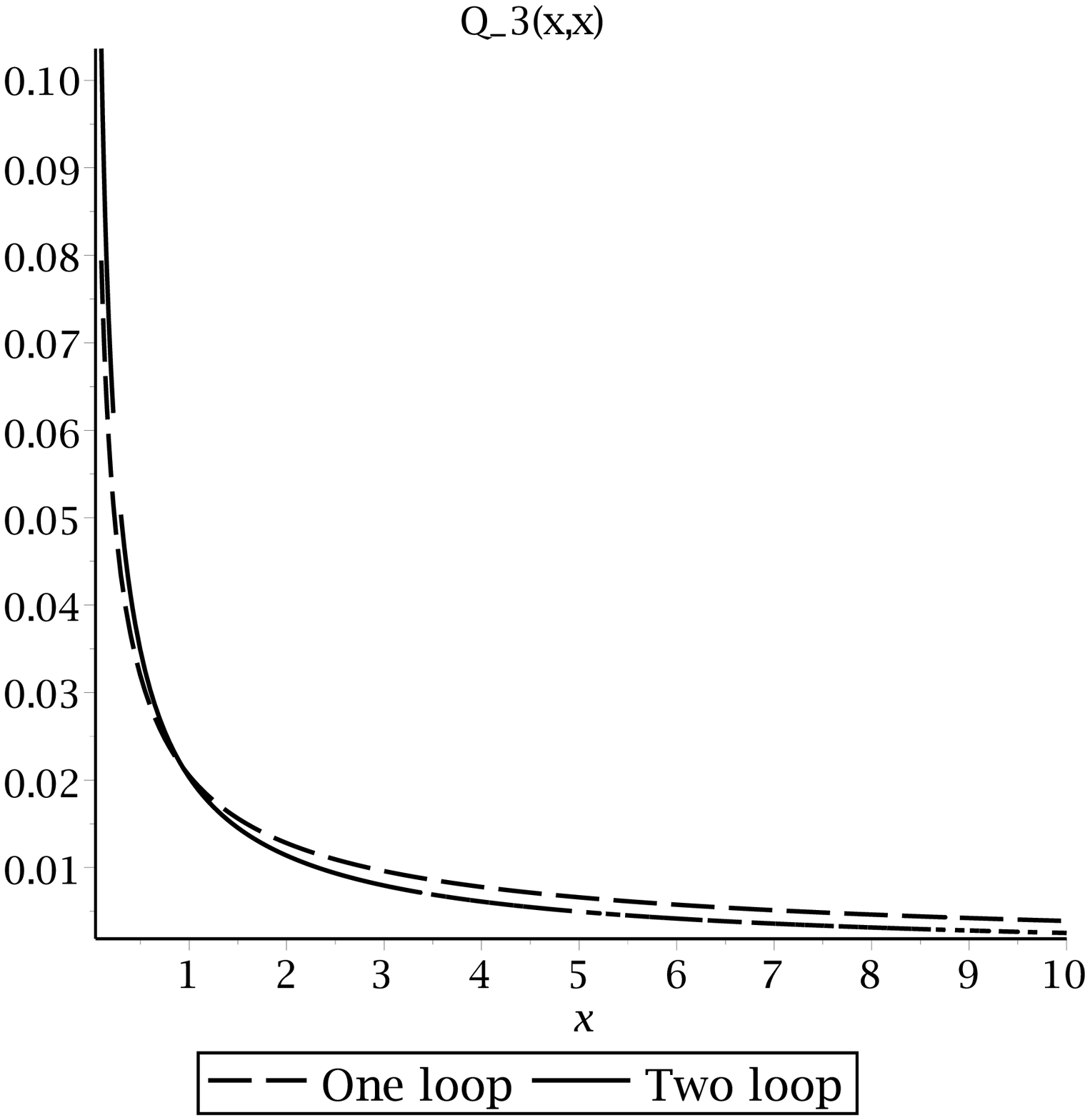}
\quad
\includegraphics[width=7.6cm,height=6cm]{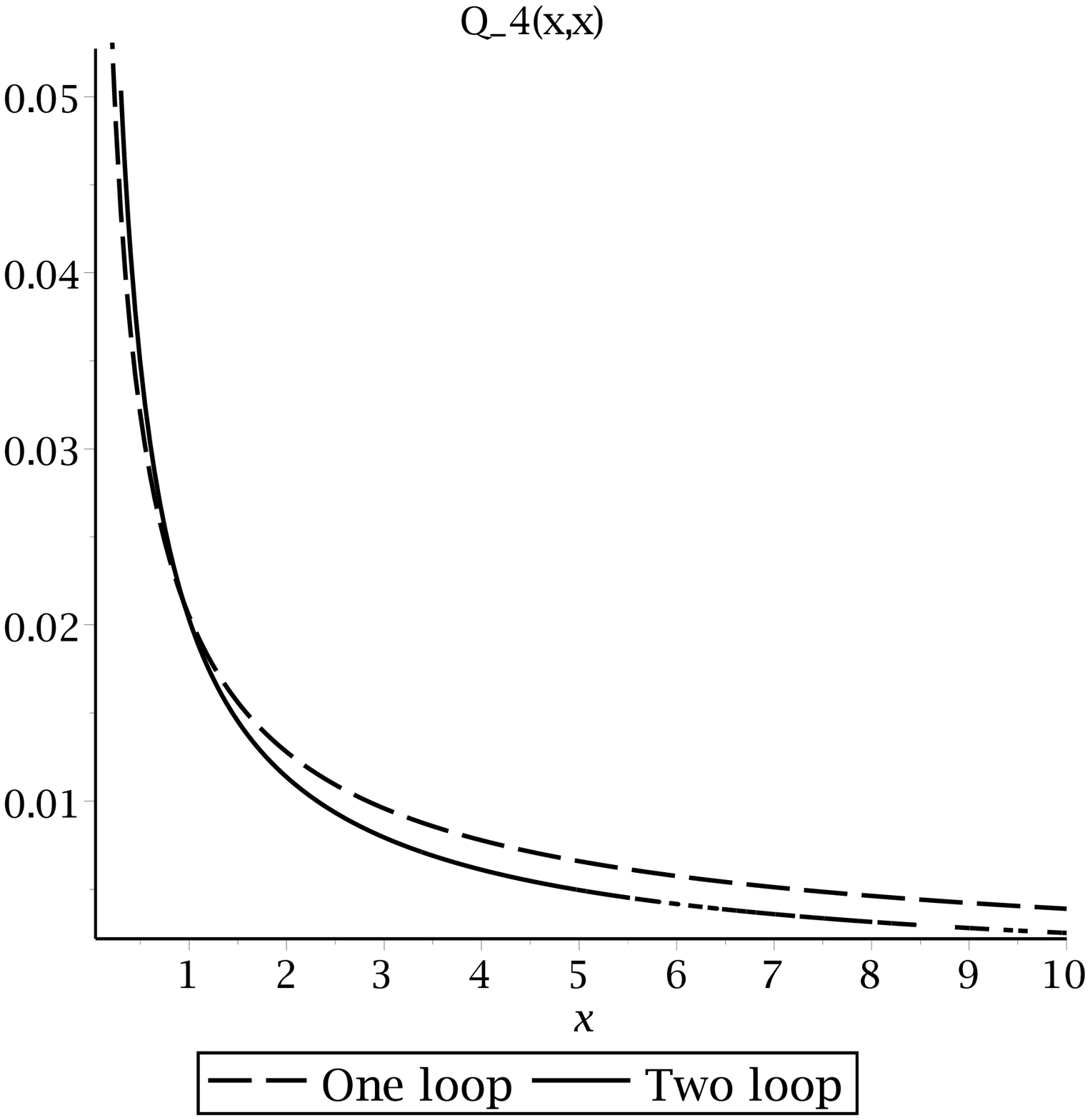}

\vspace{0.8cm}
\includegraphics[width=7.6cm,height=6cm]{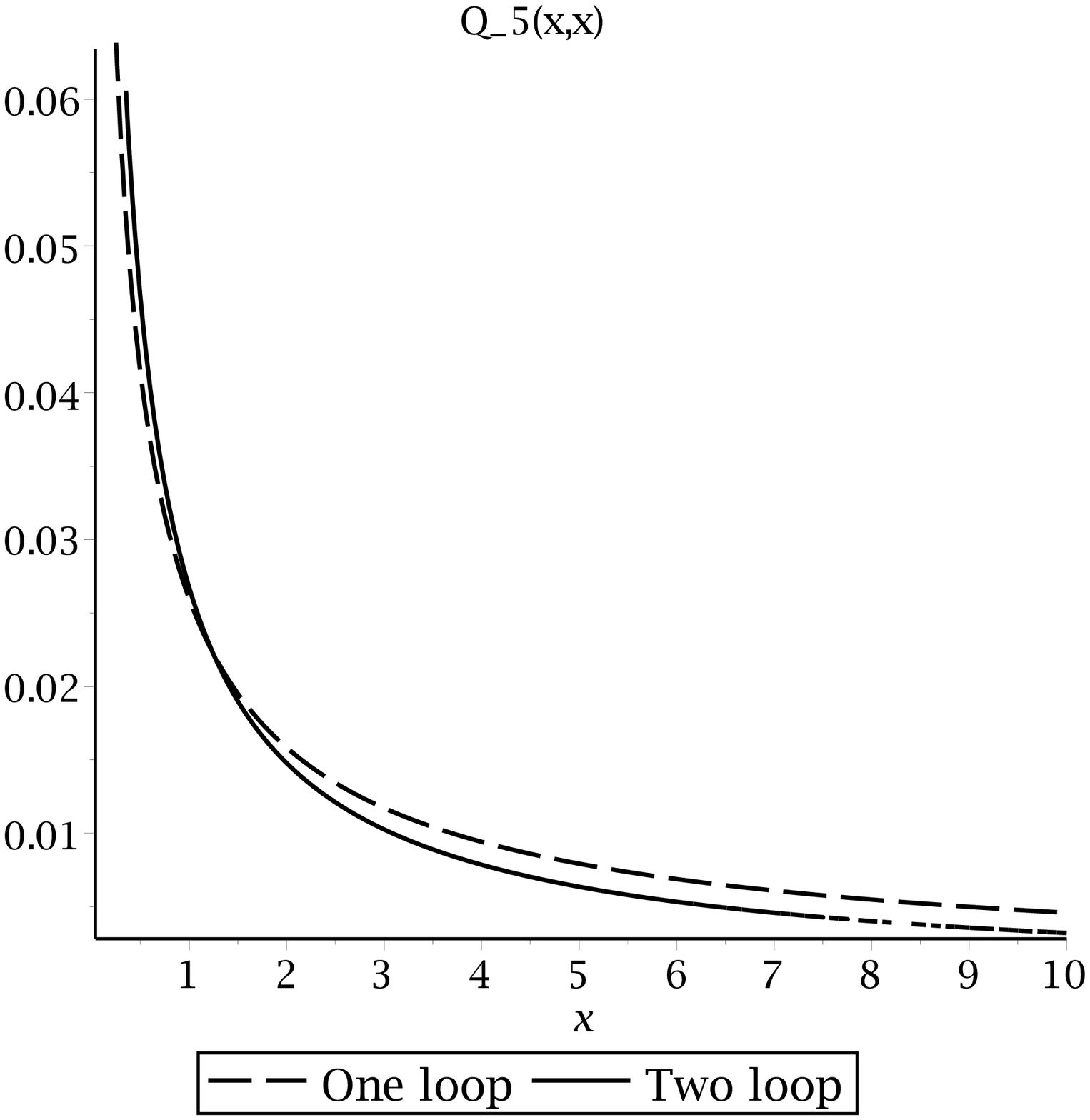}
\quad
\includegraphics[width=7.6cm,height=6cm]{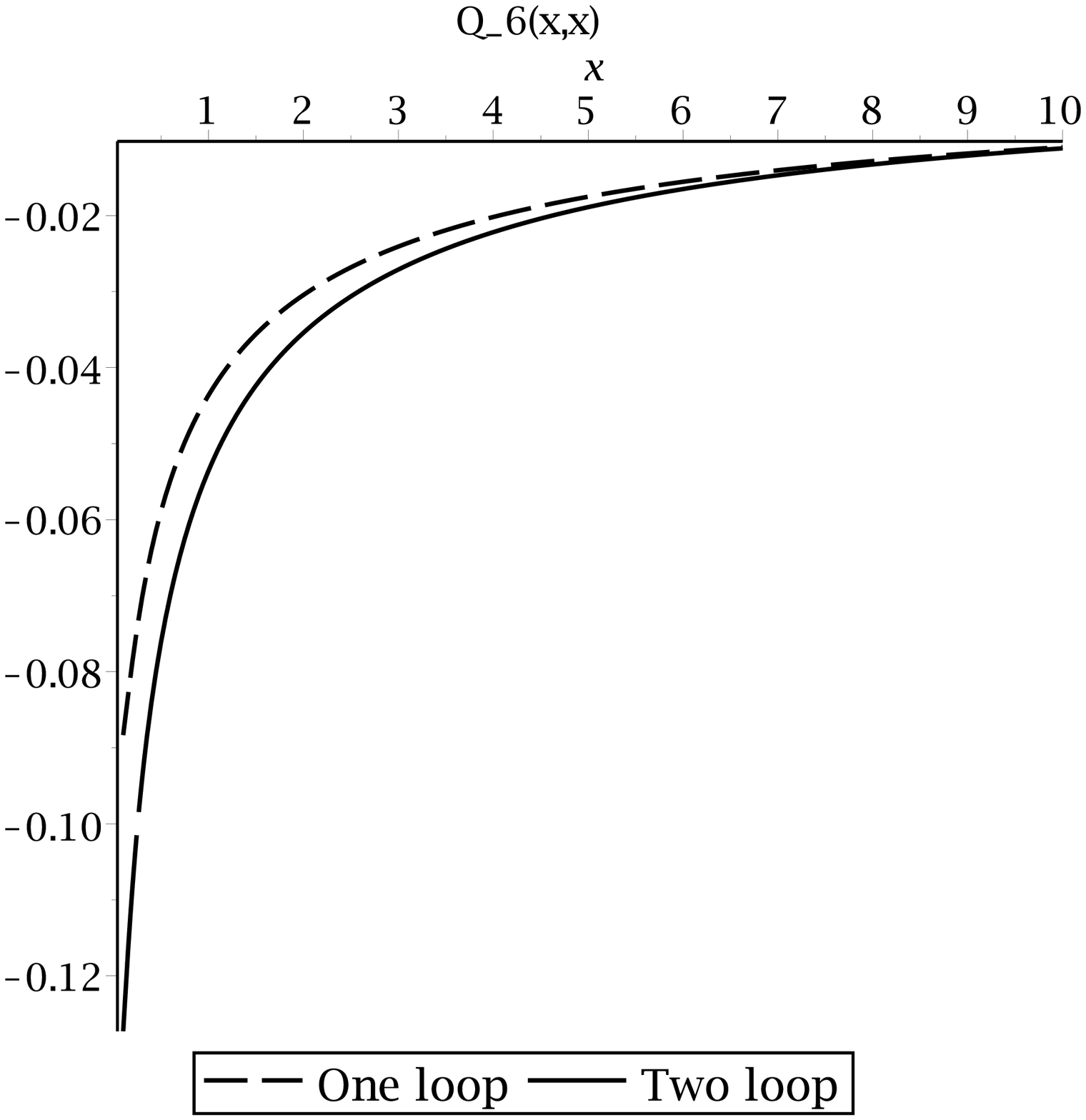}

\caption{Section $(x,x)$ of the Landau gauge channels of the quark-gluon vertex
for $\alpha_s$~$=$~$0.125$.}
\end{figure}}

{\begin{figure}
\includegraphics[width=7.6cm,height=6cm]{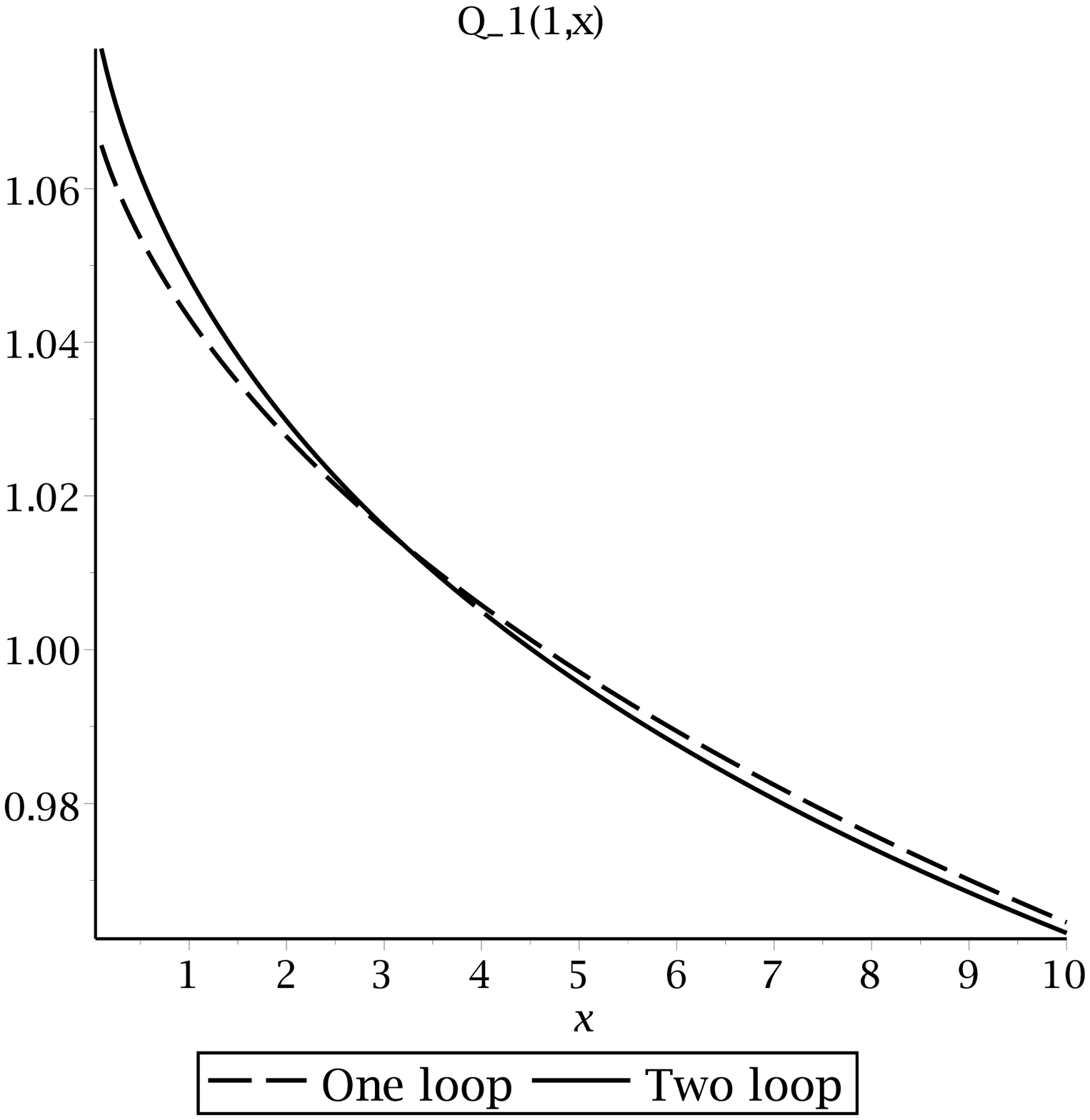}
\quad
\includegraphics[width=7.6cm,height=6cm]{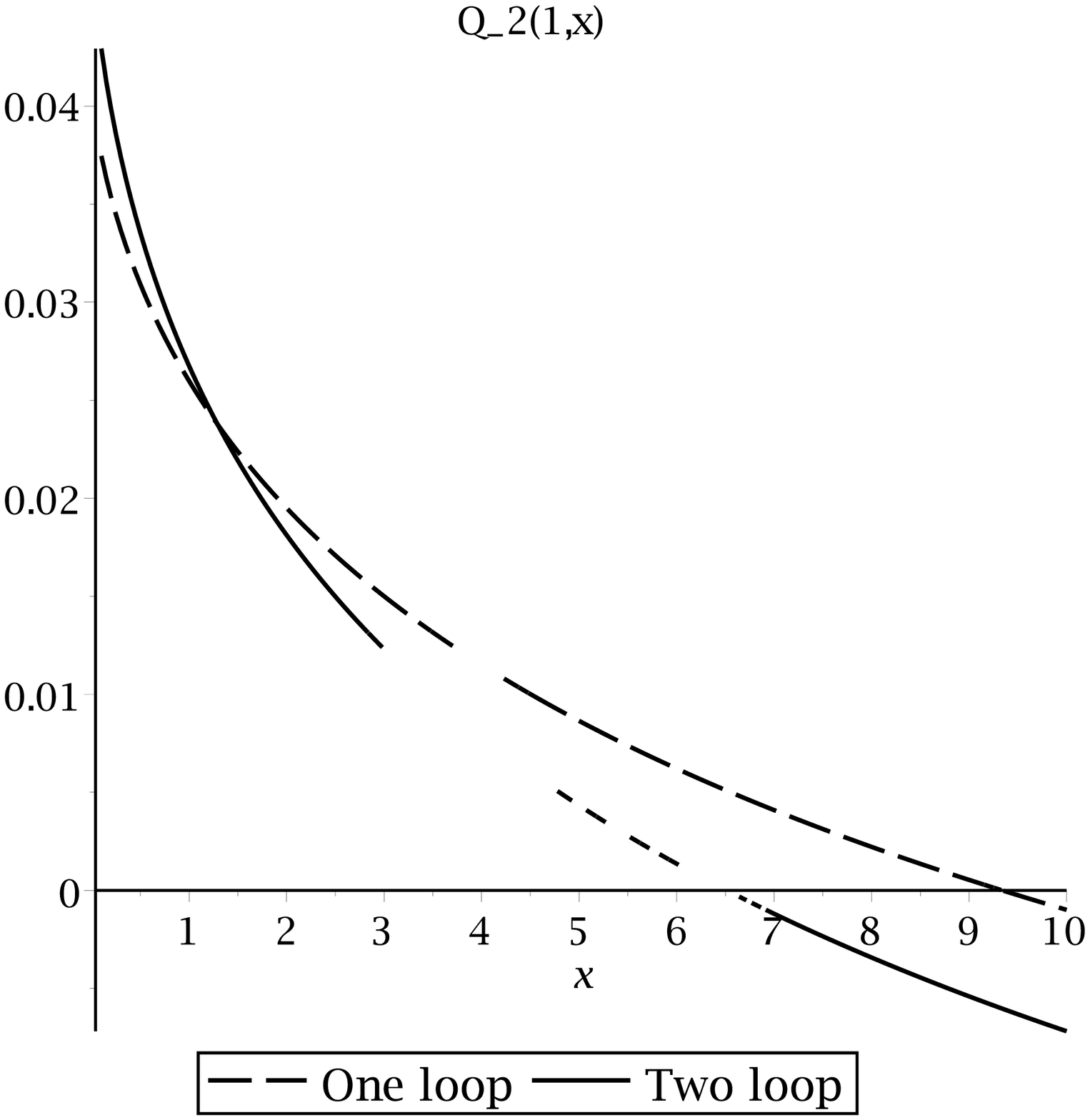}

\vspace{0.8cm}
\includegraphics[width=7.6cm,height=6cm]{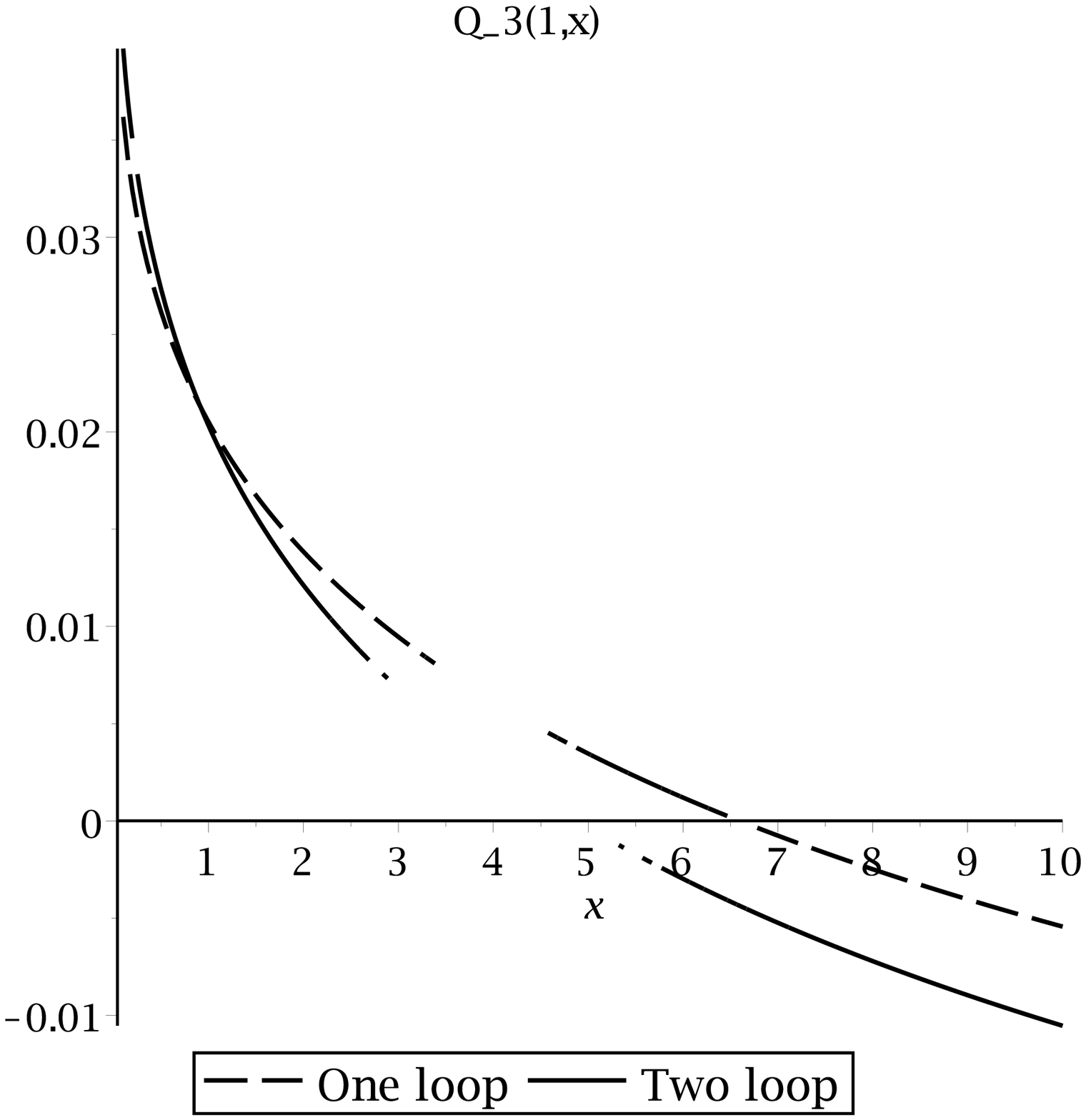}
\quad
\includegraphics[width=7.6cm,height=6cm]{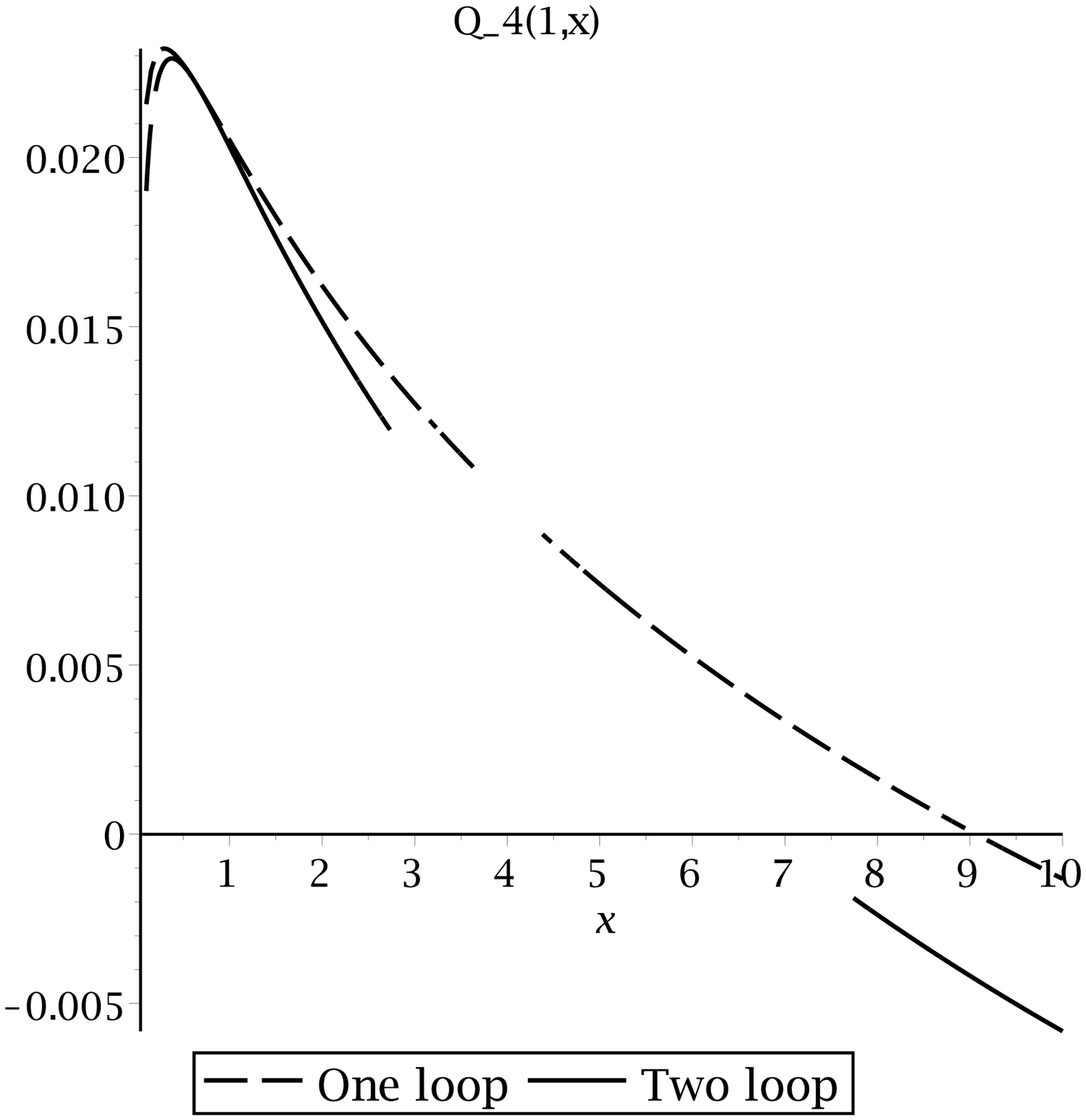}

\vspace{0.8cm}
\includegraphics[width=7.6cm,height=6cm]{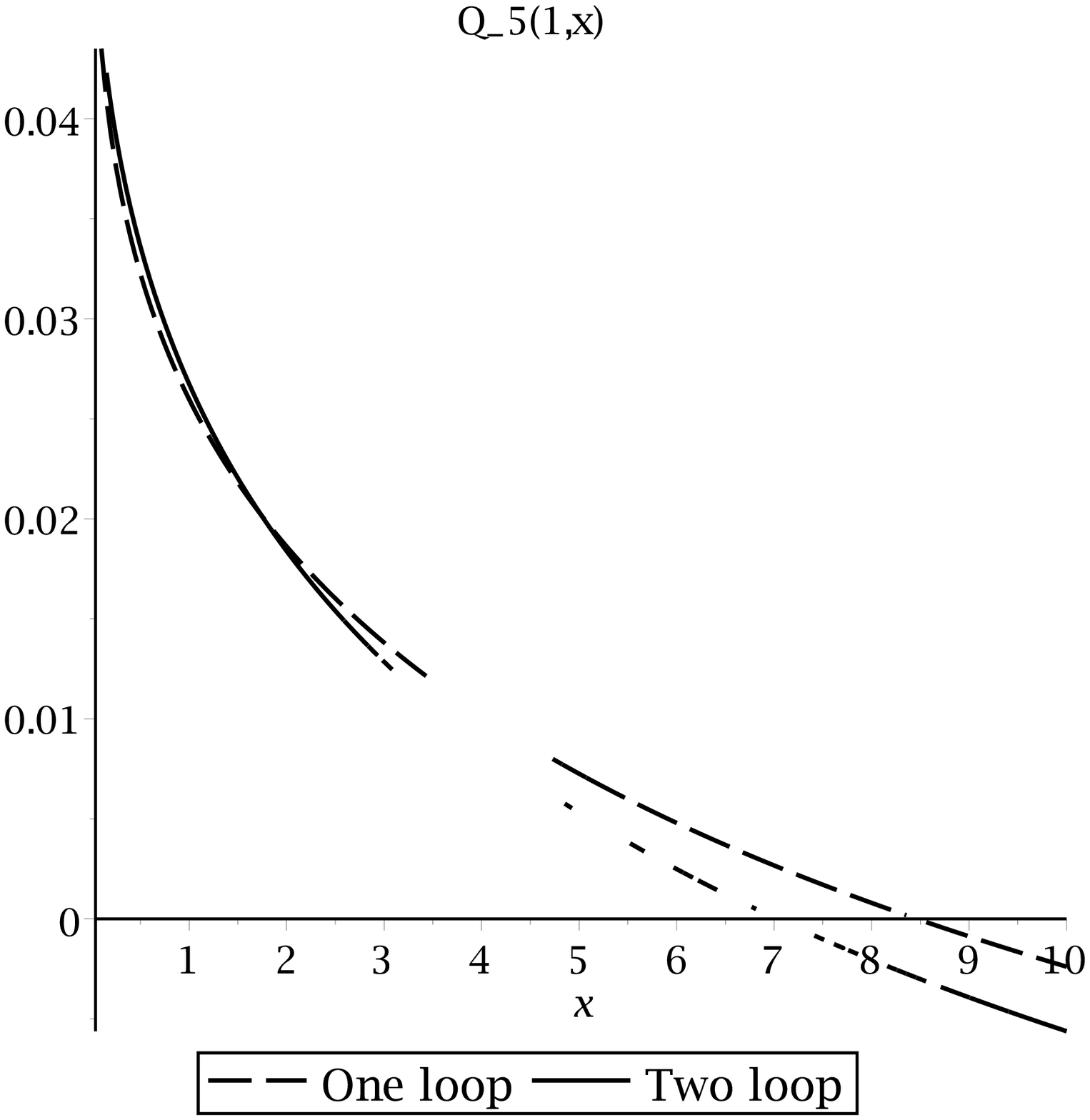}
\quad
\includegraphics[width=7.6cm,height=6cm]{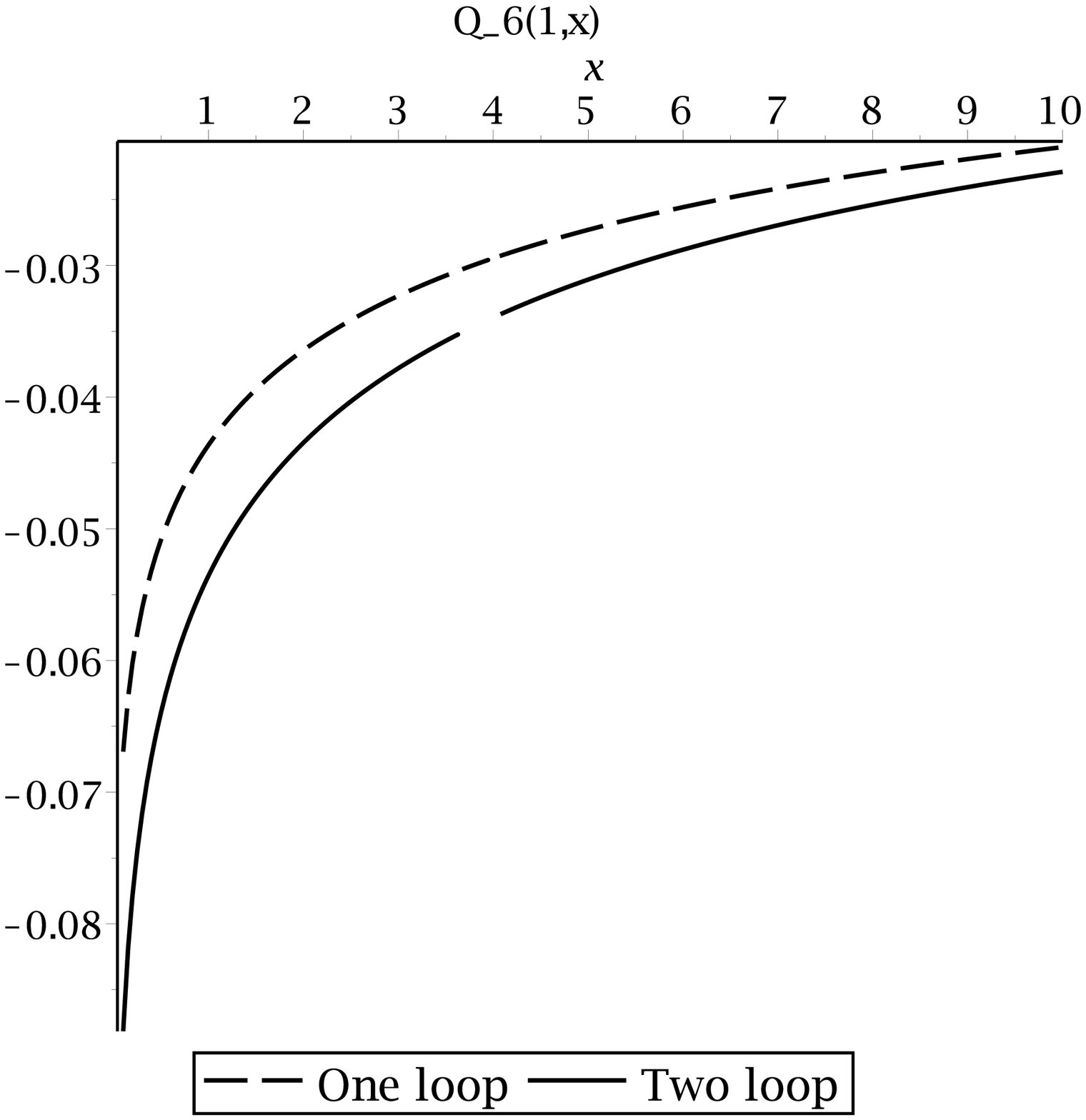}

\caption{Section $(1,x)$ of the Landau gauge channels of the quark-gluon vertex
for $\alpha_s$~$=$~$0.125$.}
\end{figure}}

\sect{Triple gluon vertex.}

Finally, we discuss the situation with the triple gluon vertex in the Landau 
gauge. Of the three cases the results for this vertex are the most involved due
in part to the number of graphs and the large number of amplitudes. For the
latter it is instructive to focus on a representative sample and then comment 
on the remainder. First we note that we have again verified that in the limit 
to the symmetric point the previous expressions given in \cite{18} are obtained
for all $\alpha$ as well as checking against \cite{56} for all amplitudes and 
for all $\alpha$. Unlike the other two vertices the integration by parts 
routine produces denominator factors of $(1-x)$, $(1-y)$ and $(x-y)$ as part of
the intermediate algebra. Such singularities could be problematic in the limit 
to the symmetric point but it is reassuring to note that they cancelled when 
the full set of diagrams was summed prior to renormalization. This provided 
another useful check. The notation used for our graphs here is 
\begin{equation}
G_k(x,y) ~=~ \Sigma^{\mbox{\footnotesize{ggg}}}_{(k)}(p,q) ~.
\end{equation}
In Figure $5$ we have plotted the section along $y$~$=$~$x$ for the six 
amplitudes which derive from the triple gluon vertex Feynman rule itself as 
these are of more interest. Hence there is a degree of symmetry in the graphs. 
This is apparent in the pairings $(1,4)$, $(2,6)$ and $(3,5)$ recalling that 
the leg with Lorentz index $\sigma$, (\ref{vertdef}), has momentum $r$ flowing 
through it and is a graphical illustration of a consistency check. By way of 
variation we have given an alternative section in Figure $6$ where the 
singularity is more evident. Of course for this section there is no analogous 
symmetry compared to the $(x,x)$ section. However, aside from this the two loop 
corrections do not differ significantly from the one loop situation for the 
particular value of $\alpha_s$ chosen. Indeed for the remaining channels the
situation is similar to the other vertices for the non-Feynman rule amplitudes 
in that the two loop corrections are not significant. Moreover, in complete
parallel with the graphs of Figure $2$ they are numerically small. 

{\begin{figure}
\includegraphics[width=7.6cm,height=6cm]{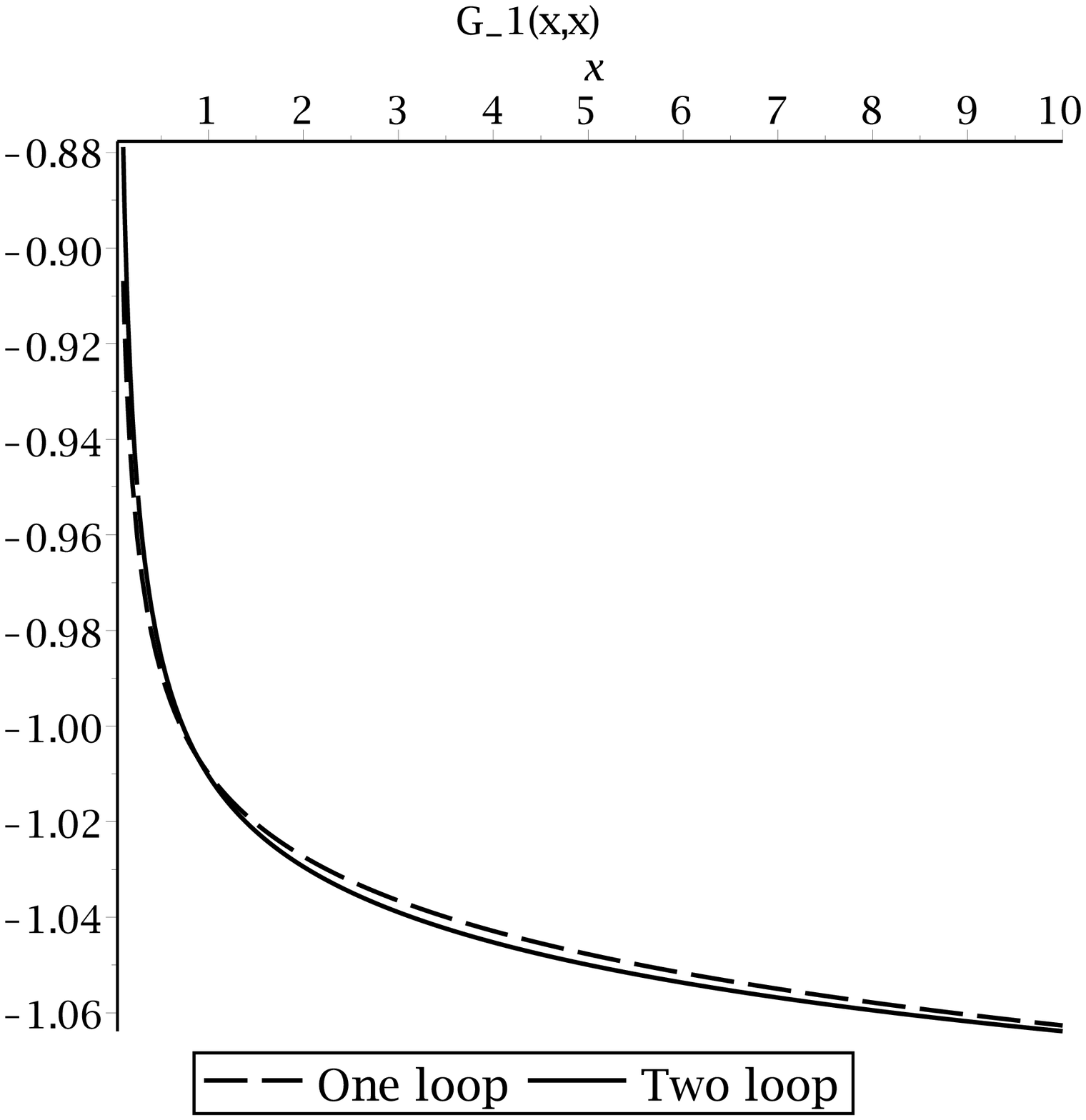}
\quad
\includegraphics[width=7.6cm,height=6cm]{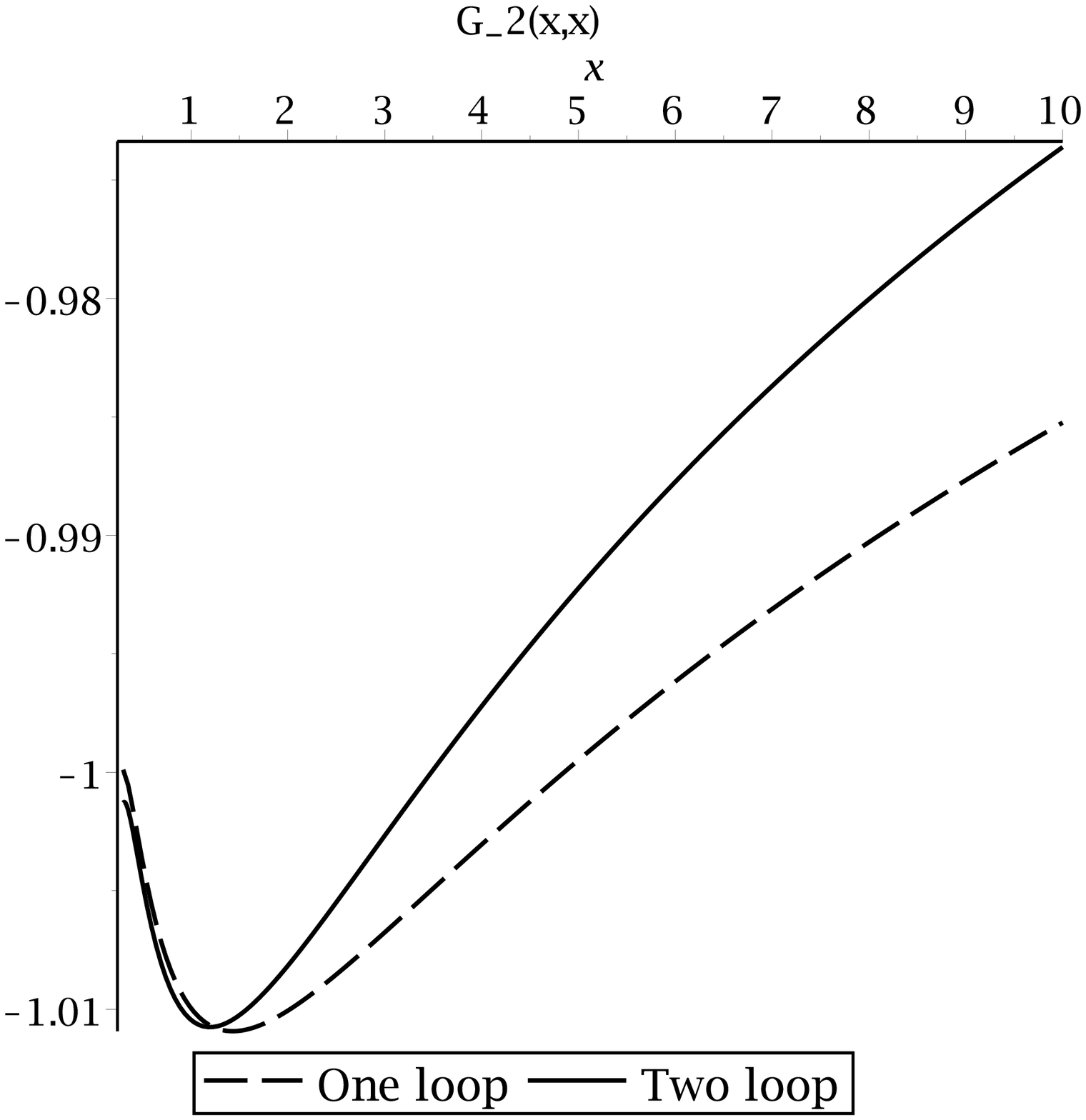}

\vspace{0.8cm}
\includegraphics[width=7.6cm,height=6cm]{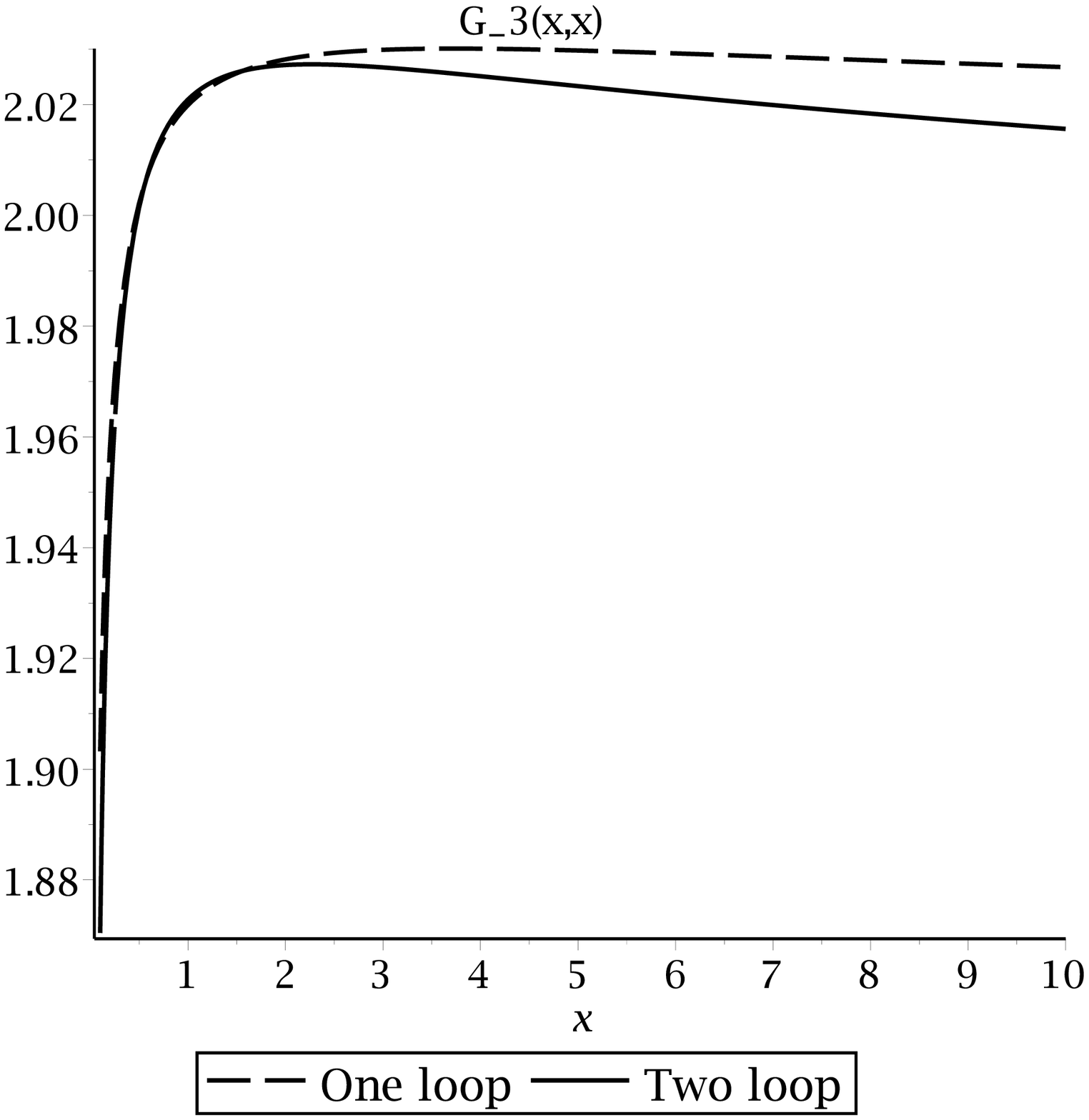}
\quad
\includegraphics[width=7.6cm,height=6cm]{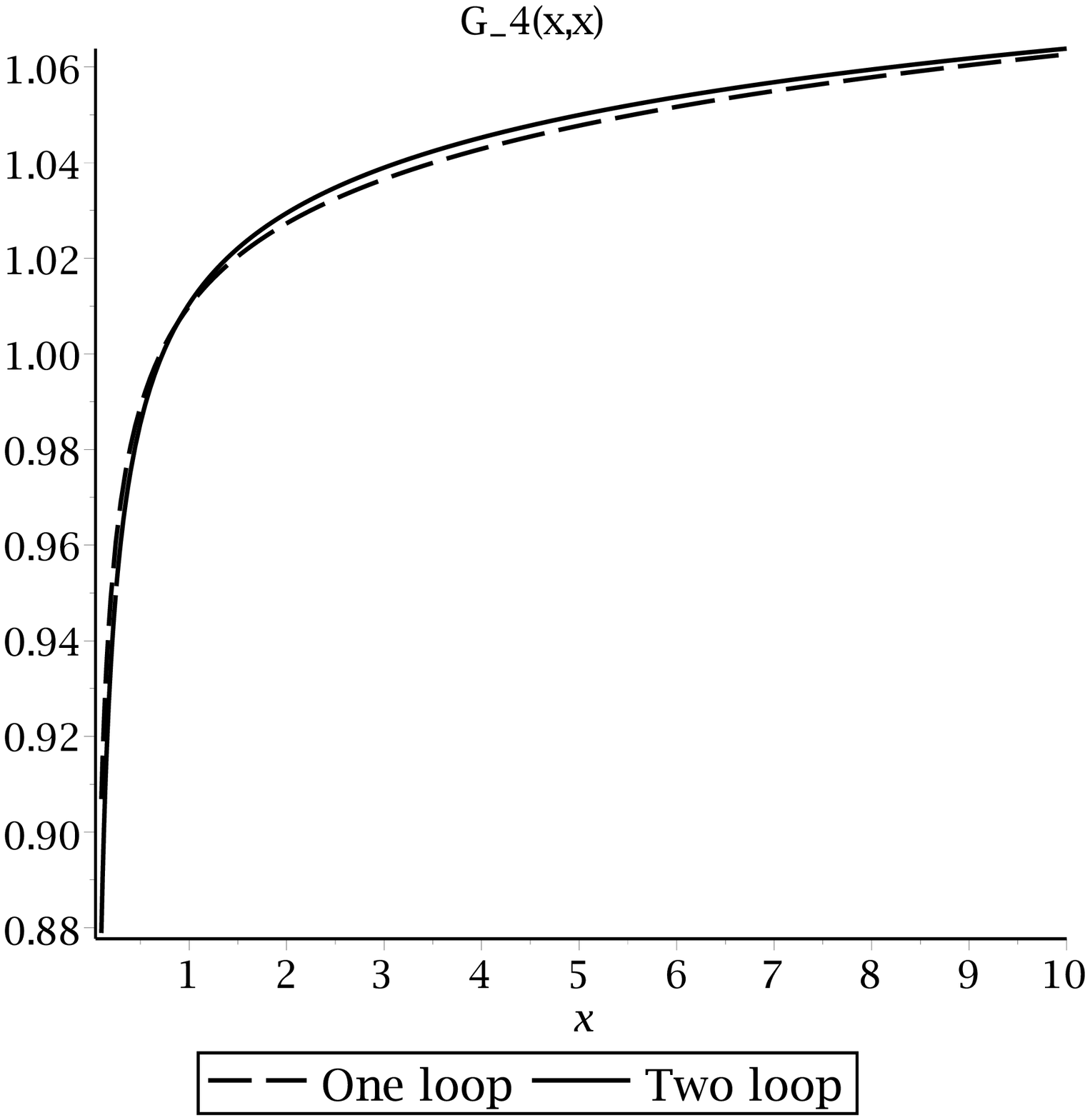}

\vspace{0.8cm}
\includegraphics[width=7.6cm,height=6cm]{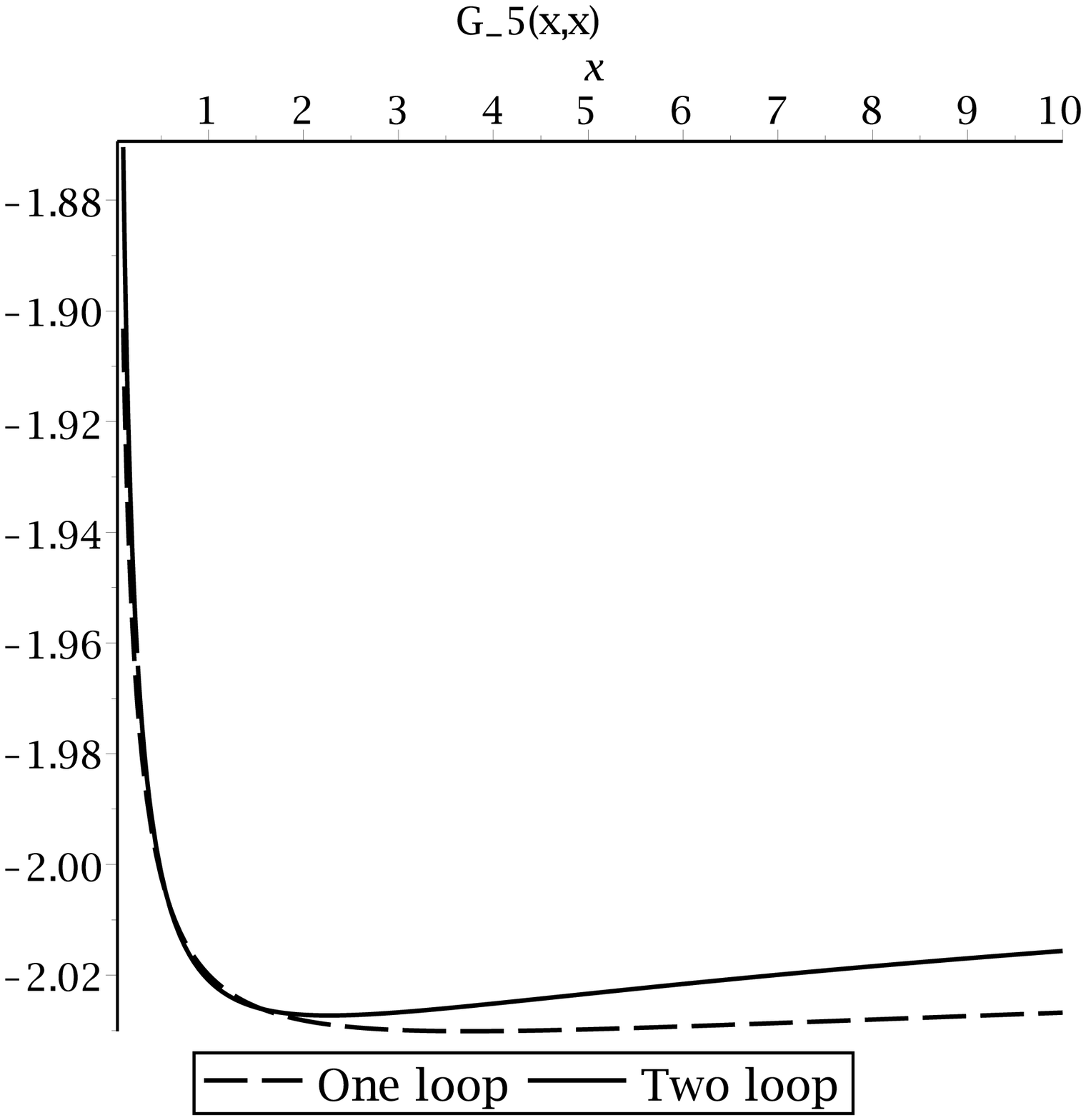}
\quad
\includegraphics[width=7.6cm,height=6cm]{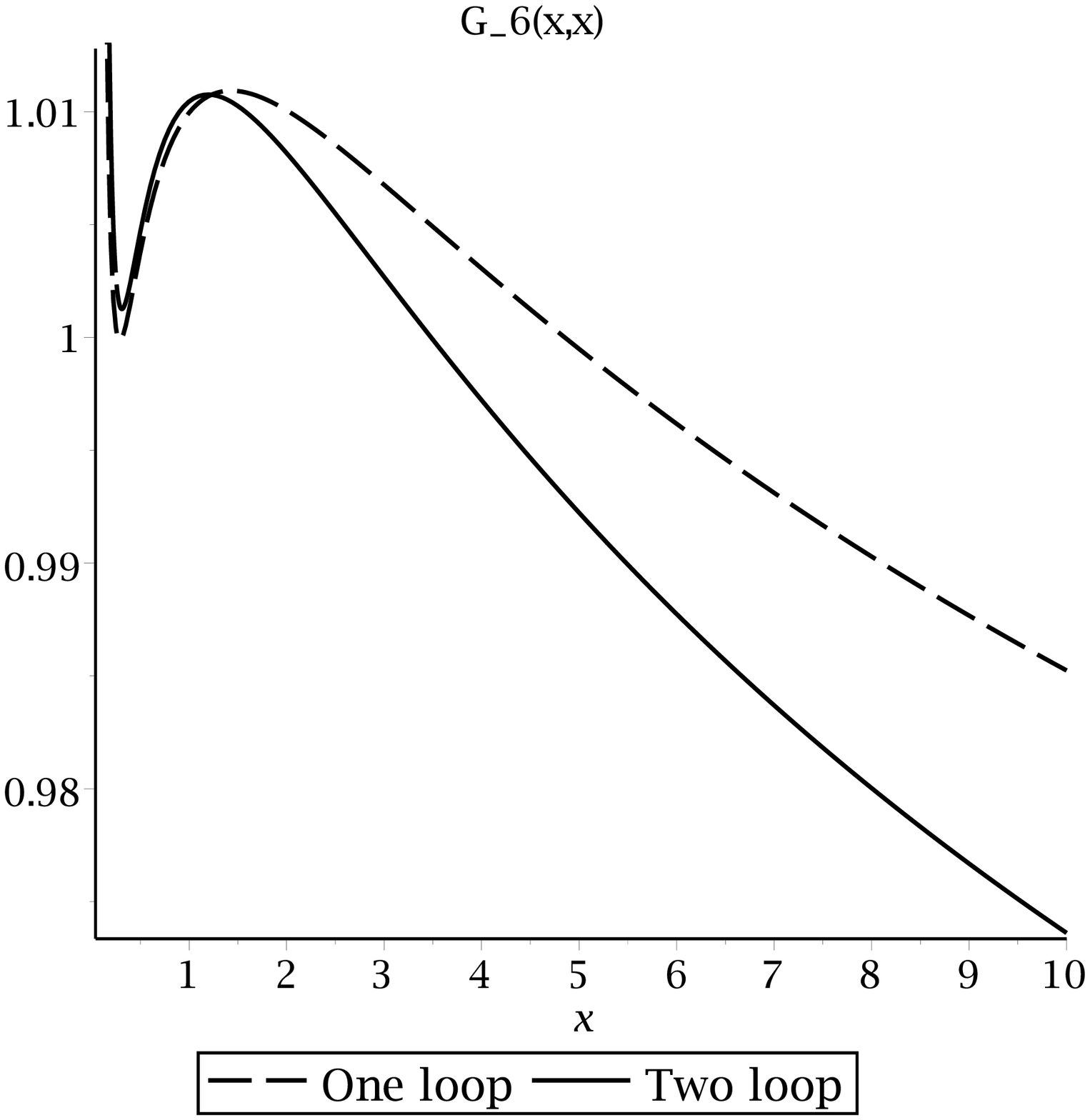}

\caption{Section $(x,x)$ of various Landau gauge channels of the triple gluon 
vertex for $\alpha_s$~$=$~$0.125$.}
\end{figure}}

{\begin{figure}
\includegraphics[width=7.6cm,height=6cm]{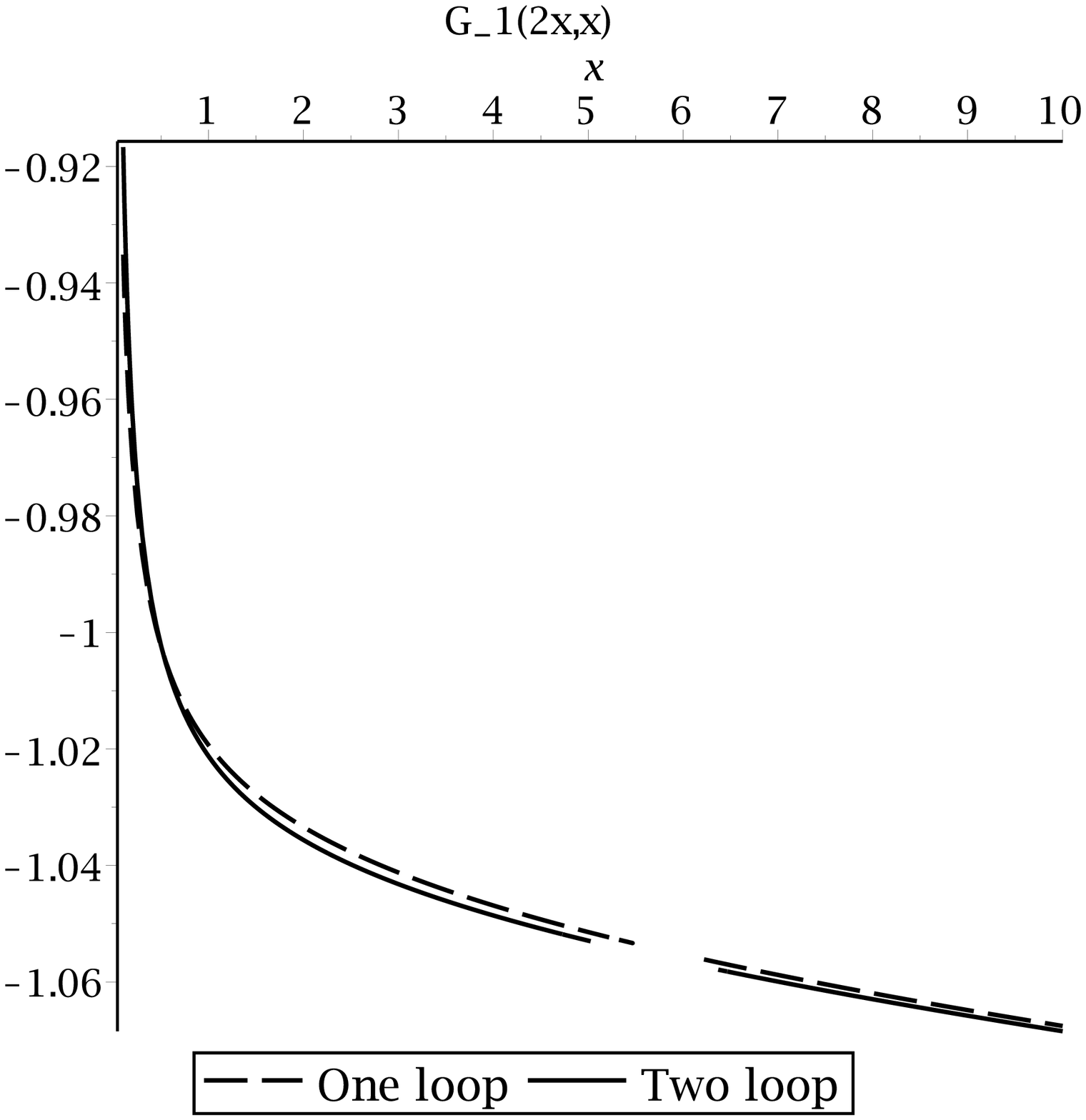}
\quad
\includegraphics[width=7.6cm,height=6cm]{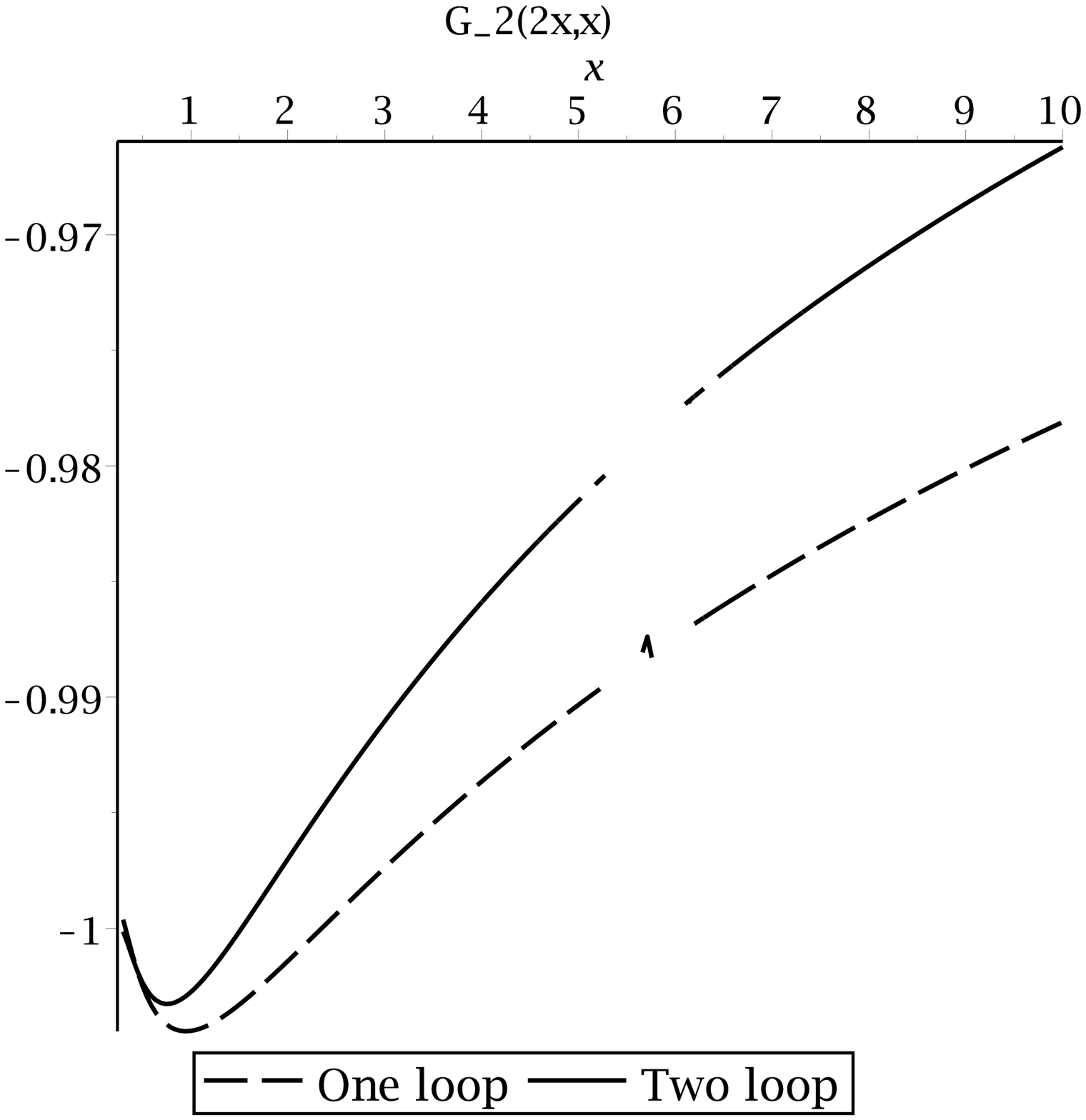}

\vspace{0.8cm}
\includegraphics[width=7.6cm,height=6cm]{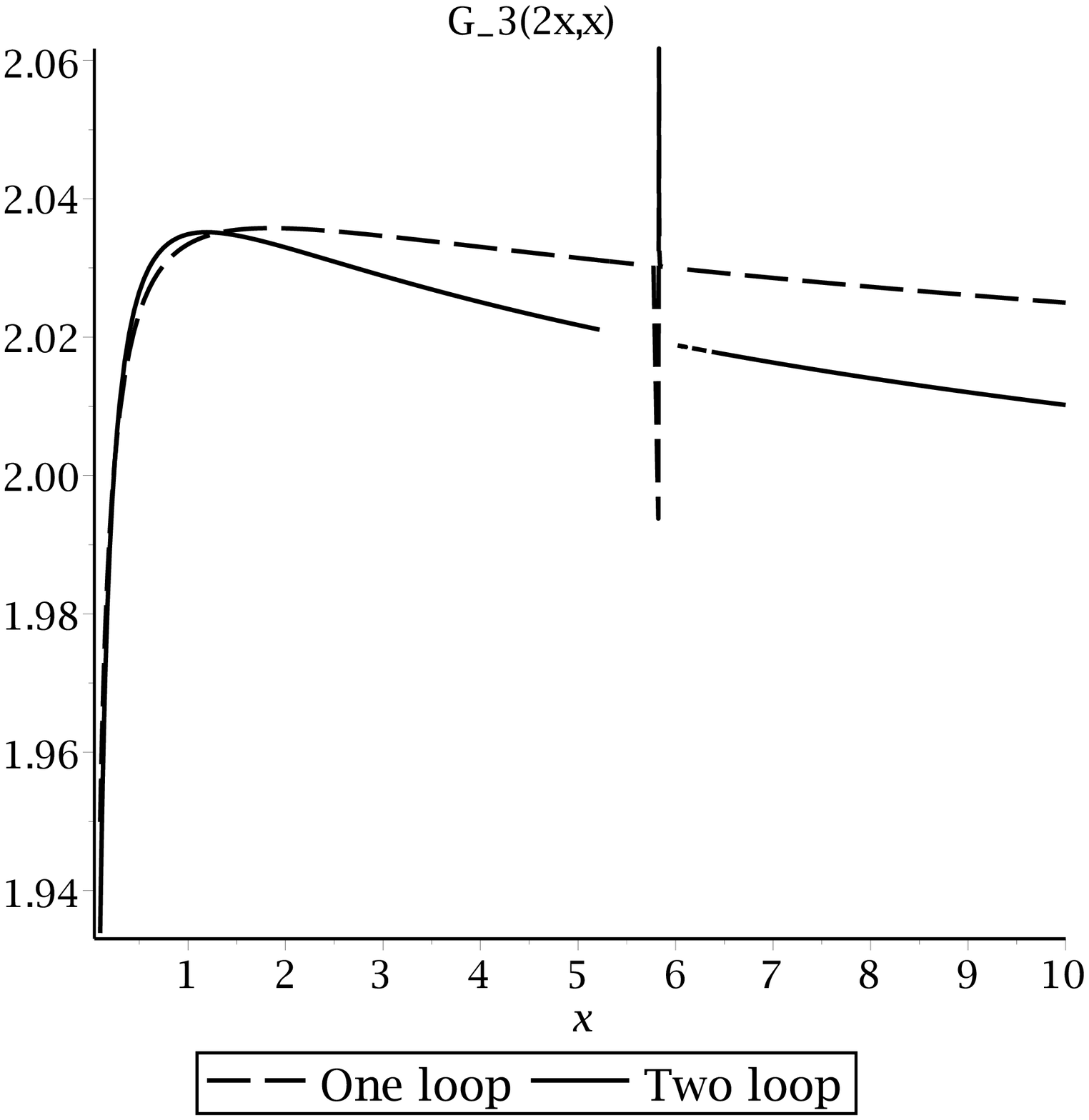}
\quad
\includegraphics[width=7.6cm,height=6cm]{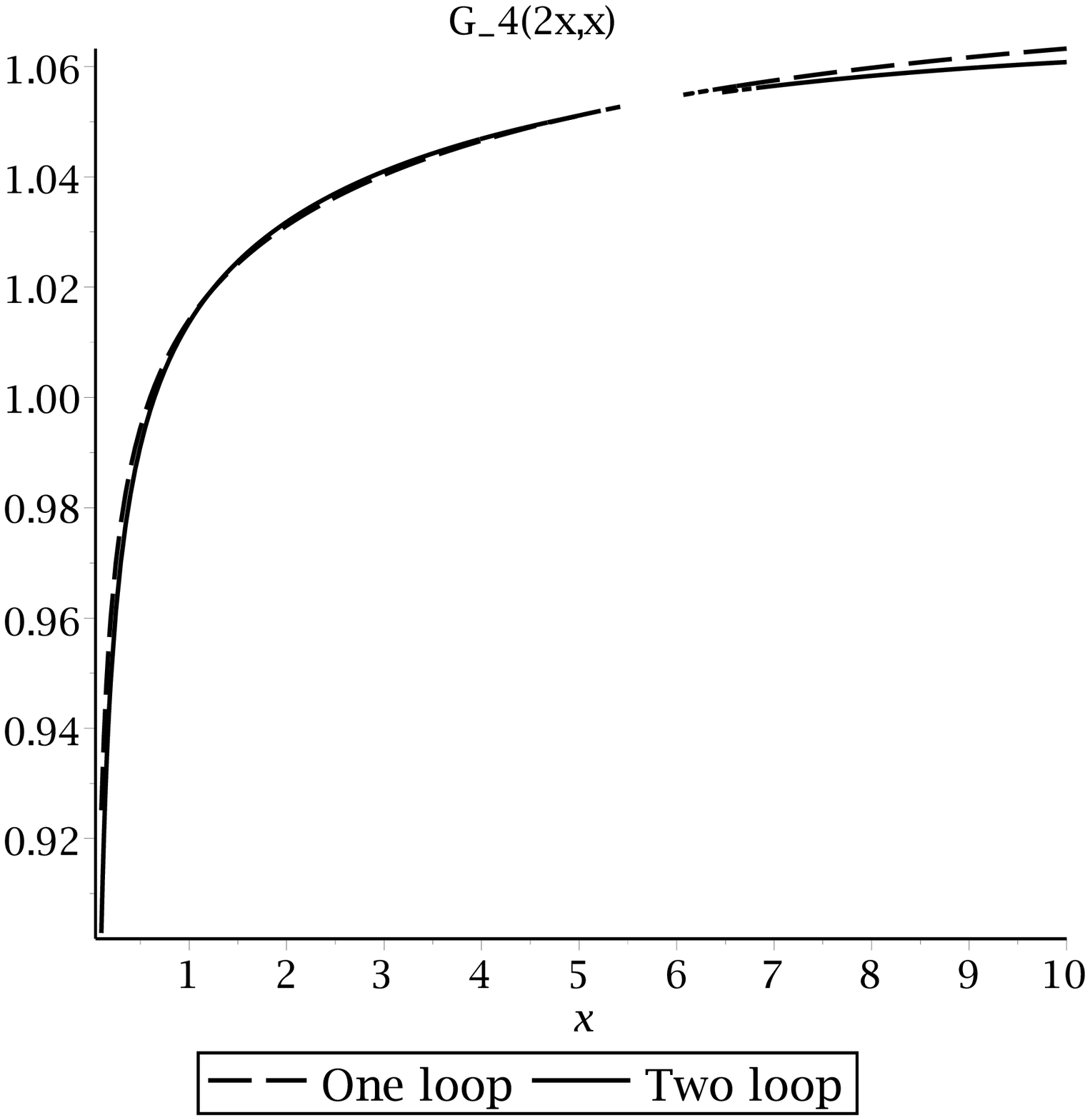}

\vspace{0.8cm}
\includegraphics[width=7.6cm,height=6cm]{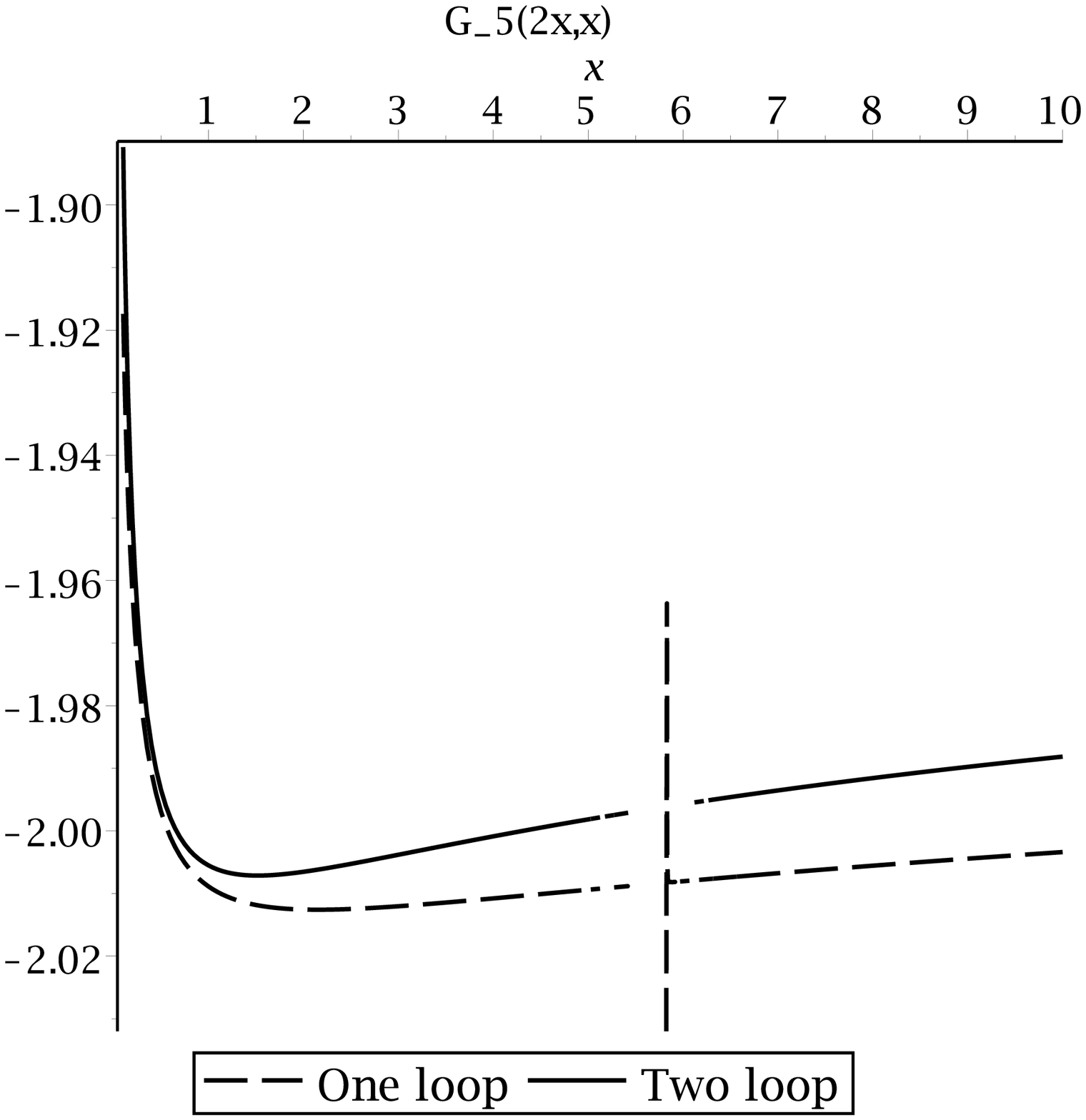}
\quad
\includegraphics[width=7.6cm,height=6cm]{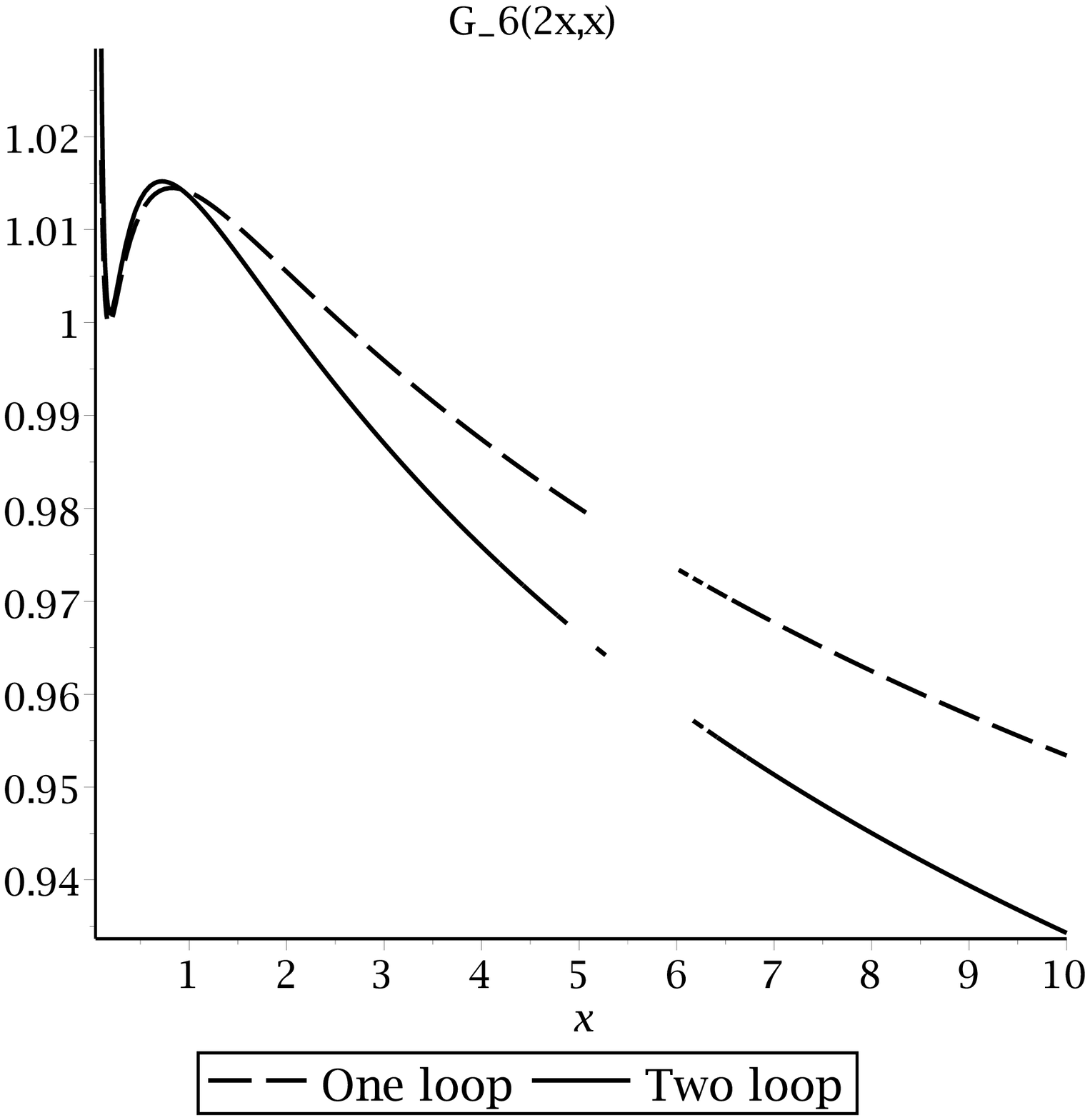}

\caption{Section $(2x,x)$ of various Landau gauge channels of the triple gluon 
vertex for $\alpha_s$~$=$~$0.125$.}
\end{figure}}

\sect{Discussion.}

We conclude with brief remarks. First, we have provided the full two loop 
off-shell vertex functions for the $3$-point vertices in QCD for a linear
covariant gauge fixing in the chiral limit. Given that recent lattice studies 
of these vertex functions are primarily to examine their behaviour in the low 
momentum regime, more precise knowledge of how the functions behave for various
external momentum configurations should prove useful for matching to high
energy behaviour and improve error analysis. As there is parallel interest in 
studying these functions using Schwinger-Dyson methods, the assumptions that 
are used for various vertex ans\"{a}tze can be explored in more detail beyond
the original and seminal work of \cite{5}. Indeed from the graphical
representations presented here it appears that the two loop corrections do not
significantly deviate from the one loop behaviour for a wide range of momenta 
scales at the particular value of the coupling constant considered. On the one 
hand this is expected from perturbation theory but one has also the potential 
now to explore at what values this breaks down. Further the channels 
corresponding to the Feynman rule of the vertex represent the dominant 
contribution. Though it is worth stressing that this is an observation based on
a two loop analysis, which should probably persist to higher orders, but it 
could be a justification for a vertex ansatz for Schwinger-Dyson analyses for 
this structure. However, it is not possible in our computation to isolate the 
quartic gluon vertex contribution and see what effect it has on the two loop 
corrections. Moreover, there is now scope to explore various definitions of 
effective strong coupling constants, and refine their running, which are based 
on the structure of $2$- and $3$-point functions and studied in, for example, 
\cite{6,7,8,9,10,11,12,13,14,15,16}. Given that our computations concentrated 
on the vertices in the chiral limit one natural extension would be to include 
massive quarks. Indeed that analysis has already been carried out at one loop 
for the quark-gluon and triple gluon vertices in \cite{21,25}. (The ghost-gluon
vertex has no quark contributions at one loop.) However, to extend those 
analyses to two loops in the completely off-shell situation would require 
knowledge of the necessary basic two loop massive master integrals. Currently 
these are not known in analytic form even for the symmetric point. In the 
interim the best that could be expected is to express the vertex functions as 
expansions in powers of the quark mass. Methods such as those developed in 
\cite{57,58} for $2$-point functions could be adapted to tackle such a problem.
However, that is beyond the scope of the current article.

\vspace{1cm}
\noindent
{\bf Acknowledgements.} The author thanks J.M. Bell and Dr M. Gorbahn for 
valuable discussions.

\appendix

\sect{Tensor basis.}

In this appendix we provide the explicit tensor bases and the projection
matrices for each vertex. As the latter are invariably quite involved, 
especially for the triple gluon vertex, we have compacted the expressions. We 
give the tensor bases for each of the vertex functions first with that for the 
ghost-gluon vertex being the two vectors 
\begin{equation}
{\cal P}^{\mbox{\footnotesize{ccg}}}_{(1) \sigma }(p,q) ~=~ p_\sigma ~~~,~~~
{\cal P}^{\mbox{\footnotesize{ccg}}}_{(2) \sigma }(p,q) ~=~ q_\sigma ~.
\end{equation} 
For the quark-gluon vertex we have 
\begin{eqnarray}
{\cal P}^{\mbox{\footnotesize{qqg}}}_{(1) \sigma }(p,q) &=& 
\gamma_\sigma ~~~,~~~
{\cal P}^{\mbox{\footnotesize{qqg}}}_{(2) \sigma }(p,q) ~=~ 
\frac{{p}_\sigma \pslash}{\mu^2} ~~~,~~~
{\cal P}^{\mbox{\footnotesize{qqg}}}_{(3) \sigma }(p,q) ~=~ 
\frac{{p}_\sigma \qslash}{\mu^2} ~, \nonumber \\
{\cal P}^{\mbox{\footnotesize{qqg}}}_{(4) \sigma }(p,q) &=& 
\frac{{q}_\sigma \pslash}{\mu^2} ~~~,~~~
{\cal P}^{\mbox{\footnotesize{qqg}}}_{(5) \sigma }(p,q) ~=~ 
\frac{{q}_\sigma \qslash}{\mu^2} ~~~,~~~
{\cal P}^{\mbox{\footnotesize{qqg}}}_{(6) \sigma }(p,q) ~=~ 
\frac{1}{\mu^2} \Gamma_{(3) \, \sigma p q} 
\end{eqnarray}
using the generalized $\gamma$-matrices. We use the convention that when a
vector is contracted with a Lorentz index on a $\gamma$-matrix, including
$\Gamma_{(n)}^{\mu_1\ldots\mu_n}$, then the vector appears in the corresponding
location in the $\gamma$-matrix string. Finally, there are fourteen independent
tensors for the triple gluon vertex which are
\begin{eqnarray}
{\cal P}^{\mbox{\footnotesize{ggg}}}_{(1) \mu \nu \sigma }(p,q) &=& 
\eta_{\mu \nu} p_\sigma ~~,~~ 
{\cal P}^{\mbox{\footnotesize{ggg}}}_{(2) \mu \nu \sigma }(p,q) ~=~ 
\eta_{\nu \sigma} p_\mu ~~,~~ 
{\cal P}^{\mbox{\footnotesize{ggg}}}_{(3) \mu \nu \sigma }(p,q) ~=~ 
\eta_{\sigma \mu} p_\nu \nonumber \\
{\cal P}^{\mbox{\footnotesize{ggg}}}_{(4) \mu \nu \sigma }(p,q) &=& 
\eta_{\mu \nu} q_\sigma ~~,~~ 
{\cal P}^{\mbox{\footnotesize{ggg}}}_{(5) \mu \nu \sigma }(p,q) ~=~ 
\eta_{\nu \sigma} q_\mu ~~,~~ 
{\cal P}^{\mbox{\footnotesize{ggg}}}_{(6) \mu \nu \sigma }(p,q) ~=~ 
\eta_{\sigma \mu} q_\nu \nonumber \\
{\cal P}^{\mbox{\footnotesize{ggg}}}_{(7) \mu \nu \sigma }(p,q) &=& 
\frac{1}{\mu^2} p_\mu p_\nu p_\sigma ~~,~~ 
{\cal P}^{\mbox{\footnotesize{ggg}}}_{(8) \mu \nu \sigma }(p,q) ~=~ 
\frac{1}{\mu^2} p_\mu p_\nu q_\sigma ~~,~~ 
{\cal P}^{\mbox{\footnotesize{ggg}}}_{(9) \mu \nu \sigma }(p,q) ~=~ 
\frac{1}{\mu^2} p_\mu q_\nu p_\sigma \nonumber \\ 
{\cal P}^{\mbox{\footnotesize{ggg}}}_{(10) \mu \nu \sigma }(p,q) &=& 
\frac{1}{\mu^2} q_\mu p_\nu p_\sigma ~~,~~ 
{\cal P}^{\mbox{\footnotesize{ggg}}}_{(11) \mu \nu \sigma }(p,q) ~=~ 
\frac{1}{\mu^2} p_\mu q_\nu q_\sigma ~~,~~ 
{\cal P}^{\mbox{\footnotesize{ggg}}}_{(12) \mu \nu \sigma }(p,q) ~=~ 
\frac{1}{\mu^2} q_\mu p_\nu q_\sigma \nonumber \\ 
{\cal P}^{\mbox{\footnotesize{ggg}}}_{(13) \mu \nu \sigma }(p,q) &=& 
\frac{1}{\mu^2} q_\mu q_\nu p_\sigma ~~,~~ 
{\cal P}^{\mbox{\footnotesize{ggg}}}_{(14) \mu \nu \sigma }(p,q) ~=~ 
\frac{1}{\mu^2} q_\mu q_\nu q_\sigma ~. 
\end{eqnarray}
In each of the three cases we have chosen $\mu$ as the common dimension scale
for simplicity and use the convention that all elements in a specific vertex 
basis have the same dimension. 

The projection matrices are straightforward to deduce once the basis tensors 
are defined. For the ghost-gluon case we have 
\begin{equation}
{\cal M}^{\mbox{\footnotesize{ccg}}} ~=~ \frac{1}{\Delta_G} \left(
\begin{array}{cc}
4 y & -~ 2 (1-x-y) \\
-~ 2 (1-x-y) & 4 x \\
\end{array}
\right) ~.
\end{equation}
To save space for the remaining cases we have chosen to define the matrices by
listing the individual elements since each projection matrix is symmetric. For
the quark-gluon vertex we define the intermediate matrix ${\cal Q}$ by 
\begin{equation}
{\cal M}^{\mbox{\footnotesize{qqg}}} ~=~ \frac{1}{4(d-2)\Delta_G^2} {\cal Q}
\end{equation}
so that the common factor is removed from the presentation of the elements. 
Thus we have 
\begin{eqnarray}
{\cal Q}_{11} &=& \left[x^2 - 2 x y - 2 x + y^2 - 2 y + 1 \right]^2 ~~,~~
{\cal Q}_{12} ~=~ -~ 4 \left[ (y - 1)^2 + x^2 - 2 (y + 1) x \right] y 
\nonumber \\
{\cal Q}_{13} &=& -~ 2 \left[ x^2 - 2 x y - 2 x + y^2 - 2 y + 1 \right] 
(x + y - 1) \nonumber \\
{\cal Q}_{14} &=& -~ 2 \left[ x^2 - 2 x y - 2 x + y^2 - 2 y + 1 \right]
(x + y - 1) \nonumber \\
{\cal Q}_{15} &=& -~ 4 \left[ (y - 1)^2 + x^2 - 2 (y + 1) x \right] x ~~,~~
{\cal Q}_{16} ~=~ 0 ~~,~~
{\cal Q}_{22} ~=~ 16 (d - 1) y^2 \nonumber \\
{\cal Q}_{23} &=& 8 (y - 1 + x) (d - 1) y ~~,~~
{\cal Q}_{24} ~=~ 8 (y - 1 + x) (d - 1) y \nonumber \\
{\cal Q}_{25} &=& -~ 4 \left[ 2 \left[(x - 2) x + (y - 1)^2 \right] 
- (x + y - 1)^2 d \right] ~~,~~
{\cal Q}_{26} ~=~ 0 \nonumber \\
{\cal Q}_{33} &=& 4 \left[ (y - 1)^2 + x^2 + 4 d x y - 2 (3 y + 1) x 
\right] ~~,~~
{\cal Q}_{34} ~=~ 4 (d - 1) (x + y - 1)^2 \nonumber \\
{\cal Q}_{35} &=& 8 (y - 1 + x) (d - 1) x ~~,~~
{\cal Q}_{36} ~=~ 0 ~~,~~
{\cal Q}_{44} ~=~ 4 \left[ (y - 1)^2 + x^2 + 4 d x y - 2 (3 y + 1) x 
\right] \nonumber \\
{\cal Q}_{45} &=& 8 (y - 1 + x) (d - 1) x ~~,~~
{\cal Q}_{46} ~=~ 0 ~~,~~
{\cal Q}_{55} ~=~ 16 (d - 1) x^2 ~~,~~
{\cal Q}_{56} ~=~ 0 \nonumber \\
{\cal Q}_{66} &=& 4 \left[ (y - 1)^2 + x^2 - 2 (y + 1) x \right] ~. 
\end{eqnarray}
Similarly defining
\begin{equation}
{\cal M}^{\mbox{\footnotesize{ggg}}} ~=~ \frac{1}{(d-2)\Delta_G^3} {\cal G}
\end{equation}
for the triple gluon vertex case the elements of the intermediate matrix 
${\cal G}$ are
\begin{eqnarray}
{\cal G}_{1 1} &=& 4 \left[ x^2 - 2 x y - 2 x + y^2 - 2 y + 1 \right]^2 y ~~,~~
{\cal G}_{1 2} ~=~ 0 ~~,~~
{\cal G}_{1 3} ~=~ 0 \nonumber \\
{\cal G}_{1 4} &=& 2 \left[ x^2 - 2 x y - 2 x + y^2 - 2 y + 1 \right]^2 
( x + y - 1 ) ~~,~~ 
{\cal G}_{1 5} ~=~ 0 ~~,~~ 
{\cal G}_{1 6} ~=~ 0 \nonumber \\
{\cal G}_{1 7} &=& -~ 16 \left[ ( y - 1 )^2 + x^2 
- 2 ( y + 1 ) x \right] y^2 
\nonumber \\
{\cal G}_{1 8} &=& -~ 8 \left[ x^2 - 2 x y - 2 x + y^2 - 2 y + 1 \right] 
( x + y - 1 ) y 
\nonumber \\
{\cal G}_{1 9} &=& -~ 8 \left[ x^2 - 2 x y - 2 x + y^2 - 2 y + 1 \right] 
( x + y - 1 ) y 
\nonumber \\
{\cal G}_{1 10} &=& -~ 8 \left[ x^2 - 2 x y - 2 x + y^2 - 2 y + 1 \right] 
( x + y - 1 ) y 
\nonumber \\
{\cal G}_{1 11} &=& -~ 4 \left[ x^2 - 2 x y - 2 x + y^2 - 2 y + 1 \right] 
( x + y - 1 )^2 
\nonumber \\
{\cal G}_{1 12} &=& -~ 4 \left[ x^2 - 2 x y - 2 x + y^2 - 2 y + 1 \right] 
( x + y - 1 )^2 
\nonumber \\
{\cal G}_{1 13} &=& -~ 16 \left[ ( y - 1 )^2 + x^2 
- 2 ( y + 1 ) x \right] x y 
\nonumber \\
{\cal G}_{1 14} &=& -~ 8 \left[ x^2 - 2 x y - 2 x + y^2 - 2 y + 1 \right] 
( x + y - 1 ) x 
\nonumber \\
{\cal G}_{2 2} &=& 4 \left[ x^2 - 2 x y - 2 x + y^2 - 2 y + 1 \right]^2 y ~~,~~
{\cal G}_{2 3} ~=~ 0 ~~,~~
{\cal G}_{2 4} ~=~ 0 
\nonumber \\
{\cal G}_{2 5} &=& 2 \left[ x^2 - 2 x y - 2 x + y^2 - 2 y + 1 \right]^2 
( x + y - 1 ) ~~,~~
{\cal G}_{2 6} ~=~ 0 
\nonumber \\
{\cal G}_{2 7} &=& -~ 16 \left[ ( y - 1 )^2 + x^2 
- 2 ( y + 1 ) x \right] y^2 
\nonumber \\
{\cal G}_{2 8} &=& -~ 8 \left[ x^2 - 2 x y - 2 x + y^2 - 2 y + 1 \right] 
( x + y - 1 ) y 
\nonumber \\
{\cal G}_{2 9} &=& -~ 8 \left[ x^2 - 2 x y - 2 x + y^2 - 2 y + 1 \right] 
( x + y - 1 ) y 
\nonumber \\
{\cal G}_{2 10} &=& -~ 8 \left[ x^2 - 2 x y - 2 x + y^2 - 2 y + 1 \right] 
( x + y - 1 ) y 
\nonumber \\
{\cal G}_{2 11} &=& -~ 16 \left[ ( y - 1 )^2 + x^2 
- 2 ( y + 1 ) x \right] x y 
\nonumber \\
{\cal G}_{2 12} &=& -~ 4 \left[ x^2 - 2 x y - 2 x + y^2 - 2 y + 1 \right] 
( x + y - 1 )^2 
\nonumber \\
{\cal G}_{2 13} &=& -~ 4 \left[ x^2 - 2 x y - 2 x + y^2 - 2 y + 1 \right] 
( x + y - 1 )^2 
\nonumber \\
{\cal G}_{2 14} &=& -~ 8 \left[ x^2 - 2 x y - 2 x + y^2 - 2 y + 1 \right] 
( x + y - 1 ) x 
\nonumber \\
{\cal G}_{3 3} &=& 4 \left[ x^2 - 2 x y - 2 x + y^2 - 2 y + 1 \right]^2 y ~~,~~
{\cal G}_{3 4} ~=~ 0 ~~,~~
{\cal G}_{3 5} ~=~ 0 
\nonumber \\
{\cal G}_{3 6} &=& 2 \left[ x^2 - 2 x y - 2 x + y^2 - 2 y + 1 \right]^2 
( x + y - 1 ) 
\nonumber \\
{\cal G}_{3 7} &=& -~ 16 \left[ ( y - 1 )^2 + x^2 - 2 
( y + 1 ) x \right] y^2 
\nonumber \\
{\cal G}_{3 8} &=& -~ 8 \left[ x^2 - 2 x y - 2 x + y^2 - 2 y + 1 \right] 
( x + y - 1 ) y 
\nonumber \\
{\cal G}_{3 9} &=& -~ 8 \left[ x^2 - 2 x y - 2 x + y^2 - 2 y + 1 \right] 
( x + y - 1 ) y 
\nonumber \\
{\cal G}_{3 10} &=& -~ 8 \left[ x^2 - 2 x y - 2 x + y^2 - 2 y + 1 \right] 
( x + y - 1 ) y 
\nonumber \\
{\cal G}_{3 11} &=& -~ 4 \left[ x^2 - 2 x y - 2 x + y^2 - 2 y + 1 \right] 
( x + y - 1 )^2 
\nonumber \\
{\cal G}_{3 12} &=& -~ 16 \left[ ( y - 1 )^2 + x^2 - 2 
( y + 1 ) x \right] x y 
\nonumber \\
{\cal G}_{3 13} &=& -~ 4 \left[ x^2 - 2 x y - 2 x + y^2 - 2 y + 1 \right] 
( x + y - 1 )^2 
\nonumber \\
{\cal G}_{3 14} &=& -~ 8 \left[ x^2 - 2 x y - 2 x + y^2 - 2 y + 1 \right] 
( x + y - 1 ) x 
\nonumber \\
{\cal G}_{4 4} &=& 4 \left[ x^2 - 2 x y - 2 x + y^2 - 2 y + 1 \right]^2 x ~~,~~
{\cal G}_{4 5} ~=~ 0 ~~,~~
{\cal G}_{4 6} ~=~ 0 
\nonumber \\
{\cal G}_{4 7} &=& -~ 8 \left[ x^2 - 2 x y - 2 x + y^2 - 2 y + 1 \right] 
( x + y - 1 ) y 
\nonumber \\
{\cal G}_{4 8} &=& -~ 16 \left[ ( y - 1 )^2 + x^2 - 2 
( y + 1 ) x \right] x y 
\nonumber \\
{\cal G}_{4 9} &=& -~ 4 \left[ x^2 - 2 x y - 2 x + y^2 - 2 y + 1 \right] 
( x + y - 1 )^2 
\nonumber \\
{\cal G}_{4 10} &=& -~ 4 \left[ x^2 - 2 x y - 2 x + y^2 - 2 y + 1 \right] 
( x + y - 1 )^2 
\nonumber \\
{\cal G}_{4 11} &=& -~ 8 \left[ x^2 - 2 x y - 2 x + y^2 - 2 y + 1 \right] ( x + y - 1 ) x 
\nonumber \\
{\cal G}_{4 12} &=& -~ 8 \left[ x^2 - 2 x y - 2 x + y^2 - 2 y + 1 \right] ( x + y - 1 ) x 
\nonumber \\
{\cal G}_{4 13} &=& -~ 8 \left[ x^2 - 2 x y - 2 x + y^2 - 2 y + 1 \right] ( x + y - 1 ) x 
\nonumber \\
{\cal G}_{4 14} &=& -~ 16 \left[ ( y - 1 )^2 + x^2 - 2 ( y + 1 ) x \right] x^2 
\nonumber \\
{\cal G}_{5 5} &=& 4 \left[ x^2 - 2 x y - 2 x + y^2 - 2 y + 1 \right]^2 x ~~,~~
{\cal G}_{5 6} ~=~ 0 
\nonumber \\
{\cal G}_{5 7} &=& -~ 8 \left[ x^2 - 2 x y - 2 x + y^2 - 2 y + 1 \right] ( x + y - 1 ) y 
\nonumber \\
{\cal G}_{5 8} &=& -~ 4 \left[ x^2 - 2 x y - 2 x + y^2 - 2 y + 1 \right] ( x + y - 1 )^2 
\nonumber \\
{\cal G}_{5 9} &=& -~ 4 \left[ x^2 - 2 x y - 2 x + y^2 - 2 y + 1 \right] ( x + y - 1 )^2 
\nonumber \\
{\cal G}_{5 10} &=& -~ 16 \left[ ( y - 1 )^2 + x^2 - 2 ( y + 1 ) x \right] x y 
\nonumber \\
{\cal G}_{5 11} &=& -~ 8 \left[ x^2 - 2 x y - 2 x + y^2 - 2 y + 1 \right] ( x + y - 1 ) x 
\nonumber \\
{\cal G}_{5 12} &=& -~ 8 \left[ x^2 - 2 x y - 2 x + y^2 - 2 y + 1 \right] ( x + y - 1 ) x 
\nonumber \\
{\cal G}_{5 13} &=& -~ 8 \left[ x^2 - 2 x y - 2 x + y^2 - 2 y + 1 \right] ( x + y - 1 ) x 
\nonumber \\
{\cal G}_{5 14} &=& -~ 16 \left[ ( y - 1 )^2 + x^2 - 2 ( y + 1 ) x \right] x^2 
\nonumber \\
{\cal G}_{6 6} &=& 4 \left[ x^2 - 2 x y - 2 x + y^2 - 2 y + 1 \right]^2 x 
\nonumber \\
{\cal G}_{6 7} &=& -~ 8 \left[ x^2 - 2 x y - 2 x + y^2 - 2 y + 1 \right] ( x + y - 1 ) y 
\nonumber \\
{\cal G}_{6 8} &=& -~ 4 \left[ x^2 - 2 x y - 2 x + y^2 - 2 y + 1 \right] ( x + y - 1 )^2 
\nonumber \\
{\cal G}_{6 9} &=& -~ 16 \left[ ( y - 1 )^2 + x^2 - 2 ( y + 1 ) x \right] x y 
\nonumber \\
{\cal G}_{6 10} &=& -~ 4 \left[ x^2 - 2 x y - 2 x + y^2 - 2 y + 1 \right] ( x + y - 1 )^2 
\nonumber \\
{\cal G}_{6 11} &=& -~ 8 \left[ x^2 - 2 x y - 2 x + y^2 - 2 y + 1 \right] ( x + y - 1 ) x 
\nonumber \\
{\cal G}_{6 12} &=& -~ 8 \left[ x^2 - 2 x y - 2 x + y^2 - 2 y + 1 \right] ( x + y - 1 ) x 
\nonumber \\
{\cal G}_{6 13} &=& -~ 8 \left[ x^2 - 2 x y - 2 x + y^2 - 2 y + 1 \right] ( x + y - 1 ) x 
\nonumber \\
{\cal G}_{6 14} &=& -~ 16 \left[ ( y - 1 )^2 + x^2 - 2 ( y + 1 ) x \right] x^2 
\nonumber \\
{\cal G}_{7 7} &=& 64 ( d + 1 ) y^3 ~~,~~
{\cal G}_{7 8} ~=~ 32 ( y - 1 + x ) ( d + 1 ) y^2 ~~,~~
{\cal G}_{7 9} ~=~ 32 ( y - 1 + x ) ( d + 1 ) y^2 
\nonumber \\
{\cal G}_{7 10} &=& 32 ( y - 1 + x ) ( d + 1 ) y^2 ~~,~~
{\cal G}_{7 11} ~=~ 16 \left[ ( x + y - 1 )^2 d + 4 x y \right] y 
\nonumber \\
{\cal G}_{7 12} &=& 16 \left[ ( x + y - 1 )^2 d + 4 x y \right] y ~~,~~
{\cal G}_{7 13} ~=~ 16 \left[ ( x + y - 1 )^2 d + 4 x y \right] y 
\nonumber \\
{\cal G}_{7 14} &=& -~ 8 \left[ 2 ( x^2 - 4 x y - 2 x + y^2 - 2 y + 1 ) 
- ( x + y - 1 )^2 d \right] ( x + y - 1 ) 
\nonumber \\
{\cal G}_{8 8} &=& 32 \left[ ( x - 2 ) x + ( y - 1 )^2 + 2 d x y \right] y ~~,~~
{\cal G}_{8 9} ~=~ 16 ( d + 1 ) ( x + y - 1 )^2 y 
\nonumber \\
{\cal G}_{8 10} &=& 16 ( d + 1 ) ( x + y - 1 )^2 y 
\nonumber \\
{\cal G}_{8 11} &=& 8 \left[ 4 d x y + x^2 + 2 x y - 2 x + y^2 - 2 y + 1 \right] ( x + y - 1 ) 
\nonumber \\
{\cal G}_{8 12} &=& 8 \left[ 4 d x y + x^2 + 2 x y - 2 x + y^2 - 2 y + 1 \right] ( x + y - 1 ) 
\nonumber \\
{\cal G}_{8 13} &=& 8 \left[ d x^2 + 2 d x y - 2 d x + d y^2 - 2 d y + d + 4 x y \right] ( x + y - 1 ) 
\nonumber \\
{\cal G}_{8 14} &=& 16 \left[ ( x + y - 1 )^2 d + 4 x y \right] x 
\nonumber \\
{\cal G}_{9 9} &=& 32 \left[ ( x - 2 ) x + ( y - 1 )^2 + 2 d x y \right] y ~~,~~
{\cal G}_{9 10} ~=~ 16 ( d + 1 ) ( x + y - 1 )^2 y 
\nonumber \\
{\cal G}_{9 11} &=& 8 \left[ 4 d x y + x^2 + 2 x y - 2 x + y^2 - 2 y + 1 \right] ( x + y - 1 ) 
\nonumber \\
{\cal G}_{9 12} &=& 8 \left[ d x^2 + 2 d x y - 2 d x + d y^2 - 2 d y + d + 4 x y \right] ( x + y - 1 ) 
\nonumber \\
{\cal G}_{9 13} &=& 8 \left[ 4 d x y + x^2 + 2 x y - 2 x + y^2 - 2 y + 1 \right] ( x + y - 1 ) 
\nonumber \\
{\cal G}_{9 14} &=& 16 \left[ ( x + y - 1 )^2 d + 4 x y \right] x ~~,~~
{\cal G}_{10 10} ~=~ 32 \left[ ( x - 2 ) x + ( y - 1 )^2 + 2 d x y \right] y 
\nonumber \\
{\cal G}_{10 10} &=& 8 \left[ d x^2 + 2 d x y - 2 d x + d y^2 - 2 d y + d + 4 x y \right] ( x + y - 1 ) 
\nonumber \\
{\cal G}_{10 10} &=& 8 \left[ 4 d x y + x^2 + 2 x y - 2 x + y^2 - 2 y + 1 \right] ( x + y - 1 ) 
\nonumber \\
{\cal G}_{10 10} &=& 8 \left[ 4 d x y + x^2 + 2 x y - 2 x + y^2 - 2 y + 1 \right] ( x + y - 1 ) 
\nonumber \\
{\cal G}_{10 10} &=& 16 \left[ ( x + y - 1 )^2 d + 4 x y \right] x ~~,~~
{\cal G}_{11 11} ~=~ 32 \left[ ( x - 2 + 2 d y ) x + ( y - 1 )^2 \right] x 
\nonumber \\
{\cal G}_{11 12} &=& 16 ( d + 1 ) ( x + y - 1 )^2 x ~~,~~
{\cal G}_{11 13} ~=~ 16 ( d + 1 ) ( x + y - 1 )^2 x 
\nonumber \\
{\cal G}_{11 14} &=& 32 ( y - 1 + x ) ( d + 1 ) x^2 ~~,~~
{\cal G}_{12 12} ~=~ 32 \left[ ( x - 2 + 2 d y ) x + ( y - 1 )^2 \right] x 
\nonumber \\
{\cal G}_{12 13} &=& 16 ( d + 1 ) ( x + y - 1 )^2 x ~~,~~
{\cal G}_{12 14} ~=~ 32 ( y - 1 + x ) ( d + 1 ) x^2 
\nonumber \\
{\cal G}_{13 13} &=& 32 \left[ ( x - 2 + 2 d y ) x + ( y - 1 )^2 \right] x ~~,~~
{\cal G}_{13 14} ~=~ 32 ( y - 1 + x ) ( d + 1 ) x^2 
\nonumber \\
{\cal G}_{14 14} &=& 64 ( d + 1 ) x^3 
\end{eqnarray}
which is much more involved. We have checked that the projection matrices of
the completely symmetric case, \cite{18}, emerge in the limit of 
$x$~$\rightarrow$~$1$ and $y$~$\rightarrow$~$1$.

\sect{Master integrals.}

In this appendix we define the various master integrals which emerge from the
application of the Laporta algorithm to the vertex functions. Several of these
are already known, \cite{30,31,59}, to the order in $\epsilon$ which is
required and for arbitrary external momentum squared. For instance, the two 
loop non-planar graph with unit propagators can be written in terms of two one
loop integrals. Therefore, we concentrate on those master integrals whose
$\epsilon$ expansion is required to an order beyond that considered in
\cite{30,31} and for the most general momentum configuration. In \cite{35}
several of these master integrals were recorded but for a less general case.
Though we have used a similar method to establish sufficient terms of the
$\epsilon$ expansion in order to write the full expressions for our vertex
functions in terms of known functions. First, we define the basic one loop
vertex function $I(\alpha,\beta,\gamma)$ by
\begin{equation}
I(\alpha,\beta,\gamma) ~=~ \int_k 
\frac{1}{(k^2)^\alpha((k-p)^2)^\beta((k+q)^2)^\gamma} 
\label{Iint}
\end{equation}
where $\int_k$~$=$~$d^dk/(2\pi)^d$. When any two of $\{\alpha,\beta,\gamma\}$
are zero or negative then the integral vanishes. If only one of the powers is 
zero or negative then one has the basic bubble integral. The master integrals
related to the latter are $I(1,1,0)$, $I(1,0,1)$ and $I(0,1,1)$. We have
highlighted all three as their $\epsilon$ expansions are different due to the
appearance of $\ln(x)$ for example in the first case. In the other non-bubble 
cases the Laporta algorithm produces $I(1,1,1)$ as the one loop master 
integral. In this case we have, \cite{30,31,59},
\begin{equation}
I(1,1,1) ~=~ -~ \frac{1}{\mu^2} \left[ \Phi_1(x,y) + \Psi_1(x,y) \epsilon 
+ \left[ \frac{\zeta(2)}{2} \Phi_1(x,y) + \chi_1(x,y) \right] 
\epsilon^2 ~+~ O(\epsilon^3) \right] ~. 
\label{I111}
\end{equation}
The first two terms of (\ref{I111}) correspond to the same terms of 
\cite{30,31}. We use a modified notation, though, where we label via subscript 
without parentheses. Also when comparing to \cite{30,31,59} we note that our 
convention defining the order of the powers on the left side of (\ref{Iint}) is
different. At $O(\epsilon^2)$ we have introduced the intermediate function
$\chi_1(x,y)$ which will be discussed later in the context of two loop masters.
Given that $I(1,1,1)$ has a large degree of symmetry one can construct various
relations between the functions which are used within our computations. We have
\begin{eqnarray}
\Phi_1\left(\frac{y}{x},\frac{1}{x}\right) &=& x \Phi_1(x,y) ~~,~~
\Phi_1\left(\frac{1}{y},\frac{x}{y}\right) ~=~ y \Phi_1(x,y) \nonumber \\
\Psi_1\left(\frac{y}{x},\frac{1}{x}\right) &=& x \left[ \Psi_1(x,y) 
+ \ln(x) \Phi_1(x,y) \right] ~~,~~
\Psi_1\left(\frac{1}{y},\frac{x}{y}\right) ~=~ 
y \left[ \Psi_1(x,y) + \ln(y) \Phi_1(x,y) \right] \nonumber \\
\chi_1\left(\frac{y}{x},\frac{1}{x}\right)
&=& x \left[ \chi_1(x,y) + \ln(x) \Psi_1(x,y) + \frac{1}{2} \ln^2(x) 
\Phi_1(x,y) \right] \nonumber \\
\chi_1\left(\frac{1}{y},\frac{x}{y}\right)
&=& y \left[ \chi_1(x,y) + \ln(y) \Psi_1(x,y) + \frac{1}{2} \ln^2(y) 
\Phi_1(x,y) \right] 
\label{funrel}
\end{eqnarray}
which can be deduced by rotating the original integral $I(1,1,1)$.

At two loops aside from the $5$-propagator master integrals of \cite{30,31}
which are already known to the order we require, and the basic two loop 
$2$-point sunset graph, it is the $4$-propagator master integrals which require
extra attention. There are two types of these and they are related to
(\ref{Iint}) when one line has a bubble subgraph. For completeness we record
each rotation of the two types. First, we have 
\begin{eqnarray}
I\left(3-\frac{d}{2},1,1\right) &=& -~ \left[ \Phi_1(x,y) 
+ \left[ \Psi_1(x,y) - \frac{1}{2} \ln(x) \Phi_1(x,y) 
- \frac{1}{2} \ln(y) \Phi_1(x,y) \right] \epsilon \right. \nonumber \\
&& \left. ~~~~+
\left[ \frac{\zeta(2)}{2} \Phi_1(x,y) + \chi_3(x,y) \right] \epsilon^2 
\right] \frac{1}{\mu^2} ~+~ O(\epsilon^3) \nonumber \\
I\left(1,3-\frac{d}{2},1\right) &=& -~ \left[
\Phi_1\left(\frac{1}{y},\frac{x}{y}\right) \right. \nonumber \\
&& \left. ~~~~
+ \left[
-~ \frac{1}{2} \ln(x) \Phi_1\left(\frac{1}{y},\frac{x}{y}\right)
- \ln(y) \Phi_1\left(\frac{1}{y},\frac{x}{y}\right)
+ \Psi_1\left(\frac{1}{y},\frac{x}{y}\right)
\right] \epsilon \right. \nonumber \\ 
&& \left. ~~~~ 
+ \left[
\ln(x) \ln(y) \Phi_1\left(\frac{1}{y},\frac{x}{y}\right)
- 2 \ln(y) \Psi_1\left(\frac{1}{y},\frac{x}{y}\right)
+ \frac{\zeta(2)}{2} \Phi_1\left(\frac{1}{y},\frac{x}{y}\right) \right. \right.
\nonumber \\ 
&& \left. \left. ~~~~~~~~
+ \chi_3\left(\frac{1}{y},\frac{x}{y}\right)
\right] \epsilon^2 ~+~ O(\epsilon^3) \right] 
\frac{1}{y\mu^2} \nonumber \\
I\left(1,1,3-\frac{d}{2}\right) &=& -~ \left[
\Phi_1\left(\frac{y}{x},\frac{1}{x}\right) \right. \nonumber \\
&& \left. ~~~~
+ \left[
- \ln(x) \Phi_1\left(\frac{y}{x},\frac{1}{x}\right)
- \frac{1}{2} \ln(y) \Phi_1\left(\frac{y}{x},\frac{1}{x}\right)
+ \Psi_1\left(\frac{y}{x},\frac{1}{x}\right)
\right] \epsilon \right. \nonumber \\
&& \left. ~~~~
+ \left[
\ln(x) \ln(y) \Phi_1\left(\frac{y}{x},\frac{1}{x}\right)
- 2 \ln(x) \Psi_1\left(\frac{y}{x},\frac{1}{x}\right)
+ \frac{\zeta(2)}{2} \Phi_1\left(\frac{y}{x},\frac{1}{x}\right)
\right. \right. \nonumber \\
&& \left. \left. ~~~~~~~~
+ \chi_3\left(\frac{y}{x},\frac{1}{x}\right)
\right] \epsilon^2 ~+~ O(\epsilon^3) \right] \frac{1}{x\mu^2}
\end{eqnarray}
and 
\begin{eqnarray}
I\left(2-\frac{d}{2},1,1\right) &=&
\frac{1}{2\epsilon} + \frac{3}{2} - \ln(x) \nonumber \\
&& + \left[ 
-~ \frac{1}{2} \Phi_1\left(\frac{y}{x},\frac{1}{x}\right) \frac{y}{x}
- \frac{1}{2} \Phi_1\left(\frac{y}{x},\frac{1}{x}\right) \frac{1}{x}
- \frac{\pi^2}{24} 
+ \frac{1}{2} \ln^2(x)
+ \frac{1}{2} \ln(x) \ln(y)
\right. \nonumber \\
&& \left. ~~~\,
+ \frac{1}{2} \Phi_1\left(\frac{y}{x},\frac{1}{x}\right)
+ \frac{9}{2}
- 3 \ln(x)
\right] \epsilon
\nonumber \\
&& + \left[
-~ \frac{3}{2} \Phi_1\left(\frac{y}{x},\frac{1}{x}\right) \frac{y}{x}
- \frac{3}{2} \Phi_1\left(\frac{y}{x},\frac{1}{x}\right) \frac{1}{x}
- \frac{\zeta(3)}{6}
- \frac{\pi^2}{8} 
+ \frac{\pi^2}{12} \ln(x) 
\right. \nonumber \\
&& \left. ~~~\,
+ \frac{1}{3} \ln^3(x)
+ \frac{1}{2} \chi_2\left(\frac{y}{x},\frac{1}{x}\right)
+ \frac{3}{2} \ln^2(x)
+ \frac{3}{2} \ln(x) \ln(y)
+ \frac{3}{2} \Phi_1\left(\frac{y}{x},\frac{1}{x}\right)
\right. \nonumber \\
&& \left. ~~~\,
+ \frac{27}{2}
- 9 \ln(x)
- \ln^2(x) \ln(y)
- \ln(x) \Phi_1\left(\frac{y}{x},\frac{1}{x}\right)
+ \ln(x) \Phi_1\left(\frac{y}{x},\frac{1}{x}\right) \frac{y}{x}
\right. \nonumber \\
&& \left. ~~~\,
+ \ln(x) \Phi_1\left(\frac{y}{x},\frac{1}{x}\right) \frac{1}{x}
\right] \epsilon^2
~+~ O(\epsilon^3) 
\nonumber \\
I\left(1,2-\frac{d}{2},1\right) &=&
\frac{1}{2\epsilon} + \frac{3}{2} - \ln(y) \nonumber \\
&& + \left[ 
-~ \frac{1}{2} \Phi_1\left(\frac{1}{y},\frac{x}{y}\right) \frac{x}{y}
- \frac{1}{2} \Phi_1\left(\frac{1}{y},\frac{x}{y}\right) \frac{1}{y}
- \frac{\pi^2}{24} 
+ \frac{1}{2} \ln(x) \ln(y)
+ \frac{1}{2} \ln^2(y)
\right. \nonumber \\
&& \left. ~~~\,
+ \frac{1}{2} \Phi_1\left(\frac{1}{y},\frac{x}{y}\right)
+ \frac{9}{2}
- 3 \ln(y)
\right] \epsilon
\nonumber \\
&& + \left[ 
-~ \frac{3}{2} \Phi_1\left(\frac{1}{y},\frac{x}{y}\right) \frac{x}{y}
- \frac{3}{2} \Phi_1\left(\frac{1}{y},\frac{x}{y}\right) \frac{1}{y}
- \frac{\zeta(3)}{6}
- \frac{\pi^2}{8} 
+ \frac{\pi^2}{12} \ln(y) 
\right. \nonumber \\
&& \left. ~~~\,
+ \frac{1}{3} \ln^3(y)
+ \frac{1}{2} \chi_2\left(\frac{1}{y},\frac{x}{y}\right)
+ \frac{3}{2} \ln(x) \ln(y)
+ \frac{3}{2} \ln^2(y)
+ \frac{3}{2} \Phi_1\left(\frac{1}{y},\frac{x}{y}\right)
\right. \nonumber \\
&& \left. ~~~\,
+ \frac{27}{2}
- 9 \ln(y)
- \ln(x) \ln^2(y)
- \ln(y) \Phi_1\left(\frac{1}{y},\frac{x}{y}\right)
+ \ln(y) \Phi_1\left(\frac{1}{y},\frac{x}{y}\right) \frac{x}{y}
\right. \nonumber \\
&& \left. ~~~\,
+ \ln(y) \Phi_1\left(\frac{1}{y},\frac{x}{y}\right) \frac{1}{y}
\right] \epsilon^2
~+~ O(\epsilon^3) 
\nonumber \\
I\left(1,1,2-\frac{d}{2}\right) &=&
\frac{1}{2\epsilon} + \frac{3}{2} - \ln(x) \nonumber \\
&& + \left[ 
-~ \frac{1}{2} \Phi_1\left(\frac{y}{x},\frac{1}{x}\right) \frac{y}{x}
- \frac{1}{2} \Phi_1\left(\frac{y}{x},\frac{1}{x}\right) \frac{1}{x}
- \frac{\pi^2}{24} 
+ \frac{1}{2} \ln^2(x)
+ \frac{1}{2} \ln(x) \ln(y)
\right. \nonumber \\
&& \left. ~~~\,
+ \frac{1}{2} \Phi_1\left(\frac{y}{x},\frac{1}{x}\right)
+ \frac{9}{2}
- 3 \ln(x)
\right] \epsilon
\nonumber \\
&& + \left[
-~ \frac{3}{2} \Phi_1\left(\frac{y}{x},\frac{1}{x}\right) \frac{y}{x}
- \frac{3}{2} \Phi_1\left(\frac{y}{x},\frac{1}{x}\right) \frac{1}{x}
- \frac{\zeta(3)}{6}
- \frac{\pi^2}{8} 
+ \frac{\pi^2}{12} \ln(x) 
\right. \nonumber \\
&& \left. ~~~\,
+ \frac{1}{3} \ln^3(x)
+ \frac{1}{2} \chi_2\left(\frac{y}{x},\frac{1}{x}\right)
+ \frac{3}{2} \ln^2(x)
+ \frac{3}{2} \ln(x) \ln(y)
+ \frac{3}{2} \Phi_1\left(\frac{y}{x},\frac{1}{x}\right)
\right. \nonumber \\
&& \left. ~~~\,
+ \frac{27}{2}
- 9 \ln(x)
- \ln^2(x) \ln(y)
- \ln(x) \Phi_1\left(\frac{y}{x},\frac{1}{x}\right)
+ \ln(x) \Phi_1\left(\frac{y}{x},\frac{1}{x}\right) \frac{y}{x}
\right. \nonumber \\
&& \left. ~~~\,
+ \ln(x) \Phi_1\left(\frac{y}{x},\frac{1}{x}\right) \frac{1}{x}
\right] \epsilon^2
~+~ O(\epsilon^3) ~. 
\end{eqnarray}
In each set we have included a new intermediate function, $\chi_2(x,y)$ and
$\chi_3(x,y)$. For the former we were able to determine it in terms of known
functions by extending the method given in \cite{35} but for a less general
momentum configuration. This was based on the irreducible two loop non-planar 
graph of \cite{31} which related the result to the known function 
$\Omega_2(x,y)$. By using the Laporta algorithm on the same irreducible graph 
we were able to relate the intermediate function $\chi_2(x,y)$ to 
$\Omega_2(x,y)$. In \cite{35} another similar intermediate function, 
$\Xi^{(1)}(x,y)$, in the notation of \cite{35}, was introduced. To assist the 
interested reader we record the relation for $\chi_2(x,y)$ in terms of this and
the more general relation to that given in \cite{35} which determines 
$\chi_2(x,y)$ in terms of known functions. We have 
\begin{eqnarray}
\chi_2(x,y) &=& 2 \left[ -~ 8 \zeta(3) 
+ \frac{\pi^2}{6} \ln(x)
- \frac{2}{3} \ln^3(x) 
- \ln(x) \Phi_1(x,y) 
+ y \ln(x) \Phi_1(x,y) \right. \nonumber \\
&& \left. ~~~ 
+ x \ln(x) \Phi_1(x,y)
+ \frac{\pi^2}{6} \ln(y) 
- \frac{2}{3} \ln^3(y)
- \ln(y) \Phi_1(x,y) 
+ y \ln(y) \Phi_1(x,y)
\right. \nonumber \\
&& \left. ~~~
+ x \ln(y) \Phi_1(x,y) 
+ 2 \Psi_1(x,y) 
- 2 y \Psi_1(x,y) 
- 2 x \Psi_1(x,y) 
+ \frac{1}{2} \Xi^{(1)}(x,y) \right]
\end{eqnarray}
with 
\begin{eqnarray}
\Xi^{(1)}(x,y) &=& 
14 \zeta(3) - \frac{\pi^2}{3} \ln(x) + \frac{4}{3} \ln^3(x) 
+ \frac{1}{2} \ln^2(x) \ln(y) + \frac{1}{2} \ln(x) \ln^2(y) 
+ \frac{3}{2} \ln(x) \Phi_1(x,y) \nonumber \\
&& -~ \frac{3}{2} \ln(x) \Phi_1(x,y) x
- \frac{3}{2} \ln(x) \Phi_1(x,y) y - \frac{\pi^2}{3} \ln(y) 
+ \frac{4}{3} \ln^3(y) + \frac{3}{2} \ln(y) \Phi_1(x,y) \nonumber \\
&& -~ \frac{3}{2} \ln(y) \Phi_1(x,y) x 
- \frac{3}{2} \ln(y) \Phi_1(x,y) y 
- \frac{3}{4} \Omega_2\left(\frac{y}{x},\frac{1}{x}\right)
+ \frac{1}{4} \Omega_2\left(\frac{1}{x},\frac{y}{x}\right) \nonumber \\
&& -~ \frac{3}{4} \Omega_2\left(\frac{x}{y},\frac{1}{y}\right)
+ \frac{1}{4} \Omega_2\left(\frac{1}{y},\frac{x}{y}\right) 
+ \frac{1}{4} \Omega_2(x,y) + \frac{1}{4} \Omega_2(y,x) - 3 \Psi_1(x,y)
\nonumber \\
&& +~ 3 \Psi_1(x,y) x + 3 \Psi_1(x,y) y ~.
\end{eqnarray}
This reduces to the expression given in \cite{35} for that specific momentum
configuration. For our other intermediate functions $\chi_1(x,y)$ and 
$\chi_3(x,y)$ we were not able to evaluate them in terms of known functions. 
However, it transpires that within the computation of the vertex functions they
always appear in one combination which is their difference. In this case we 
managed to determine this in terms of known functions from \cite{30,31}. 
Briefly the method involves considering the $\epsilon$ expansion of the Feynman
parameter representation of say $I(1,1,1)$ and $I(3-d/2,1,1)$ and exploiting 
various properties of $I(1,1,1)$ noted in \cite{30,31}. This produces 
\begin{equation}
\chi_3(x,y) ~=~ \chi_1(x,y) + \Phi_2(x,y) - \frac{1}{2} \ln(xy) \Psi_1(x,y) 
+ \frac{1}{4} \left[ \ln^2(x) + \ln^2(y) \right] \Phi_1(x,y)
\end{equation}
which we have included within our computational setup. Finally, we note that in
\cite{55} a similar situation arose at the fully symmetric point in that a 
particular combination of harmonic polylogarithms always appeared in the final 
vertex functions. It is interesting that for the more general case here the
difference in the corresponding functions could be determined. 

Finally, for completeness we record the explicit forms of each of the functions
which appear in the final vertex functions and were derived in \cite{30,31}. 
Using the same notation as \cite{30,31}, at one loop we have
\begin{equation}
\Phi_1(x,y) ~=~ \frac{1}{\lambda} \left[ 2 \mbox{Li}_2(-\rho x)
+ 2 \mbox{Li}_2(-\rho y)
+ \ln \left( \frac{y}{x} \right)
\ln \left( \frac{(1+\rho y)}{(1+\rho x)} \right)
+ \ln(\rho x) \ln(\rho y) + \frac{\pi^2}{3} \right] ~.
\end{equation}
The functions $\lambda(x,y)$ and $\rho(x,y)$ are defined by, \cite{30,31},
\begin{equation}
\lambda(x,y) ~=~ \sqrt{\Delta_G} ~~~,~~~
\rho(x,y) ~=~ \frac{2}{[1-x-y+\lambda(x,y)]} 
\end{equation}
where the former equation connects notation. At two loops the two main
functions are, \cite{30,31},
\begin{eqnarray}
\Phi_2(x,y) &=& \frac{1}{\lambda} \left[ 6 \mbox{Li}_4(-\rho x)
+ 6 \mbox{Li}_4(-\rho y)
+ 3 \ln \left( \frac{y}{x} \right)
\left[ \mbox{Li}_3(-\rho x) - \mbox{Li}_3(-\rho y) \right] \right. \nonumber \\
&& \left. ~~~
+ \frac{1}{2} \ln^2 \left( \frac{y}{x} \right)
\left[ \mbox{Li}_2(-\rho x) + \mbox{Li}_2(-\rho y) \right] 
+ \frac{1}{4} \ln^2(\rho x) \ln^2(\rho y) 
\right. \nonumber \\
&& \left. ~~~
+ \frac{\pi^2}{2} \ln(\rho x) \ln(\rho y) 
+ \frac{\pi^2}{12} \ln^2 \left( \frac{y}{x} \right)
+ \frac{7\pi^4}{60} \right] 
\end{eqnarray}
and
\begin{eqnarray}
\Omega_2(x,y) &=& 6 \mbox{Li}_3(-\rho x) + 6 \mbox{Li}_3(-\rho y)
+ 3 \ln \left( \frac{y}{x} \right)
\left[ \mbox{Li}_2(-\rho x) - \mbox{Li}_2(-\rho y) \right] \nonumber \\
&& -~ \frac{1}{2} \ln^2 \left( \frac{y}{x} \right)
\left[ \ln(1+\rho x) + \ln(1+\rho y) \right] 
\nonumber \\
&& +~ \frac{1}{2} \left[ \pi^2 + \ln(\rho x) \ln(\rho y) \right]
\left[ \ln(\rho x) + \ln(\rho y) \right] ~. 
\end{eqnarray}
The other function which appears in the above master integrals, $\Psi_1(x,y)$, 
is not present in the final vertex functions. We reproduce it here, 
\cite{30,31},
\begin{eqnarray}
\Psi_1(x,y) &=& -~ \frac{1}{\lambda} \left[ 
4 \mbox{Li}_3 \left( - \frac{\rho x(1+\rho y)}{(1-\rho^2 xy)} \right) 
+ 4 \mbox{Li}_3 \left( - \frac{\rho y(1+\rho x)}{(1-\rho^2 xy)} \right) 
- 4 \mbox{Li}_3 \left( - \frac{xy \rho^2}{(1-\rho^2 xy)} \right) 
\right. \nonumber \\
&& \left. ~~~~~~~
+ 2 \mbox{Li}_3 \left( \frac{x \rho(1+\rho y)}{(1+\rho x)} \right) 
+ 2 \mbox{Li}_3 \left( \frac{y \rho(1+\rho x)}{(1+\rho y)} \right) 
- 2 \mbox{Li}_3 ( \rho^2 xy ) 
- 2 \zeta(3)
\right. \nonumber \\
&& \left. ~~~~~~~
- 2 \ln (y) \mbox{Li}_2 \left( \frac{x \rho(1+\rho y)}{(1+\rho x)} \right) 
- 2 \ln (x) \mbox{Li}_2 \left( \frac{y \rho(1+\rho x)}{(1+\rho y)} \right) 
- \frac{2}{3} \ln^3 \left( 1-\rho^2 xy \right)
\right. \nonumber \\
&& \left. ~~~~~~~
+ \frac{2}{3} \ln^3 \left( 1+\rho x \right)
+ \frac{2}{3} \ln^3 \left( 1+\rho y \right)
+ 2 \ln(\rho) \ln^2 \left( 1-\rho^2 x y \right)
\right. \nonumber \\
&& \left. ~~~~~~~
- 2 \ln(1-\rho^2 xy ) \left[ \ln(\rho x) \ln(\rho y)
+ \ln \left( \frac{y}{x} \right)
\ln \left( \frac{(1+\rho y)}{(1+\rho x)} \right)
\right. \right. \nonumber \\
&& \left. \left. ~~~~~~~~~~~~~~~~~~~~~~~~~~~~~~
+ 2 \ln(1+\rho x) \ln(1+\rho y) + \frac{\pi^2}{3} \right] 
\right. \nonumber \\
&& \left. ~~~~~~~
+ \frac{1}{2} \ln \left( xy \rho^2 \right) \left[ \ln(\rho x) \ln(\rho y)
+ \ln \left( \frac{y}{x} \right)
\ln \left( \frac{(1+\rho y)}{(1+\rho x)} \right)
- \ln^2 \left( \frac{(1+ \rho x)}{(1+\rho y)} \right) 
\right. \right. \nonumber \\
&& \left. \left. ~~~~~~~~~~~~~~~~~~~~~~~~~~~
+ \frac{2\pi^2}{3} 
\right] \right] ~. 
\end{eqnarray}
This has a considerably more involved structure compared to $\Phi_1(x,y)$,
$\Phi_2(x,y)$ and $\Omega_2(x,y)$.


\begin{thebibliography}{99}
\bibitem{1} C.D. Roberts \& A.G. Williams, Prog. Part. Nucl. Phys. {\bf 33} 
(1994), 477.
%%CITATION = HEP-PH 9403224;%%
\bibitem{2} R. Alkofer \& L. von Smekal, Phys. Rept. {\bf 353} (2001), 281. 
%%CITATION = HEP-PH 0007355;%%
\bibitem{3} P. Maris \& C.D. Roberts, Int. J. Mod. Phys. {\bf E12} (2003), 297.
%%CITATION = NUCL-PH 0301049;%%
\bibitem{4} W. Celmaster \& R.J. Gonsalves, Phys. Rev. {\bf D20} (1979), 1420.
%%CITATION = PHRVA,D20,1420;%% 
\bibitem{5} J.S. Ball \& T.W. Chiu, Phys. Rev. {\bf D22} (1980), 2550; Phys. 
Rev. {\bf D23} (1981), 3085.
%%CITATION = PHRVA,D22,2550;%% 
\bibitem{6} A. Cucchieri \& T. Mendes, PoS LAT2007 (2007), 297.
%%CITATION = 0710.0412;%%
\bibitem{7} I.L. Bogolubsky, E.M. Ilgenfritz, M. M\"{u}ller-Preussker \& A.
Sternbeck, PoS LAT2007 (2007), 290.
%%CITATION = 0710.1968;%%
\bibitem{8} A. Maas, Phys. Rev. {\bf D75} (2007), 116004.
%%CITATION = 0704.0722;%%
\bibitem{9} A. Sternbeck, L. von Smekal, D.B. Leinweber \& A.G. Williams,
PoS LAT2007 (2007), 304.
%%CITATION = 0710.1982;%%
\bibitem{10} I.L. Bogolubsky, E.M. Ilgenfritz, M. M\"{u}ller-Preussker \& A.
Sternbeck, Phys. Lett. {\bf B676} (2009), 69.
%%CITATION = 0901.0736;%%
\bibitem{11} A. Cucchieri \& T. Mendes, Phys. Rev. Lett. {\bf 100} (2008),
241601.
%%CITATION = 0712.3517;%%
\bibitem{12} A. Cucchieri \& T. Mendes, Phys. Rev. {\bf D78} (2008), 094503.
%%CITATION = 0804.2371;%%
\bibitem{13} O. Oliveira \& P.J. Silva, Phys. Rev. {\bf D79} (2009), 031501.
%%CITATION = 0809.0258;%%
\bibitem{14} Ph. Boucaud, J.P. Leroy, A.L. Yaounac, J. Micheli, O. P\`{e}ne \&
J. Rodr\'{\i}guez-Quintero, JHEP {\bf 0806} (2008), 099.
%%CITATION = 0803.2161;%%
\bibitem{15} C.S. Fischer, A. Maas \& J.M. Pawlowski, Annals Phys. {\bf 324}
(2009), 2408.
%%CITATION = 0810.1987;%%
\bibitem{16} G. 't Hooft, Nucl. Phys. {\bf B61} (1973), 455.
%%CITATION = NUPHA,B197,477;%%
\bibitem{17} W.A. Bardeen, A.J. Buras, D.W. Duke \& T. Muta, Phys. Rev.
{\bf D18} (1978), 3998.
%%CITATION = PHRVA,D18,3998;%%
\bibitem{18} J.A. Gracey, Phys. Rev. {\bf D84} (2011), 085011.
%%CITATION = 1108.4806;%%
\bibitem{19} P. Pascual \& R. Tarrach, Nucl. Phys. {\bf B174} (1980), 123;
Nucl. Phys. {\bf B181} (1981), 546.
%%CITATION = NUPHA,B174,123;%% 
\bibitem{20} A.I. Davydychev, P. Osland \& O.V. Tarasov, Phys. Rev. {\bf D54} 
(1996), 4087; Phys. Rev. {\bf D59} (1999), 109901(E).
%%CITATION = HEP-PH 9605348;%%
\bibitem{21} A.I. Davydychev, P. Osland \& L. Saks, JHEP {\bf 0108} (2001), 
050.
%%CITATION = HEP-PH 0105072;%%
\bibitem{22} M. Binger \& S.J. Brodsky, Phys. Rev. {\bf D74} (2006), 054016.
%%CITATION = HEP-PH 0602199;%%
\bibitem{23} A.I. Davydychev, P. Osland \& O.V. Tarasov, Phys. Rev. {\bf D58} 
(1998), 036007.
%%CITATION = HEP-PH 9801380;%%
\bibitem{24} A.I. Davydychev \& P. Osland, Phys. Rev. {\bf D59} (1998), 014006.
%%CITATION = HEP-PH 9806522;%%
\bibitem{25} A.I. Davydychev, P. Osland \& L. Saks, Phys. Rev. {\bf D63} 
(2000), 014022.
%%CITATION = HEP-PH 0008171;%%
\bibitem{26} K.G. Chetyrkin \& A. R\'{e}tey, hep-ph/0007088.
%%CITATION = HEP-PH 0007088;%%
\bibitem{27} K.G. Chetyrkin \& T. Seidensticker, Phys. Lett. {\bf B495} (2000),
74.
%%CITATION = HEP-PH 0008094;%%
\bibitem{28} N. Ahmadiniaz \& C. Schubert, Nucl. Phys. {\bf B869} (2013), 417.
%%CITATION = 1210.2331;%%
\bibitem{29} S. Laporta, Int. J. Mod. Phys. {\bf A15} (2000), 5087.
%%CITATION = HEP-PH 0207004;%%
\bibitem{30} N.I. Ussyukina \& A.I. Davydychev, Phys. Atom. Nucl. {\bf 56}
(1993), 1553.
%%CITATION = HEP-PH 9307327;%%
\bibitem{31} N.I. Ussyukina \& A.I. Davydychev, Phys. Lett. {\bf B332} (1994),
159.
%%CITATION = HEP-PH 9402223;%%
\bibitem{32} E. Remiddi \& J.A.M. Vermaseren, Int. J. Mod. Phys. {\bf A15}
(2000), 725.
%%CITATION = HEP-PH 9905237;%%
\bibitem{33} T.G. Birthwright, E.W.N. Glover \& P. Marquard, JHEP {\bf 0409}
(2004), 042.
%%CITATION = HEP-PH 0407343;%%
\bibitem{34} J. Ablinger, J. Bl\"{u}mlein \& C. Schneider, J. Math. Phys.
{\bf 52} (2011), 102301.
%%CITATION = 1105.6063;%%
\bibitem{35} M. Gorbahn \& S. J\"{a}ger, Phys. Rev. {\bf D82} (2010), 114001.
%%CITATION = 1004.3997;%%
\bibitem{36} A.L. Blum, M.Q. Huber, M. Mitter \& L. von Smekal, Phys. Rev. 
{\bf D89} (2014), 061703.
%%CITATION = 1401.0713;%%
\bibitem{37} G. Eichmann, R. Williams, R. Alkofer \& M. Vujinovic,
Phys. Rev. {\bf D89} (2014), 105014.
%%CITATION = 1402.1365;%%
\bibitem{38} A.D. Kennedy, J. Math. Phys. {\bf 22} (1981), 1330.
%%CITATION = JMAPA,22,1330;%%
\bibitem{39} A. Bondi, G. Curci, G. Paffuti \& P. Rossi, Ann. Phys. {\bf 199}
(1990), 268.
%%CITATION = APNYA,199,268;%%
\bibitem{40} A.N. Vasil'ev, S.\'{E}. Derkachov \& N.A. Kivel, Theor. Math.
Phys. {\bf 103} (1995), 487.
%%CITATION = TMPHA,103,487;%%
\bibitem{41} A.N. Vasil'ev, M.I. Vyazovskii, S.\'{E}. Derkachov \& N.A. Kivel,
Theor. Math. Phys. {\bf 107} (1996), 441.
%%CITATION = TMPHA,107,441;%%
\bibitem{42} A.N. Vasil'ev, M.I. Vyazovskii, S.\'{E}. Derkachov \& N.A. Kivel,
Theor. Math. Phys. {\bf 107} (1996), 710.
%%CITATION = TMPHA,107,710;%%
\bibitem{43} C. Studerus, Comput. Phys. Commun. {\bf 181} (2010), 1293.
%%CITATION = 0912.2546;%%
\bibitem{44} C.W. Bauer, A. Frink \& R. Kreckel, cs/0004015.
%%CITATION = CS 0004015;%%
\bibitem{45} J.A.M. Vermaseren, math-ph/0010025. 
%%CITATION = MATH-PH 0010025;%%
\bibitem{46} M. Tentyukov \& J.A.M. Vermaseren, Comput. Phys. Commun. {\bf 181}
(2010), 1419.
%%CITATION = HEP-PH 0702279;%%
\bibitem{47} P. Nogueira, J. Comput. Phys. {\bf 105} (1993), 279. 
%%CITATION = JCTPA,105,279;%% 
\bibitem{48} S.A. Larin \& J.A.M. Vermaseren, Phys. Lett. {\bf B303} (1993),
334.
%%CITATION = PHLTA,B303,334;%%
\bibitem{49} D.J. Gross \& F.J. Wilczek, Phys. Rev. Lett. {\bf 30}
(1973), 1343.
%%CITATION = PRLTA,30,1343;%%
\bibitem{50} H.D. Politzer, Phys. Rev. Lett. {\bf 30} (1973), 1346.
%%CITATION = PRLTA,30,1346;%%
\bibitem{51} W.E. Caswell, Phys. Rev. Lett. {\bf 33} (1974), 244.
%%CITATION = PRLTA,33,244;%%
\bibitem{52} D.R.T. Jones, Nucl. Phys. {\bf B75} (1974), 531.
%%CITATION = NUPHA,B75,531;%%
\bibitem{53} O.V. Tarasov \& A.A. Vladimirov, Sov. J. Nucl. Phys. {\bf 25}
(1977), 585.
%%CITATION = SJNCA,25,585;%%
\bibitem{54} E. Egorian \& O.V. Tarasov, Theor. Math. Phys. {\bf 41} (1979), 
863.
%%CITATION = TMPHA,41,863;%%
\bibitem{55} L.G. Almeida \& C. Sturm, Phys. Rev. {\bf D82} (2010), 054017.
%%CITATION = 1004.4613;%%
\bibitem{56} J.M. Bell, unpublished.
\bibitem{57} A.I. Davydychev, V.A. Smirnov \& J.B. Tausk, Nucl. Phys.
{\bf B410} (1993) 325.
%%CITATION = HEP-PH/9307371;%%
\bibitem{58} F.A. Behrends, A.I. Davydychev, V.A. Smirnov \& J.B. Tausk, Nucl. 
Phys. {\bf B439} (1995), 536.
%%CITATION = HEP-PH/9410232;%%
\bibitem{59} A.I. Davydychev, J. Phys. {\bf A25} (1992), 5587.
%%CITATION = JPAGB,A25,5587;%%
\end{thebibliography}
\end{document}